\documentclass[longauth]{aa}  %{emulateapj}  [,referee]  referee referee

\newcommand*{\aas}{\ensuremath{A\&A}}

\newcommand*{\mass}{\ensuremath{M_\star\ }}
\newcommand*{\msun}{\ensuremath{M_\odot\ }}

\newcommand*{\PaA}{Pa\ensuremath{\alpha}}
\newcommand*{\PaB}{Pa\ensuremath{\beta}}
\newcommand*{\PaG}{Pa\ensuremath{\gamma}}
\newcommand*{\Ha}{H\ensuremath{\alpha}}
\newcommand*{\Hb}{H\ensuremath{\beta}}

\newcommand*{\PaD}{Pa\ensuremath{\delta}}
\newcommand*{\BrG}{Br\ensuremath{\gamma}}

\newcommand*{\xFeIIa}{[Fe\,{\scshape ii}]\,\ensuremath{\lambda1.257 \mu m}}

\newcommand*{\xFeIIc}{[Fe\,{\scshape ii}]\,\ensuremath{\lambda1.64 \mu m}}
\newcommand*{\xHtwo}{H$_2$\,\ensuremath{\lambda2.121 \mu m}}  % 2.12125
\newcommand*{\Htwo}{H$_2$}

\newcommand*{\xFeII}{[Fe\,{\scshape ii}]\,\ensuremath{\lambda1.257 \mu m}}

\newcommand*{\xSII}{[S\,{\scshape ii}]\,\ensuremath{\lambda\lambda6717\text{-}6731}}
\newcommand*{\xOII}{[O\,{\scshape ii}]\,\ensuremath{\lambda\lambda3726\text{-}3729}}

\newcommand*{\xOIII}{[O\,{\scshape iii}]\,\ensuremath{\lambda5007 \AA}}
\newcommand*{\xSIII}{[S\,{\scshape iii}]\,\ensuremath{\lambda9530 \AA}}

\newcommand*{\xNII}{[N\,{\scshape ii}]\,\ensuremath{\lambda6583 \AA}}
\newcommand*{\xPII}{[P\,{\scshape ii}]\,\ensuremath{\lambda1.19 \mu m }}

\newcommand*{\xCI}{[C\,{\scshape i}]\,\ensuremath{\lambda9850 \AA }}

\newcommand*{\PII}{[P\,{\scshape ii}]}

\newcommand*{\SIII}{[S\,{\scshape iii}]}
\newcommand*{\SII}{[S\,{\scshape ii}]}
\newcommand*{\FeII}{[Fe\,{\scshape ii}]}

\newcommand*{\OIII}{[O\,{\scshape iii}]}
\newcommand*{\OII}{[O\,{\scshape ii}]}
\newcommand*{\OI}{[O\,{\scshape i}]}
\newcommand*{\CI}{[C\,{\scshape i}]}
\newcommand*{\NII}{[N\,{\scshape ii}]}
\newcommand*{\SiVI}{[Si\,{\scshape vi}]}
\newcommand*{\HeI}{He\,{\scshape i}}

\newcommand{\ffrac}[2]{\ensuremath{\frac{\displaystyle #1}{\displaystyle #2}}}

\usepackage{graphicx}

\usepackage[colorlinks=true,
            linkcolor=red,
            urlcolor=red,
            citecolor=blue]{hyperref}

\usepackage{txfonts}

\newcommand{\myemail}{antonello.calabro@inaf.it}

\usepackage{multirow,array}
\usepackage[normalem]{ulem}

\usepackage{amsmath,bm}

\usepackage{tikz,xcolor,hyperref}

\definecolor{lime}{HTML}{A6CE39}
\DeclareRobustCommand{\orcidicon}{%
    \begin{tikzpicture}
    \draw[lime, fill=lime] (0,0) 
    circle [radius=0.16] 
    node[white] {{\fontfamily{qag}\selectfont \tiny ID}};
    \draw[white, fill=white] (-0.0625,0.095) 
    circle [radius=0.007];
    \end{tikzpicture}
    \hspace{-2mm}
}

\foreach \x in {A, ..., Z}{%
    \expandafter\xdef\csname orcid\x\endcsname{\noexpand\href{https://orcid.org/\csname orcidauthor\x\endcsname}{\noexpand\orcidicon}}
}

\begin{document}

%\setlength{\parskip}{0.1pt}


\title{Near-infrared emission line diagnostics for AGN from the local Universe to $z \sim 3$}

\author{
%Antonello Calabr{\`o}\orcidA{}\inst{1}
Antonello Calabr{\`o}\inst{1}
\and Laura Pentericci\inst{1}
\and Anna Feltre\inst{2}
\and Pablo Arrabal Haro\inst{3}
\and Mario Radovich\inst{4}
\and Lise-Marie Seill{\'e}\inst{5}
\and Ernesto Oliva\inst{2}
\and Emanuele Daddi\inst{6}
\and Ricardo Amor{\'i}n\inst{7,8}
\and Micaela B. Bagley\inst{9}
\and Laura Bisigello\inst{4,10}
\and V{\'e}ronique Buat\inst{5}
\and Marco Castellano\inst{1}
\and Nikko J. Cleri\inst{11,12}
\and Mark Dickinson\inst{3}
\and Vital Fern\'{a}ndez\inst{7}
\and Steven L. Finkelstein\inst{9}
\and Mauro Giavalisco\inst{13}
\and Andrea Grazian\inst{4}
\and Nimish P. Hathi\inst{14}
\and Michaela Hirschmann\inst{15}
\and St{\'e}phanie Juneau\inst{3}
\and Jeyhan S. Kartaltepe\inst{16}
\and Anton M. Koekemoer\inst{14}
\and Ray A. Lucas\inst{14}
\and Casey Papovich\inst{17,18}
\and Pablo G. P{\'e}rez-Gonz{\'a}lez\inst{19}
\and Nor Pirzkal\inst{20}
\and Paola Santini\inst{1}
\and Jonathan Trump\inst{21}
\and Alexander de la Vega\inst{22}
\and Stephen M. Wilkins\inst{23,24}
\and L. Y. Aaron {Yung}\inst{25,26}
\and Paolo Cassata\inst{10}
\and Raphael A. S. Gobat\inst{27}
\and Sara Mascia\inst{1}
\and Lorenzo Napolitano\inst{1}
\and Benedetta Vulcani\inst{4}
}

% Antonello Calabrò, Laura Pentericci, Anna Feltre, Pablo Arrabal Haro, Mario Radovich, Lise Marie Seillé, Ernesto Oliva, Emanuele Daddi, Ricardo Amorín, Micaela B. Bagley, Laura Bisigello, Véronique Buat, Marco Castellano, Nikko Cleri, Mark Dickinson, Vital Fernández, Steven Finkelstein, Mauro Giavalisco, Andrea Grazian, Nimish Hathi, Michaela Hirschmann, Stéphanie Juneau, Jeyhan S. Kartaltepe, Anton Koekemoer, Ray A. Lucas, Casey Papovich, Pablo Pérez-González, Nor Pirzkal, Paola Santini, Jonathan Trump, Alexander de la Vega, Stephen Wilkins, L.Y. Aaron Yung, Paolo Cassata, Raphael Gobat, Sara Mascia, Lorenzo Napolitano, Benedetta Vulcani 
% A. Calabrò, L. Pentericci, A. Feltre, P. Arrabal Haro, M. Radovich, L. Seillé, E. Oliva, E. Daddi, R. Amorín, M. Bagley, L. Bisigello, V. Buat, M. Castellano, N. Cleri, M. Dickinson, V. Fernández, S. Finkelstein, M. Giavalisco, A. Grazian, N. Hathi, M. Hirschmann, S. Juneau, J. Kartaltepe, A. Koekemoer, R. A. Lucas, C. Papovich, P. Pérez-González, N. Pirzkal, P. Santini, J. Trump, A. de la Vega, S. Wilkins, L.Y. A. Yung, P. Cassata, R. Gobat, S. Mascia, L. Napolitano, B. Vulcani 

\institute{INAF - Osservatorio Astronomico di Roma, via Frascati 33, 00078, Monte Porzio Catone, Italy (\myemail)
\and INAF - Osservatorio Astrofisico di Arcetri, Largo E. Fermi 5, I-50125, Firenze, Italy % 2
\and NSF's National Optical-Infrared Astronomy Research Laboratory, 950 N. Cherry Ave., Tucson, AZ 85719, USA % 3
\and INAF - Osservatorio Astronomico di Padova, Vicolo dell'Osservatorio 5, I-35122, Padova, Italy % 4
\and Aix Marseille Univ, CNRS, CNES, LAM Marseille, France % 5
\and Universit{\'e} Paris-Saclay, Universit{\'e} Paris Cit{\'e}, CEA, CNRS, AIM, 91191, Gif-sur-Yvette, France % 6
\and Instituto de Investigaci\'{o}n Multidisciplinar en Ciencia y Tecnolog\'{i}a, Universidad de La Serena, Raul Bitr\'{a}n 1305, La Serena 2204000, Chile % 7
\and Departamento de Astronom\'{i}a, Universidad de La Serena, Av. Juan Cisternas 1200 Norte, La Serena 1720236, Chile % 8
\and Department of Astronomy, The University of Texas at Austin, Austin, TX, USA % 9
\and Dipartimento di Fisica e Astronomia ``G.Galilei'', Universit\'a di Padova, Via Marzolo 8, I-35131 Padova, Italy % 10
\and Department of Physics and Astronomy, Texas A\&M University, College Station, TX, 77843-4242 USA % 11
\and George P.\ and Cynthia Woods Mitchell Institute for Fundamental Physics and Astronomy, Texas A\& M University, College Station, TX, 77843-4242 USA % 12
\and University of Massachusetts Amherst, 710 North Pleasant Street, Amherst, MA 01003-9305, USA % 13
\and Space Telescope Science Institute, 3700 San Martin Dr., Baltimore, MD 21218, USA % 14
\and Institute of Physics, Laboratory of Galaxy Evolution, Ecole Polytechnique F\'{e}d\'{e}rale de Lausanne (EPFL), Observatoire de Sauverny, 1290 Versoix, Switzerland % 15
% \and NSF's NOIRLab, 950 N. Cherry Ave., Tucson, AZ 85719, USA % 15 
\and Laboratory for Multiwavelength Astrophysics, School of Physics and Astronomy, Rochester Institute of Technology, 84 Lomb Memo-
rial Drive, Rochester, NY 14623, USA % 16
\and Department of Physics and Astronomy, Texas A\&M University, College Station, TX, 77843-4242 USA % 17
\and George P. and Cynthia Woods Mitchell Institute for Fundamental Physics and Astronomy, Texas A\&M University, College Station, TX, 77843-4242 USA % 18
\and Centro de Astrobiolog\'{\i}a (CAB), CSIC-INTA, Ctra. de Ajalvir km 4, Torrej\'on de Ardoz, E-28850, Madrid, Spain % 19
\and ESA/AURA Space Telescope Science Institute % 20
\and Department of Physics, 196 Auditorium Road, Unit 3046, University of Connecticut, Storrs, CT 06269, USA % 21
\and Department of Physics and Astronomy, University of California, 900 University Ave, Riverside, CA 92521, USA % 22
\and Astronomy Centre, University of Sussex, Falmer, Brighton BN1 9QH, UK % 23
\and Institute of Space Sciences and Astronomy, University of Malta, Msida MSD 2080, Malta % 24
\and NASA Postdoctoral Fellow % 25
\and Astrophysics Science Division, NASA Goddard Space Flight Center, 8800 Greenbelt Rd, Greenbelt, MD 20771, USA % 26
\and Instituto de F\'{i}sica, Pontificia Universidad Cat\'{o}lica de Valpara\'{i}so, Casilla, 4059, Valpara\'{i}so, Chile % 27
}
\date{Submitted to A\&A}
%\and 18, INAF-Osservatorio Astronomico di Roma, via di Frascati 33, 00078, Monte Porzio Catone, Italy

\abstract 
{
Optical rest-frame spectroscopic diagnostics are usually employed to distinguish between star formation and AGN-powered emission. However, this method is biased against dusty sources, hampering a complete census of the AGN population across cosmic epochs. To mitigate this effect, it is crucial to observe at longer wavelengths in the rest-frame near-infrared (near-IR), which is less affected by dust attenuation and can thus provide a better description of the intrinsic properties of galaxies. AGN diagnostics in this regime have not been fully exploited so far, due to the scarcity of near-infrared observations of both AGNs and star-forming galaxies, especially at redshifts higher than $0.5$.
Using Cloudy photoionization models, we identify new AGN - star formation diagnostics based on the ratio of bright near-infrared emission lines, namely [SIII] $9530$\AA, [CI] $9850$\AA, [PII] $1.188 \mu m$, [FeII] $1.257 \mu m$, and [FeII] $1.64 \mu m$ to Paschen lines (either Pa$\gamma$ or Pa$\beta$), providing simple, analytical classification criteria. % With the [FeII]/Pa$\beta$ and [FeII]/[PII] ratios we also control for the contribution of shocks. 
We apply these diagnostics to a sample of $64$ star-forming galaxies and AGNs at $0\leq z \leq 1$, and $65$ sources at $1\leq z \leq 3$ recently observed with JWST-NIRSpec in CEERS. 
We find that the classification inferred from the near-infrared is broadly consistent with the optical one based on the BPT and the [SII]/H$\alpha$ ratio. However, in the near-infrared, we find $\sim 60 \%$ more AGNs than in the optical ($13$ instead of $8$), with $5$ sources classified as 'hidden' AGNs, showing a larger AGN contribution at longer wavelengths, possibly due to the presence of optically thick dust. 
The diagnostics we present provide a promising tool to find and characterize AGNs from $z=0$ to $z\simeq3$ with low and medium-resolution near-IR spectrographs in future surveys.
}


\titlerunning{\footnotesize near-IR rest-frame AGN diagnostics}
\authorrunning{A.Calabr\`o et al.}
 \maketitle
\section{Introduction}\label{introduction}

Optical spectroscopy of galaxies has been for decades one of the most efficient tools for identifying Active Galactic Nuclei (AGN) in the Universe. 
This has been possible thanks to a combination of favorable factors. 
First, optical spectroscopy gives access to the rest-frame Ultraviolet (UV) to optical emission of galaxies from $z=0$ to $z\simeq6$. This spectral range is full of bright emission lines sensitive to the properties of the ionizing sources, and that are exploited to distinguish between ionization driven by an AGN or by young O and B stars (see \Citealt{kewley19} for a review). 
The best-known star-formation vs AGN diagnostic is the BPT diagram, first introduced by \citet{baldwin81}, which informs the physical nature of the emitting source by comparing the \NII/\Ha\ and \OIII/\Hb\ emission line ratios. Similar diagnostics adopt \SII/\Ha\ or \OI/\Ha\ instead of the ionized nitrogen \citep{veilleux87,kewley01}.
A second reason is represented by the higher multiplexity and sensitivity reached by optical spectrographs compared to instruments sensitive to other regions of the electromagnetic spectrum. Finally, the atmosphere is mostly transparent in the visible band and thus the observations can be performed mostly from the ground. 

Spectroscopy also has the advantage of differentiating the spectral type of AGN and measuring its level of activity, distinguishing between Type 2 and Type 1. 
The first are characterized by narrow (FWHM $< 1000$ km/s) emission lines coming from the Narrow Line Region (NLR), and identified through the BPT and similar diagrams. Type 1, in addition to AGN-like narrow emission line ratios, also show broad component (FWHM $> 1000$ km/s) permitted lines. These are generated in gas clouds orbiting closer to the accretion disk and dubbed as Broad Line Region (BLR), which in Type 2 AGNs are hidden by the dusty torus (\Citealt{antonucci93}, but see also \Citealt{elitzur14} for alternative explanations).

These advantages have led in the past decades to find statistical samples of spectroscopically confirmed Type 2 and Type 1 AGNs, both in the local Universe, thanks to the SDSS survey \citep[e.g.,][]{kauffmann03}, and at higher redshifts, thanks to large spectroscopic campaigns conducted with multi-object spectrographs (MOS) from the ground, like zCOSMOS with VIMOS at the VLT \citep{mignoli13,mignoli19}, the FMOS survey at Subaru \citep{kartaltepe15,kashino19}, the MOSDEF survey with MOSFIRE at Keck \citep{coil15}, and with slitless grism spectroscopy with the Hubble Space Telescope \citep{trump11,backhaus22}. 

Despite all these efforts, the census of AGNs from optical observations is not complete. Indeed, most theoretical models predict a black-hole accretion rate density that is higher by up to one order of magnitude compared to observational measurements (e.g., \Citealt{vito18} and references therein) at all redshifts from $0$ to $\sim7$, thus suggesting that we are missing a significant population of optically obscured AGNs. 
One physical reason for this is that the rest-frame UV and optical emission might still be severely affected by dust attenuation. Indeed, the accretion disk onto the supermassive black-hole (SMBH), powering the AGN emission, typically lies in the central part of a galaxy where the dust attenuation might be higher \citep{goddard17,wang17,greener20,yung21}. 
This effect is more severe for more massive SMBHs, which are hosted in more massive and dustier galaxies. For example, A$_{UV}$ is of the order of $\geq 3$ magnitudes in galaxies with M$_\star > 10^{10}$ M$_\odot$ \citep{pannella15}, and can also reach A$_{UV}\simeq 15$ mag in more luminous AGNs according to \citet{yung21}, with optical lines being highly suppressed by factors of $5$ to $1000$ assuming the \citet{calzetti00} law. 
There might be also systems in which the dust geometry is resembling that of a mixed model, that is, a homogeneous mixture of dust and emitting gas \citep{calzetti96}, rather than the typical dust-screen model. In this mixed configuration, we can find A$_V$ rising to 30 mag toward the center, as observed in local Ultra Luminous Infrared galaxies (ULIRGs) and massive starbursts at $0.5 \lesssim z \lesssim 1$ \citep{calabro18,calabro19}. 
For the most compact objects, the line of sight toward the central black-hole might be completely dominated by dust residing in the host galaxy ISM at scales of $\leq 1$ kpc \citep{calabro18}.

To mitigate the problem of dust attenuation, a possible solution is to observe at longer wavelengths in the rest-frame near-infrared (near-IR), where AGN radiation is less susceptible to ISM attenuation, and where our line of sight is less obstructed by dust from the BLR and NLR themselves. In the case of a mixed model, Paschen lines can recover $30\%$ to $40\%$ more of the total star-formation of the galaxy compared to \Ha\ and \Hb\ \citep{calabro18}. Since near-IR lines provide a more unbiased view of the systems, the presence of AGNs can be assessed with higher reliability, and their intrinsic properties studied more easily. 

This approach has successfully discovered hidden broad-line AGNs, a type of obscured AGNs with narrow Balmer lines but with a broad ($>1000$ km/s) component in \PaA\ and \PaB\ \citep{riffel06,onori17,lamperti17}. \citet{denbrok22} and \citet{ricci22} have shown that \PaA\ and \PaB\ are better tracers of black-hole mass compared to \Ha. 
Similarly, metal emission lines at $\lambda_{rest} > 1 \mu m$ might be better diagnostics than optical ones, especially for identifying Type 2 AGNs, given that they have typically larger column densities of dust than Type 1 AGNs (\Citealt{mignoli19} and references therein). 
\citet{lamassa19} have described the limitations of the optical BPT analysis, as they find that one-half of mid-infrared color selected AGNs from the WISE survey have optical emission line ratios typical of star-forming galaxies rather than of AGNs, which can be due to the highly obscured AGN clouds. \citet{rodriguezardila04} and \citet{colina15} have proposed \xHtwo/\BrG\ versus \xFeIIa/\PaB\ and \xFeIIc/\BrG\ (respectively), as useful alternatives to standard optical AGN diagnostics, and successfully applied them to study obscured AGNs in local galaxy samples. 
Therefore, going to longer wavelengths could potentially reveal a hidden population of BL and NL AGNs, and also better characterize their physical properties. 

In this paper, we add to previous efforts in the field by exploring new AGN and star-formation (SF) diagnostics in the near-infrared. To be applied to large samples of galaxies, we need to identify relatively bright emission lines at $\lambda_\text{rest-frame} \gtrsim 0.8 \mu m$, and to cover a variety of ISM conditions of the galaxies in terms of metallicity, gas density, and ionization. This can be done with the help of Cloudy photoionization models \citep{ferland17}, from which we can easily simulate the properties of emission lines emitted by gas clouds (representative of the galaxy ISM) irradiated by different types of ionizing sources, and then compare them to the observational results. 

First, this analysis is applied at $z\sim0$, exploiting the near-IR observations and catalogs performed in the last decades and publicly available in the literature.
Then, thanks to more recent spectroscopic facilities, we use these diagnostics even beyond the Local Universe. In particular, we apply them to a sample of starburst galaxies at intermediate redshifts ($0.5<z<0.9$) observed through Magellan/FIRE. Finally, we exploit the new data coming from JWST/NIRspec, which after the first light and becoming operational in 2022, has started to take spectra from low to high-resolution mode ($30<$ R $<2700$) in the spectral range from $0.8$ to $5.2 \mu m$. The JWST mission \citep{gardner06,gardner23} is thus opening a new window by shedding light on the near-IR rest-frame properties of galaxies at high redshift up to at least $ z \sim 3$. In this paper, we use data from the Cosmic Evolution Early Release Science Survey (CEERS) survey (Finkelstein et al., in prep.), which has collected so far among the largest samples of galaxies at 'cosmic noon' and is thus suited to analyze a sizeable sample of AGNs at $z>1$. 

The paper is organized as follows. In Section \ref{observations_data}, we describe the dataset of local star-forming galaxies and AGNs, and then present the data reduction, redshift measurement and sample selection of higher redshift sources observed with Magellan/FIRE and JWST/NIRSpec. In Section \ref{CLOUDY}, we describe the Cloudy photoionization models that we adopt to study the emission line properties of star-forming galaxies and AGNs, and display these models in the classical BPT diagram.  
We show our results in Section \ref{results}. First, we present new near-IR diagnostic diagrams that can be used to distinguish between AGN and star-forming driven ionization, providing new separation lines in analytical form. Then, we apply these diagnostics to observations from $z=0$ to $z=3$, and compare the near-IR classification to the optical one. 
We discuss our results in Section \ref{discussion}, suggesting ideas to improve the statistics of AGNs with forthcoming spectroscopic surveys.
In our analysis, we adopt a \citet{chabrier03} initial mass function (IMF) and, unless stated otherwise, we assume a cosmology with $H_{0}=70$ $\rm km\ s^{-1}Mpc^{-1}$, $\Omega_{\rm m} = 0.3$, $\Omega_\Lambda = 0.7$.

\section{Observations and data}\label{observations_data}

Given the wide range of cosmic epochs that we want to explore, from the local Universe to redshift $\sim3$, different instruments and data are required.  In this section, we thus describe the various datasets that we use to study the properties of galaxies and AGNs with the same near-IR diagnostics. We start by introducing the new JWST/NIRSpec spectra coming from the CEERS survey, which provide physical information for galaxies at $1<z<3$.  At intermediate redshifts ($0.5 < z < 1$) we use near-IR ground-based spectra taken with the FIRE spectrograph at the Magellan telescope. We supplement this dataset with public spectral compilations of local star-forming galaxies and AGNs observed with the Spex-IRTF near-IR spectrograph and several optical instruments.

\subsection{JWST/NIRSpec data from CEERS}\label{CEERS}

\begin{figure*}[ht!]
    \centering
    \includegraphics[angle=0,width=1.\linewidth,trim={0.1cm 0.1cm 0.1cm 0.7cm},clip]{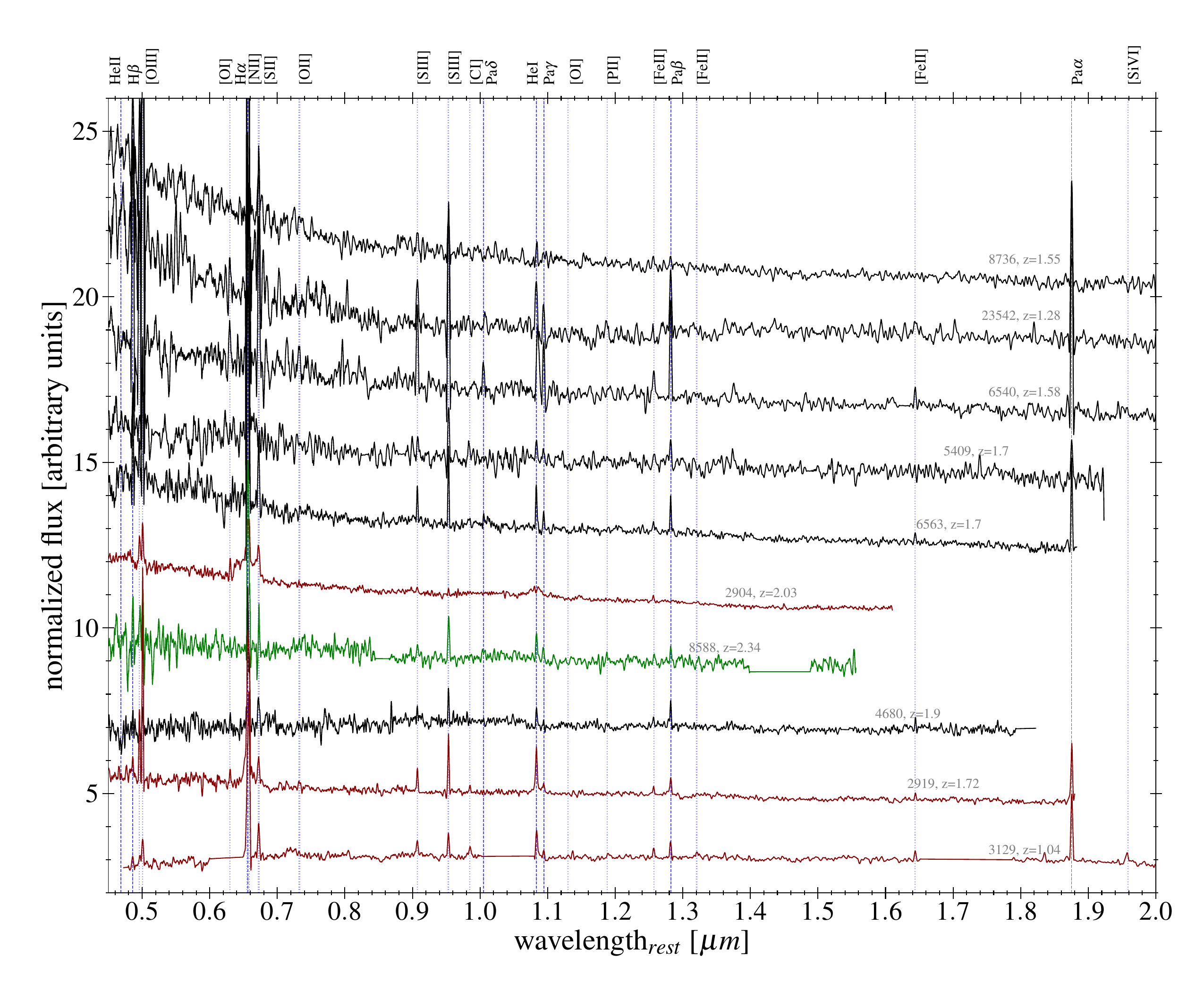}
    \caption{Rest-frame spectra of $10$ sources observed by JWST/NIRSpec in CEERS. The spectra are normalized to their median flux and offset relative to each other to help the visualization. They are colored in black, red, and green, according to whether they are identified, respectively, as optical+near-IR star-forming, optical+near-IR AGNs, or hidden AGNs (i.e., optical SF and near-IR AGNs), according to the diagnostic diagrams presented in this paper (see Section \ref{nearIR_diagnostics_highz}). The CEERS ID and the spectroscopic redshift of each source are written besides each spectrum. The emission lines that can be identified in the range $0.4 \mu m < \lambda < 2 \mu m$ are highlighted above the top axis (see also Table \ref{table_lines}).  
    }\label{spectra_example}
\end{figure*}

The data analyzed here are taken from the Cosmic Evolution Early Release Science Survey (CEERS; ERS 1345, PI: S. Finkelstein), a public survey to study galaxy evolution from $z\sim 1$ 
up to the reionization epoch in the Extended Groth Strip (EGS; \Citealt{noeske07}) field, which is one of the five fields covered by the Cosmic Assembly Near-Infrared Deep Extragalactic Legacy Survey \citep[CANDELS;][]{grogin11,koekemoer11}. The main features of the CEERS program are presented in \citet{arrabalharo23a} and \citet{finkelstein23}, and will be described in more detail in Finkelstein et al. (in prep.).

The galaxies analyzed in this work were observed with NIRSpec \citep{jakobsen22} during the CEERS epoch 2 observations (December 2022). The spectra were taken with the MOS configuration, which requires precise placement of the sources inside Micro Shutter Arrays \citep[MSA;][]{ferruit22} with size $0.2 \arcsec$ $\times 0.46 \arcsec$. These NIRSpec observations were divided into 6 different MSA pointings, and performed with the G140M/F100LP, G235M/F170LP, and G395M/F290LP medium resolution (R $\sim 1000$) gratings (herewith denoted with "M"), and with the prism in low-resolution mode (R $\sim100$).
We use only the M-gratings observations in this paper, to ensure a complete deblending of the \OIII + H$\beta$ and \NII + H$\alpha$ emission line triplets in the optical. 
With this spectroscopic setup, we obtain an almost continuous wavelength coverage from $\sim 1$ to $\sim 5.2\ \mu m$.

The MSA apertures were made of 3-shutter slitlets, and a 3-point nodding pattern was performed along the direction of the longest slitlet side to enable background subtraction, shifting the pointing by a shutter length in each direction. The sources were observed by integrating $14$ groups in three exposures (one for each nod position) in NRSIRS2 readout mode, totaling 3107s of integrated time per grating for each MSA pointing.

The targeted sources were chosen among existing catalogs in EGS \citep{stefanon17,kodra23}, for which photometric redshifts were already determined in all cases, while for a small subset, the spectroscopic redshifts were also available from 3D-HST grism spectroscopy \citep{brammer12}. The final position of the open shutters was defined with the help of the MSA Planning Tool (MPT) to maximize the number of observed targets, preferentially selecting galaxy candidates at $z>1$. A more exhaustive description of NIRSpec observations in CEERS is provided by \citet{arrabalharo23a}. The details of the aperture position angle (PA), targeted area, and target priority criteria will be fully described in Finkelstein et al. (in prep.).

\subsubsection{CEERS spectral reduction}\label{spectral_reduction}

The spectral reduction of NIRSpec observations was performed by the CEERS collaboration, as described in \citet{arrabalharo23b} and further explained in Arrabal Haro et al.\ (in prep.). We highlight here the main processing steps based on the JWST Pipeline \citep{bushouse22}\footnote{\url{https://jwst-pipeline.readthedocs.io/en/latest/index.html}, version 1.8.5} and on the Calibration Reference Data System (CRDS) mapping 1061.

First, raw images are corrected for the detector 1/$f$ noise, dark current, and bias, and for the ``snowball'' contamination, which appears as trails in the images, likely produced by high-energy cosmic rays. This yields count-rate maps, from which two-dimensional (2D) spectra are created for each source.
We apply to these the flat-field correction, background subtraction (using the 3-nod pattern), and photometric and wavelength calibration. The resulting spectra are also resampled to correct for spectral trace distortions. In the pipeline, a slit loss correction is already performed by default in the \texttt{pathloss} step.

Then, the intermediate background subtracted 2D spectra (one for each nod position) are combined to create the final 2D spectrum of the source. The one-dimensional (1D) spectra are also extracted using customized apertures that are visually determined by CEERS members by collapsing the spectra along the wavelength direction and identifying with high confidence the continuum along the spatial direction.
An additional step is performed on the reduced 2D spectra to mask  possible hot pixels, detector gaps, and reduction artifacts that are not associated with the main continuum trace or with an emission line.   
The noise spectrum is automatically calculated by the \textit{JWST} pipeline, and then corrected as described in \citet{arrabalharo23b} to take into account the spectral resampling performed in the previous stages.

\subsubsection{Residual flux scaling and final spectra}\label{slitloss}

Although slit loss correction is performed during the main processing with the JWST pipeline (Section \ref{spectral_reduction}), we further check whether additional flux corrections are needed. To this aim, we convolve the previously calibrated, slit-loss-corrected 1D spectra with broadband filters and compare them to the photometric data available for each target through ground-based near-IR imaging, and/or space-based instruments such as HST/WFC3, Spitzer/IRAC, and JWST/NIRCam. This is possible because the continuum is detected with a S/N $>3$ per pixel for all our NIRSpec spectra. In particular, for all the selected sources we have HST, IRAC, or ground-based photometry, ensuring at least one band available within each grating, and a maximum number of $14$ broadband filters available from $\lambda = 1 \mu m$ to $5 \mu m$. 

In most cases, the shape of the spectrum is consistent with the broadband photometry across the entire wavelength range $1 \mu m < \lambda < 5 \mu m$, indicating that the relative flux calibration between different gratings is generally reliable. We thus multiply the spectra by a single scaling factor to match the photometric data when needed. 
For a few sources ($9 \%$), a different scaling correction factor should be applied to some gratings to match the photometry available in that wavelength range. 
These issues are not related to a specific grating, or to a particular MSA pointing. We find instead that almost all of these peculiar sources have at least one wavelength gap in the spectrum, which might be responsible for creating jumps in the flux density between different wavelength ranges. These calibration issues will be better described in the survey paper (Finkelstein et al. in prep.), and fixed in future spectroscopic reductions. % For one source, the discrepancy might arise instead from the fact that the observed region could have a slightly different spectrum from the whole galaxy. However, in this case, the correction factor is $< 10\%$ and much smaller than in sources with gaps.  

In Fig. \ref{spectra_example}, we show as an example the fully calibrated spectra (normalized to their median flux and converted to a rest-frame reference) of $10$ galaxies chosen among our CEERS sample at different redshifts. This highlights the variety of wavelength coverage, continuum shapes, and spectral properties that we are able to probe.

\subsubsection{Spectral analysis and sample selection}\label{sample_selection}

%%%%%%% ----------------------------------------------------------------------

The spectral analysis is performed as follows. 
In the first stage, we visually check all the 1D spectra 
looking for emission lines from $1\mu m$ to $5 \mu m$. 
In this spectral range, there are subsets of lines that are intrinsically brighter for most galaxies, either star-forming or AGNs, such as the H$\beta$ + \OIII\ triplet and H$\alpha$ + \NII\ triplet in the optical, and the \HeI + Pa$\gamma$ doublet and the Pa$\beta$ + \FeII\ triplet in the near-IR.
Driven by the visual identification of these groups of lines, we assign a first spectroscopic redshift guess. 

We are interested in building a sample of galaxies in the redshift range between $1$ and $3$. 
Therefore, in this phase, we select those galaxies in which the spectrum covers at least the \Hb\ + \OIII\ triplet and the \PaB\ + \FeII\ triplet. For the final sample, in addition to the two above triplets, we require the identification of the \Ha\ + \NII\ triplet and the \xSIII\ line, as the presence of all these lines are essential requirements for the analysis of optical and near-IR diagnostics presented in this paper.
With these criteria, we automatically select $65$ systems between $z\sim1$ and $z\sim3$, among which $27$ galaxies have $z < 1.75$ and thus also have \PaA\ detected in the spectrum.
This constitutes our final selected CEERS sample for all the forthcoming analyses.  We also note that a few sources (10), despite being at the requested redshift, have been removed, as one of the emission line groups required above falls in detector gaps and is not available.

\begin{table}[ht!]
\centering{\textcolor{blue}{Typical lines observed in star-forming galaxies and AGNs.}}
\renewcommand{\arraystretch}{1.5} 
\vspace{-0.2cm}
\small
\begin{center} { %\small
\begin{tabular}{ | m{2cm} | m{3.5cm} |} 
  \hline
  \textbf{ion} & \boldsymbol{$\lambda_{rest} [\mu m]$} \\ 
  \hline
  \Hb & $0.4861$ \\
  \OIII & $0.4959$, $0.5007$ \\
  \OI & $0.6300$ \\
  \NII & $0.6548$, $0.6583$ \\
  \Ha & $0.6563$ \\
  \SII & $0.6716$, $0.6731$ \\
  \OII & $0.7331$, $0.7320$ \\
  \SIII & $0.9069$, $0.9531$ \\
  \CI & $0.985026$ \\
  \PaD & $1.0049$ \\
  \HeI & $1.083$ \\
  \PaG & $1.0938$ \\
  \PII & $1.18828$ \\
  \PaB & $1.2818$ \\
  \FeII & $1.257$, $1.32$ \\
  \FeII & $1.64$ \\
  \PaA & $1.875$ \\
  \SiVI & $1.962$ \\
  \Htwo & $2.121$ \\
  \BrG & $2.1655$ \\
  \hline
\end{tabular} }
\end{center}

\vspace{-0.1cm}
\caption{\small Table with the list of optical and near-IR rest-frame lines and doublets that we are able to detect in our spectra for star-forming galaxies and AGNs between $z=0$ and $z=3$, and that we consider in this paper. 
}
\vspace{-0.6cm}
\label{table_lines}
\end{table}

\subsection{Line flux measurements}\label{mpfit}

\begin{figure*}[ht!]
    \centering
    \includegraphics[angle=0,width=1\linewidth,trim={0.1cm 5cm 2.5cm 0cm},clip]{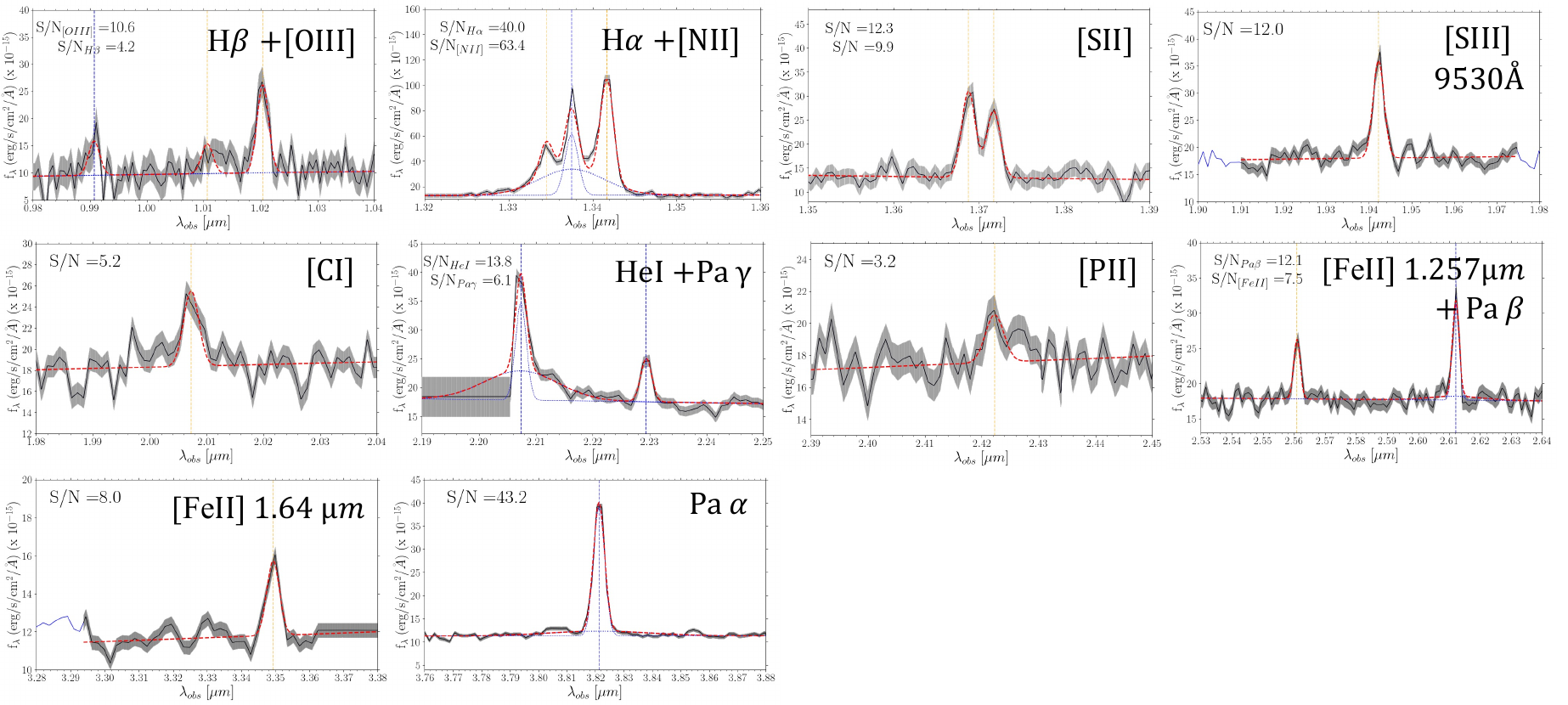}
    \caption{Figure displaying the spectrum of the broad-line AGN with CEERS ID 3129 at $z=1.04$, around emission lines detected in the full spectral range of NIRSpec, in increasing order of wavelength. The noise spectrum is represented with gray error bars around the observed flux (black line). The emission lines are fitted with multiple gaussian components, as described in the text. Blue and orange vertical dashed lines highlight the gaussian central wavelength of permitted and forbidden lines, respectively. For broad lines, both the narrow and broad components are shown with a dashed and dotted blue line, respectively. The global best fit profile is drawn with a red dashed line. The S/N of the detection is highlighted in the top left corner of each panel, while the name of each line is written in the top right corner (see also Table \ref{table_lines}). 
    }\label{fit_lines_3129}
\end{figure*}

\begin{figure}[ht!]
    \centering
    \includegraphics[angle=0,width=1\linewidth,trim={0.1cm 0cm 2cm 2cm},clip]{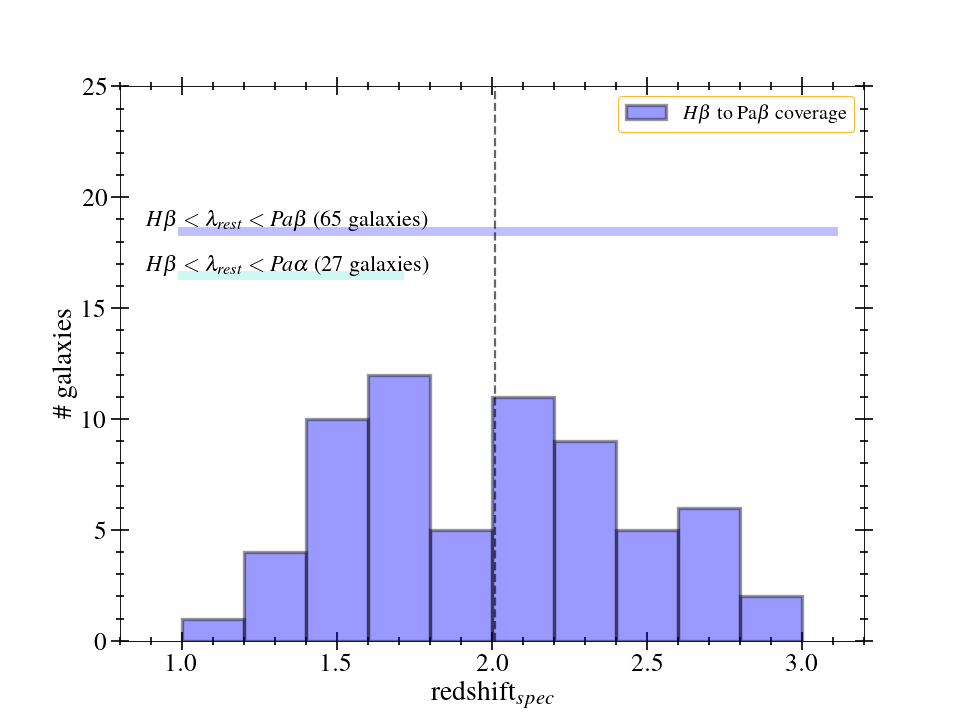}
    \includegraphics[angle=0,width=1\linewidth,trim={0.1cm 0cm 2cm 2cm},clip]{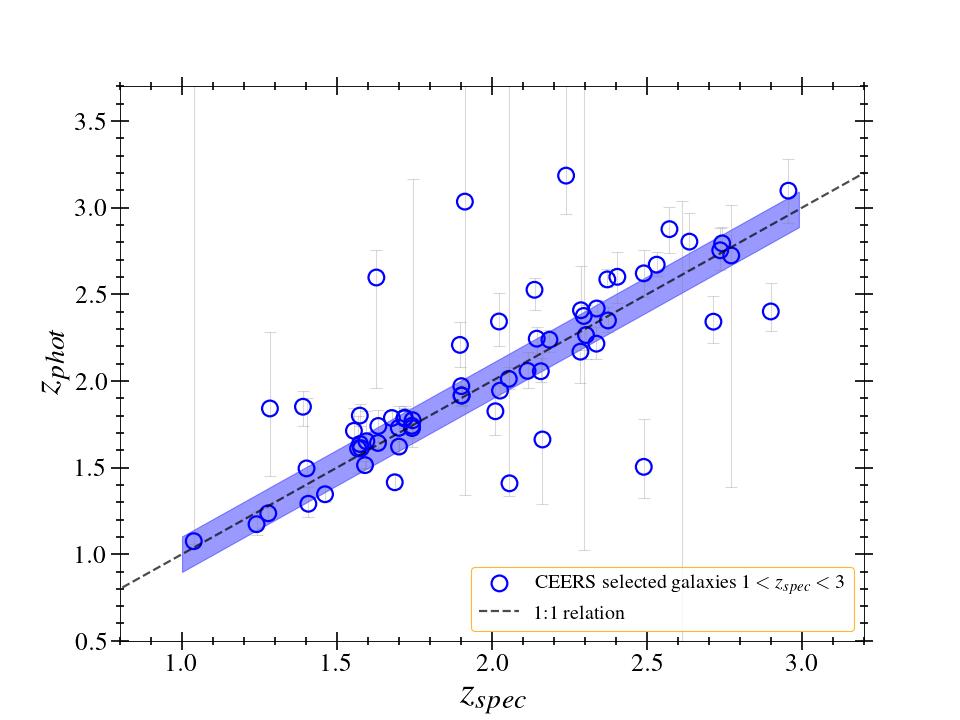}
    \caption{\textit{Top :} Spectroscopic redshift distribution of emission line galaxies in the CEERS survey in the redshift range between $1$ and $3$, where both optical and rest-frame near-IR lines can be detected, from \Hb\ to \PaB. The vertical dashed black line highlights the median redshift of our selected sample ($z_\text{median}=2.0$). 
    \textit{Bottom :} Comparison between the spectroscopic redshifts estimated from CEERS spectra and previously available photometric redshifts \citep{kodra23}.
    }\label{histogram_redshift}
\end{figure}

\begin{figure}[h!]
    \centering
    \includegraphics[angle=0,width=1\linewidth,trim={0.1cm 0cm 2cm 2cm},clip]{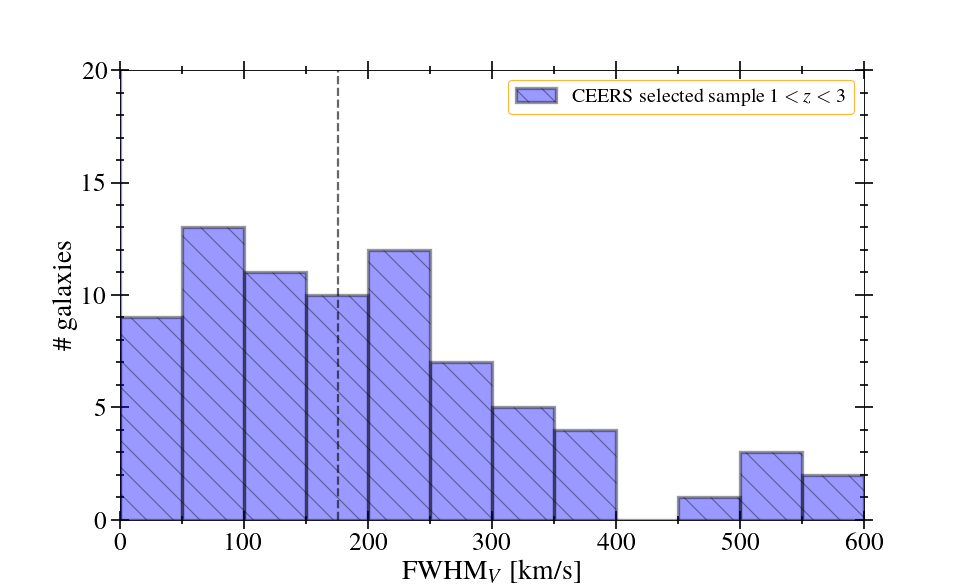}
    \caption{Distribution of intrinsic FWHM widths (i.e., deconvolved from instrumental broadening) for CEERS galaxies selected at $1<z<3$ as described in the text. The FWHM is estimated from the \Ha\ line. If \Ha\ is not detected, we use the highest S/N line in the spectrum, that is, \xSIII\ or \xOIII. The line width is a variable parameter in the fit but is assumed to be the same value for all narrow lines, with a tolerance of $50$ km/s. The median value of the sample (FWHM$_\text{median} \simeq 175$ km/s) is represented with a vertical dashed black line. 
    }\label{FWHM_distribution}
\end{figure}

In addition to the emission lines considered for the first redshift guess and sample selection, we can also identify in many cases other intrinsically fainter lines. The full subset of lines relevant for this work (i.e., from $\lambda_\text{rest} = 0.4 \mu m$ to $ \lambda_\text{rest} = 2 \mu m$) and that we can detect for at least a subset of galaxies, is summarized in Table \ref{table_lines}. 

For a more accurate determination of the spectroscopic redshift and for the measurement of emission line fluxes, we use the $\chi^2$ minimization based MPFIT \footnote{GitHub Repository: stsci.tools/lib/stsci/tools/nmpfit.py} routine \citep{markwardt09} to fit all the lines listed in Table \ref{table_lines}, assuming single Gaussian functions on top of a linear continuum. We also assume for all the lines a single redshift $z$ and the same line velocity width $\sigma_v$, which are first determined by fitting the highest S/N emission line in the whole spectral range, usually H$\alpha$, or \xSIII\ in case H$\alpha$ is not present. The H$\beta$ + \OIII\ and H$\alpha$ + \NII\ triplets are fitted simultaneously with the same underlying continuum, fixing the ratio between the two components of \NII\ and \OIII\ to $0.338$ and $0.3356$, respectively \citep{storey88}. In addition, we fit together the following couples of emission lines lying close in wavelength: \HeI\ and \PaG, \PaB\ and \xFeIIa, without any constraint on their flux ratios. 

After the first estimation of $z$ and $\sigma_v$, for the remaining lines we leave a tolerance of $100$ km/s for both parameters. This provides in all cases a good fit, which we control by requiring a statistical significance of the fit of $95\%$, as determined from the reduced $\chi^2$ ($\chi^2_{red}$). The quality of the relative wavelength calibration is excellent, as determined inside the CEERS collaboration, with precision at the level of less than $5\%$ (Arrabal Haro et al., in prep). As a further test, we compared our redshifts to those determined by fitting the lines in each grating separately. These measurements were done using the methodology described in \citet{fernandez23}, for the complete sample. We find that they are consistent within $1$ standard deviation, suggesting that the wavelength calibration is robust across the three gratings and the full wavelength range of NIRSpec. %  (Vital Fernandez, private communication). 

In addition to the spectroscopic redshift and the line velocity width, this MPFIT procedure yields emission line fluxes, underlying continuum, equivalent widths (EW), and their corresponding uncertainties for all the lines \footnote{A catalog with ID, coordinates, and emission line measurements for the CEERS galaxy sample selected in this work at $1<z<3$ is available in electronic form at the CDS and on the GitHub page \url{https://github.com/Anthony96/Line_measurements_nearIR.git}.}. While for the first line we require a high detection threshold (S/N $\geq 5$) to secure the redshift and $\sigma_v$, for the remaining lines we require a S/N $\geq 2$ to be detected. This choice was also adopted in \citet{calabro19} to improve the detection of fainter lines while preserving its reliability, given that the fit is performed with a restricted number of free parameters compared to a blind search.
For non-detected lines, we adopt a $1\sigma$ upper limit.  
In Figure \ref{fit_lines_3129} we show as an example of our procedure the fit performed for all the emission lines of the source ID 3129 (a broad-line AGN at $z=1.04$) that are relevant to this work.
We show instead in Fig. \ref{histogram_redshift}-top the spectroscopic redshift distribution of our selected targets. They span the entire range from $z \simeq 1$ to $z \simeq 3$, with a median redshift $z_\text{median}=2$ and $1\sigma$ dispersion of $\sim 0.5$. 

We also compare the spectroscopic redshifts $z_{spec}$ to the photometric estimates $z_{phot}$ available from the latest EGS catalog assembled by \citet{kodra23}. The $z_{phot}$ values are derived by fitting stellar templates to the available UV to near-IR photometry with a Hierarchical Bayesian method. We refer to \citet{kodra23} for the detailed photometric analysis. 
We find in general a good consistency, with a median absolute deviation of $0.11$. %, of the same order as the typical photometric redshift uncertainties ($=0.1$). 
We find a $\sim 10\%$ fraction of catastrophic outliers with $|z_{phot} - z_{spec}| > 0.5$ , but no systematic under- or over-estimations across all the explored redshifts (Fig. \ref{histogram_redshift}-bottom). We also note that the outliers tend to have larger uncertainties of $z_{phot}$ compared to the whole sample.

In Fig. \ref{FWHM_distribution}, we show the distribution of the FWHM line velocity widths from the combined \Ha, \xSIII, and \xOIII\ lines.
Given the FWHM resolution of our NIRSpec spectroscopic setting ($\simeq 300$ km/s) most of the lines appear well resolved. By subtracting in quadrature the instrumental resolution from the observed quantity, we calculate the intrinsic FWHM line widths. We find a median intrinsic FWHM$_V$ of $200$ km/s, with standard deviation $\sigma$ of $130$ km/s and a tail of galaxies with higher FWHM up to $\sim 550$ km/s. 
Furthermore, we find $3$ objects for which all the permitted lines, including the Balmer, Paschen, and \HeI\ lines, require an additional broader component in the fit, with FWHM $> 1000$ km/s, well above the maximum found for the rest of the sample in Fig. \ref{FWHM_distribution}. 

For the Balmer and Paschen emission lines, we apply an underlying stellar absorption correction, rescaling their fluxes upwards. We apply for this correction the absorption equivalent widths (EW$_{abs}$) estimated from theoretical stellar population models, namely EW$_{abs}=$ $3.6$, $2.7$, $2.7$, $2.5$, and $2$ \AA\ for \Hb, \Ha, \PaG, \PaB, and \PaA, respectively. We note that these values for the optical lines are consistent with \citet{dominguez13}. In our CEERS galaxies, they produce an increase of the line fluxes of, respectively, $3.4\%$, $0.5\%$, $3.7\%$, $2.1\%$, and $1\%$ on average, thus would not alter significantly the results of this paper. 

Finally, using all the hydrogen recombination lines available (typically those from the Balmer and Paschen series), we also check the quality of the final calibrated spectra by comparing their emission line ratios to the theoretical predictions, considering a variety of dust attenuation laws. While a more detailed analysis of dust attenuation will be presented in a follow-up paper (Calabro et al. 2023b, in prep.), a representative test is shown in Appendix \ref{appendix0}. This test suggests that Balmer and Paschen line fluxes across a large range of wavelengths ($\sim 0.4$ to $\sim 2 \mu m$) are physically meaningful and overall consistent with the predictions of typical dust attenuation models, even when the lines reside in different gratings, suggesting that their relative calibration can be trusted at the first order. 

\subsection{Main physical properties of the CEERS sample}\label{massSFR}

\begin{figure}[h!]
    \centering
    \includegraphics[angle=0,width=1\linewidth,trim={0.1cm 0cm 3cm 2cm},clip]{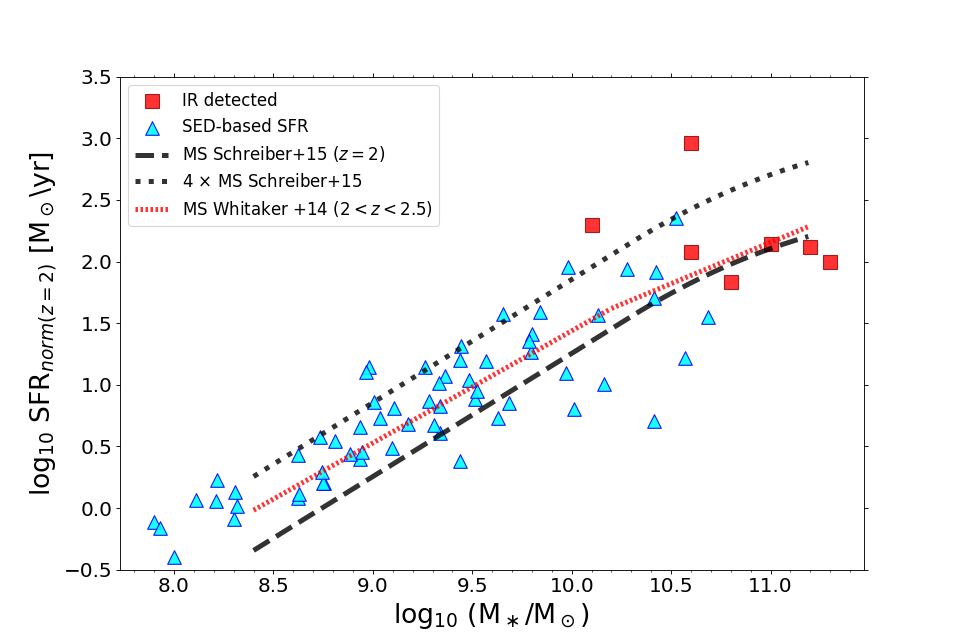}
    \caption{Diagram showing the stellar mass \mass - SFR diagram of CEERS galaxies selected in this work between redshift $1$ and $3$. The \mass and SFR are taken from \citet{stefanon17} as described in the text (cyan triangles) or from Le Bail et al. (2023, in prep.) if Herschel detected (red squares). The sample is representative of the star-forming Main Sequence (MS) at the median redshift of the sample ($z_\text{median}=2$). The black dotted line represents the starburst limit, that is, a factor of $4$ above the MS of \citet{schreiber15}. The SFRs shown in this plot are normalized to $z=2$ assuming that they scale with redshift as ($1+z$)$^{2.8}$ \citep{sargent14}.
    }\label{mainsequence}
\end{figure}

The $65$ galaxies selected in Section \ref{sample_selection} show a variety of physical properties. They have H-band (F160W) magnitudes ranging between $H_\text{AB} = 21$ and $27.2$, with a median value of $24.2$. Considering previous catalogs published in the CANDELS collaboration \citep{stefanon17}, we can also study the distribution of other main properties of the sample, including the stellar mass and the SFR. We consider here the stellar masses and SFRs obtained through fitting HST, CFHT, and Spitzer photometry (from band $u$ to IRAC channel 4) with BC03 stellar population models, assuming a Chabrier IMF, exponentially declining SFH ($\tau$ models), and a solar metallicity \citep{acquaviva12}, taking also into account the contribution from emission lines to the observed photometry. 
While different assumptions in the SED fitting may be applied to specific objects, we aim here at investigating the global properties of the sample that can be compared to other galaxies. Nonetheless, the stellar masses that we consider are in agreement with the median stellar masses reported by \citet{stefanon17}, which combine estimates from different groups using different prescriptions for the fit. 

Representing these quantities in Fig. \ref{mainsequence}, we notice that our sample is fairly representative of the star-forming main sequence at $z\sim2$ \citep{schreiber15,whitaker14}, with stellar masses \mass ranging between $\log$ M/M$_\odot$ $=7.8$ and $11.3$ (median \mass $\simeq 10^{9.5}$ \msun). 
A small subset of CEERS galaxies is also detected at S/N $\geq 5$ with Herschel and may host dust-obscured star formation, thus we use in these cases the SFRs inferred through fitting the infrared and radio photometry (from Spitzer $24 \mu m$ flux to VLA $1.4$ GHz flux) as described in Le Bail et al. (2023 in prep), which adopts a similar methodology to \citet{jin18}. This subset lies in the upper right part of the diagram with higher stellar masses and higher SFRs, and total infrared luminosities L$_\text{IR}$ (integrated in the range $8$-$1000 \mu m$) $\gtrsim 10^{12}$ L$_\odot$.

\subsection{Magellan/FIRE spectra for starburst galaxies at $z \sim 0.7$}\label{Magellan_starbursts}

To increase the statistics of the sample and the redshift range, we also consider $6$ systems at intermediate redshifts between $0.5$ and $0.9$ (z$_\text{median} \simeq 0.7$) with both optical and near-IR rest-frame coverage. These systems are presented in \citet{calabro18} and were originally selected in the COSMOS field as infrared bright galaxies representative of a population of highly star-forming galaxies with  $\log$ (M$_\ast$/M$_\odot$) $> 10$  and $\log$ (L$_\text{IR}$/L$_\odot$) between $11.5$ and $12.5$. 

Their near-IR spectra were observed with the FIRE echelle spectrograph \citep{simcoe13}, mounted at the Magellan $6.5$m Baade Telescope, in two observing runs in 2017 and 2018 (P.I. P. Cassata and R. Gobat). Thanks to their spectral coverage, from $z$ to the K band, they include at least the \Ha\ and the \PaB\ emission lines.
The optical spectra are instead taken with the VIMOS spectrograph \citep{lefevre03} and are publicly available from the zCOSMOS survey \citep{lilly07}, and they cover the missing rest-frame range $0.3<\lambda<0.6\mu m$ that includes the \xOIII\ and \Hb\ lines. 

The observations, data reduction, and final calibrated spectra are fully described in \citet{calabro19}. 
The flux measurements are also presented in the above paper for the Balmer and Paschen lines, while here we also measure with the same approach the other metal lines, including \xOIII, \xNII, \xSIII, \xFeIIa, \xCI, and \xPII. For the subset considered here, we are able at least to detect \PaB\ with S/N $\geq3$ and set upper limits on \SIII, \FeII, \PII, or \CI\ in the worst case. 
Analyzing these sources in the classical BPT diagram, $5$ are found to be consistent with star-forming models, while one source (dubbed BPT AGN) lies beyond the maximum separation line of \citet{kewley01}.

\subsection{A comparison sample of local starbursts and AGNs}\label{local_galaxies_AGNs}

Local samples of galaxies and AGNs can provide a useful benchmark to test and apply near-IR rest-frame diagnostics. 
Spectral compilations in the local Universe were made by \citet{riffel06}, who present one of the most extensive NIR spectral atlases of AGNs. This atlas includes optical/UV and X-ray AGNs found from previous surveys, and it is representative of the entire class of AGNs at $z\sim0$, with different degrees of nuclear activity. It comprises $27$ UV/optically selected sources at $z < 0.1$, including $12$ broad line AGNs (Type 1) and $15$ narrow line AGNs (Type 2), with the addition of $4$ starburst galaxies that do not show evidence of nuclear activity through either broad line components, coronal lines, or AGN-like line ratios in optical diagnostics.

To increase the size of the star-forming sample for comparison, we also consider the optical and near-IR spectra of $16$ infrared luminous spiral galaxies presented more recently by \citet{riffel19}. These galaxies are all at $z < 0.05$ and were originally selected from the IRAS Revised Bright Galaxy Sample with L$_\text{FIR} > 10 ^{10.10}$ L$_\odot$. Their nature (AGN or star-forming) was previously determined through optical spectroscopy: 
$8$ systems are purely star-forming (HII) galaxies, $3$ are LINERs, $1$ is a Type 1 AGN, and $4$ systems have intermediate properties between LINER/HII or Type 2 AGN/HII. 

The near-IR spectra of all these sources are taken using the Spex slit spectrograph \citep{rayner03}, mounted at the NASA 3m IRTF telescope at Mauna Kea. 
For the details of the spectroscopic observations, data reduction, and line measurements, we refer to \citet{riffel06,riffel19}. 
In the case of Type 1 AGNs, they decompose the broad and narrow component of permitted lines, so we only take their narrow component for our purposes. 

In addition, we consider the sample of $11$ systems from the compilation made more recently by \citet{onori17}, and for which we have a similar rest-frame spectral coverage to the data presented before and the same subset of detected emission lines from $0.4$ to $2 \mu m$ rest-frame. These systems are selected as X-ray emitters from the Swift/Burst Alert Telescope 70-month catalog, and include obscured (type 2) and intermediate type AGNs, and starburst galaxies at $z<0.1$. 
Part of this sample ($8$ systems) is observed at high-resolution ($4000 < R < 17000$) with X-shooter at the VLT. 
For the remaining part of this sample ($3$ galaxies), similarly high resolution spectra ($R>4000$) are taken at LBT with the single slit NIR spectrograph LUCI \citep{seifert03}. 
We refer to \citet{onori17} for a complete description of the observations, data reduction, and line measurements. In particular, given the high resolution, they perform a fit with three-Gaussian components, which should be representative of the narrow component (that we use here), and eventually a broad component and an outflow component. 
Taking the forbidden line fluxes and the narrow component of permitted lines, it is possible to classify spectroscopically the nature of each source in the classical BPT diagram. Following this procedure, almost all the \citet{onori17} sources ($9$) are AGNs ($1$ Type 1, and $8$ Type 2), while $1$ galaxy is classified as star-forming and an additional $1$ lies in the composite BPT region.

In total, we assemble a final local sample of $58$ sources, including $14$ Type-1 AGN, $26$ Type-2 AGN, $13$ starburst (purely star-forming) galaxies, $3$ LINERs, and $5$ sources with mixed line properties. 
Among the many optical and near-IR lines available in the literature for these galaxies, we select those in Table \ref{table_lines}, which we will use in the following sections as AGN diagnostics and which can be detected also with NIRSpec in the CEERS sample. 
In all cases, we directly take the flux measurements available in the above papers.

We note that star-forming systems considered at $z<1$ do not cover such a large variety of physical conditions as in the CEERS sample. They rather represent a subset of massive and dusty starbursts, forming stars at higher rates and possibly with higher efficiency than average galaxies in this redshift range. Despite this, they are very useful to constrain the maximum star-forming line ratios as they likely probe extreme ISM conditions among the star-forming galaxy population and the highest metallicities. 

\section{Emission line modeling with Cloudy}\label{CLOUDY}

\begin{figure}[t!]
    \centering
    \includegraphics[angle=0,width=0.95\linewidth,trim={0.1cm 0.1cm 0.1cm 0.1cm},clip]{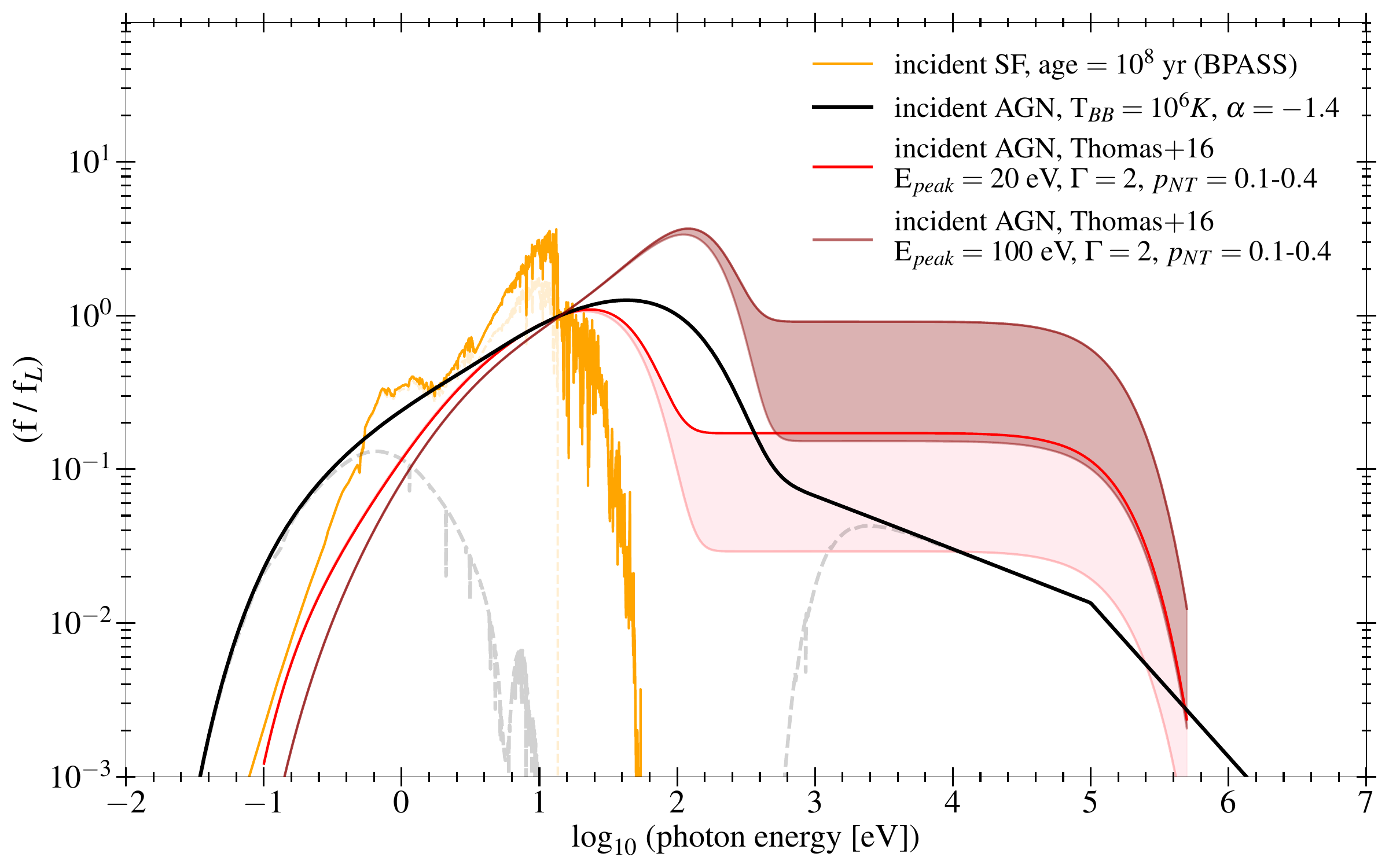}
    \includegraphics[angle=0,width=0.95\linewidth,trim={0.1cm 0.1cm 0.1cm 0.1cm},clip]{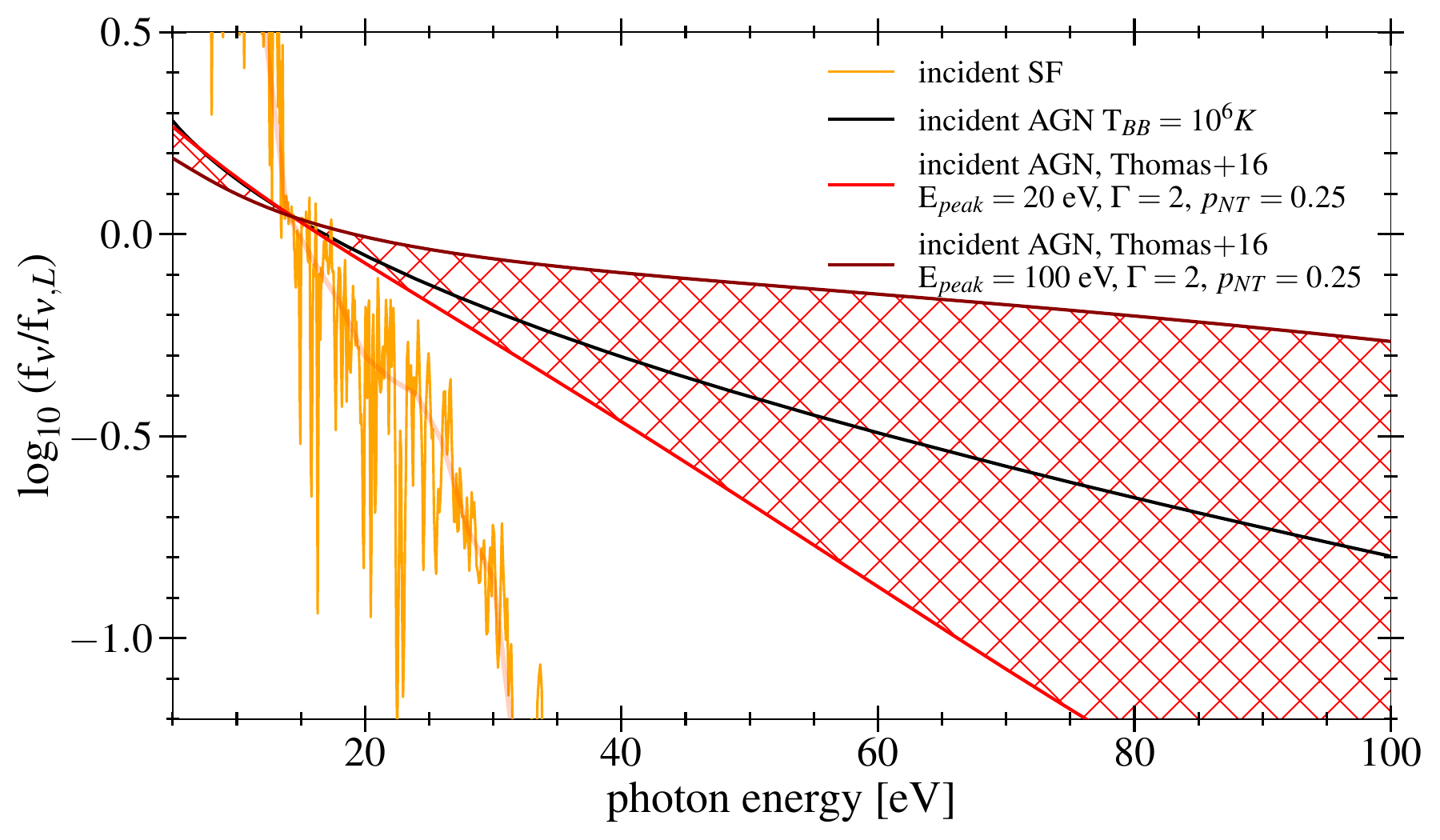} 
    \caption{\textit{Top:} Diagram with the shape of the ionizing radiation field (energy units, normalized to the flux at the Lyman limit) assumed in CLOUDY to model the stellar emission (orange curve) and the AGN emission (black, red, and darkred curves, as described in the legend). For these representative curves we have assumed $\log (U) = -2$, solar metallicity, and a density of $10^3 cm^{-3}$ in all cases. In the AGN models from \citet{thomas16}, we identify with the same color (red or darkred) the family with a common E$_\text{peak}$, in which the parameter p$_{NT}$ increases toward the top from the lighter to the darker line. The dashed shaded lines indicate the transmitted continuum. \textit{Bottom:} Zoom-in of the top panel in the energy range between $5$ and $100$ eV, but with a linear scale on the x-axis and the flux density f$_\nu$ (normalized to the flux density at the Lyman limit) on the y-axis. This version of the plot is inspired by \citet{feltre16}. The area between the two AGN shapes of \citet{thomas16} with the same p$_{NT}=0.25$ and varying E$_\text{peak}$ (in the range $20$-$100$ eV) is filled with a chequered red pattern. % Same as in the top panel, but with the The inset in the top right is inspired by \citet{feltre16}.
    }\label{ionizing_spectra}
\end{figure}

\begin{table}[h!]
\centering{\textcolor{orange}{Star-forming models}}
\renewcommand{\arraystretch}{1.5} 
\vspace{-0.2cm}
\small
\begin{center} { %\small
\begin{tabular}{ | m{1.4cm} | m{1.5cm}| m{1.5cm} | m{2.8cm} |} 
  \hline
  \textbf{SFH} & \textbf{log(U)} & \boldsymbol{$\log n_H/cm^{-3}$} & \boldsymbol{$Z_{gas}$} and \boldsymbol{$Z_{stars}$} [\boldsymbol{$Z_\odot$}] \\ % \textbf{log n$_H$/cm$^{-3}$ & \textbf{Z} \\
  \hline
  \multirow{2}{*}{\parbox[t]{\hsize}{ constant, with age $=10^8$ yr } } & \multirow{2}{*}{\parbox[t]{\hsize}{ $-4$,$-3.5$,$-3$, $-2.5$,$-2$, $-1.5$,$-1$ } } & \multirow{2}{*}{\parbox[t]{\hsize}{$2$, $3$, $4$} } & \multirow{2}{*}{\parbox[t]{\hsize}{$0.05$, $0.1$, $0.15$, $0.2$, $0.3$, $0.4$, $0.5$, $0.7$, $1$, $2$} } \\
   & & & \\
  \hline
\end{tabular} }
\end{center}

\centering{\textcolor{blue}{AGN models (Risaliti et al. 2000)}}
\renewcommand{\arraystretch}{1.5}  % # 2 if you want to increase cell height
\vspace{-0.2cm}
\small
\begin{center} { %\small
\begin{tabular}{ | m{0.7cm} | m{1.2cm} | m{1.4cm}| m{1.5cm} | m{2.cm} |} 
  \hline
  \boldsymbol{$T_{BB}$} \textbf{[K]} & \boldsymbol{$\alpha$} \textbf{index} & \textbf{log(U)} & \boldsymbol{$\log n_H/cm^{-3}$} & \boldsymbol{$Z_{gas}$} [\boldsymbol{$Z_\odot$}] \\ % \textbf{log n$_H$/cm$^{-3}$ & \textbf{Z} \\
  \hline
  \multirow{2}{*}{\parbox[t]{\hsize}{ $10^6$ } } & \multirow{2}{*}{\parbox[t]{\hsize}{ $-1.2$,$-1.4$, $-1.7$,$-2.0$ } } & \multirow{2}{*}{\parbox[t]{\hsize}{ $-4$,$-3.5$,$-3$, $-2$,$-1$ } } & \multirow{2}{*}{\parbox[t]{\hsize}{$2$, $3$, $4$} } & \multirow{2}{*}{\parbox[t]{\hsize}{$0.3$, $0.4$, $0.5$, $0.7$, $1$, $2$, $3$ } } \\
   & & & & \\
  \hline
\end{tabular} }
\end{center}

\centering{\textcolor{red}{AGN models (Thomas et al. 2016, 2018)}}
\renewcommand{\arraystretch}{1.5}  % # 2 if you want to increase cell height
\vspace{-0.2cm}
\small
\begin{center} { %\small
\begin{tabular}{ | m{0.9cm} | m{1.2cm} | m{0.25cm} | m{1.4cm}| m{1.5cm} | m{1.2cm} |} 
  \hline
  \boldsymbol{$E_{peak}$} \textbf{[eV]} & \boldsymbol{$p_{NT}$} & \boldsymbol{$\Gamma$} & \textbf{log(U)} & \boldsymbol{$\log n_H/cm^{-3}$} & \boldsymbol{$Z_{gas}$} [\boldsymbol{$Z_\odot$}] \\ % \textbf{log n$_H$/cm$^{-3}$ & \textbf{Z} \\
  \hline
  \multirow{2}{*}{\parbox[t]{\hsize}{ $20$-$100$ } }  & \multirow{2}{*}{\parbox[t]{\hsize}{ $0.1$, $0.25$, $0.4$ } } & \multirow{2}{*}{\parbox[t]{\hsize}{ $2.0$ } } & \multirow{2}{*}{\parbox[t]{\hsize}{ $-4$, $-3.5$, $-3$, $-2$, $-1$ } } & \multirow{2}{*}{\parbox[t]{\hsize}{$2$, $3$, $4$} } & \multirow{2}{*}{\parbox[t]{\hsize}{$0.3$, $0.4$, $0.5$, $0.7$, $1$, $2$, $3$ } } \\
   & & & & & \\
  \hline
\end{tabular} }
\end{center}

\centering{\textbf{Solar abundances and depletion factors for all the models}}
\renewcommand{\arraystretch}{1.5}  % # 2 if you want to increase cell height
\vspace{-0.2cm}
\small
\begin{center} { %\small
\begin{tabular}{ | m{1.6cm} | m{3.cm}| m{3.cm} |} 
\hline
 Element & 12+$\log$ (X/H)$_{tot}$  & $\delta$ (X) \\ 
  \hline
  \textbf{C} & $8.55$ & $-0.2$  \\ 
  \textbf{N} & $7.97$ &  -  \\ 
  \textbf{O} & $8.87$ & $-0.2$  \\ 
  \textbf{S} & $7.27$ & $-0.2$  \\ 
  \textbf{Mg} & $7.58$ & $-1.2$  \\ 
  \textbf{Si} & $7.55$ & $-0.9$  \\ 
  \textbf{P} & $5.57$ & $-0.5$  \\ 
  \textbf{Cl} & $5.27$ & $0$  \\ 
  \textbf{Ti} & $4.93$   & $-2.3$  \\ 
  \textbf{Mn} & $5.53$ & $-1.2$  \\ 
  \textbf{Fe} & $7.51$ & $-1.5$  \\ 
  \textbf{Zn} & $4.65$ & $-0.35$  \\ 
  % \textbf{log n$_H$/cm$^{-3}$ & \textbf{Z} \\
  \hline
  \hline
\end{tabular} }
\end{center}

\vspace{-0.1cm}
\caption{Table with the full range of parameters used in Cloudy for constructing the grid of star-forming models and AGN models explored in this work. The AGN models are made following both the approach of \citet{risaliti00}, which is based on the standard 'table AGN' command in Cloudy, and of \citet{thomas16}, with the custom SED shapes injected in pyCloudy through the 'table SED' command. The solar abundances and the median depletion factors of metals from the gas phase onto dust grains are set according to \citet{savage96}. The depletion factors are in agreement with those estimated by \citet{jenkins09} for a depletion strength F$^\ast = 0.5$. 
}
\label{table_models}
\end{table}

To provide a theoretical basis for interpreting the observations and to build a common ground from which both the optical and the near-IR emission lines of galaxies can be understood, we build a set of photoionization models with Cloudy. We use the python package pyCloudy v.0.9.11 \footnote{\url{https://github.com/Morisset/pyCloudy/tree/0.9.11} }\ %\footnote{\doi{10.5281/zenodo.6558656}}, %\url{https://github.com/Morisset/pyCloudy/tree/0.9.11}{https://github.com/Morisset/pyCloudy/tree/0.9.11}}, 
which runs with version 17.01 of the Cloudy code \citep{ferland17}. This code assumes a symmetric distribution of gas around or in front of an ionizing source, and then calculates the radiated emission after it is processed by the gas layer. In principle, it is able to predict the emergent continuum and line intensities over the entire electromagnetic spectrum. However, we focus here on emission lines spanning from the optical rest-frame ($\sim 4000$ \AA) to the near-IR rest-frame ($\sim 2 \mu m$), which can be simultaneously observed with NIRSpec in the redshift range $1 \lesssim z \lesssim 3$. In the following subsections, we describe the star-forming and AGN models adopted in this work. 

\subsection{Star-formation models}\label{SF_models} 

For a theoretical understanding of the radiation emitted by star-forming galaxies, we consider an ionization-bounded, spherically symmetric shell of gas surrounded by a population of young (O and B type) stars. We note that this is a simplified picture, where a single shell of gas is representative of the ensemble of HII regions in a galaxy and the emergent spectrum is representative of the whole galaxy emission. 
Regarding the incident ionizing spectrum, we consider SEDs derived with BPASS stellar population models \citep{eldridge17}, assuming a Salpeter IMF with an upper mass limit of $300$ M$_\odot$, stellar metallicity equal to the gas phase metallicity of the surrounding gas cloud, and continuous star-formation history with age ranging from $10^7$ to $10^9$ years. For representation purposes, we show the prediction relative to a single age of $100$ Myr, as other ages would not produce significant variations of the emission line ratios studied in this paper. 

We specify the brightness of the incident radiation field, varying the ionization parameter $\log U$ between $-4$ and $-1$ (see Table \ref{table_models}-top panels), while we assume a hydrogen number density $n_H$ for the gas cloud ranging from $10^2$ to $10^4$ $cm^{-3}$. 
The gas-phase metallicity Z$_\text{gas}$ varies from $10\%$ solar to two times solar, where the solar abundance of each element is set as in \citet{savage96}. 
All the elements are scaled together with metallicity (i.e., by the same factor compared to solar), except nitrogen, which follows the prescription of \citet{feltre16} to take into account secondary production.
The helium abundance is set as $-1$ in logarithmic scale \citep{dopita06}. %  $0.2485 + 1.7756 \times $ Z as in \citet{bressan12}. 

We consider that some elements in the gas phase are depleted onto dust grains. 
The dust depletion is defined with the standard logarithmic notation system as $\delta (X) = [X/H] \equiv \log(X/H)_g - \log (X/H)_c$, where $c$ refers to the cosmic abundance of an element and $g$ to its gas phase component. As shown by \citet{savage96} and other works, $\delta (X)$ can vary significantly as a function of the interstellar environment and grain composition. It is relatively well studied in our Galaxy and nearby systems, such as the Large and Small Magellanic Clouds, while it is still poorly constrained at higher redshifts. In \citet{jenkins09}, they parametrize the strength of the dust depletion with a factor F$^\ast$ varying between $0$ (minimum depletion) and $1$ (maximum strength), and it does not depend on the particular element. For this work, we consider average depletion coefficients as reported in \citet{savage96}, which are generally consistent with those estimated from \citet{jenkins09} for a medium depletion strength (F$^\ast=0.5$). 
The solar abundances and depletion factors for all the elements are listed in Table \ref{table_models}-bottom panel. We can distinguish between non-refractive elements with $\delta (X)$ close to $0$, such as S, O, and C, and refractive elements with a more negative $\delta (X)$, such as Fe and Si.

\subsection{AGN models}\label{AGN_models}

The narrow-line region (NLR) of an AGN is modeled as follows. 
Regarding the ionizing source, the continuum emission from the AGN accretion disk is modeled using multiple prescriptions. First, we use the built-in 'AGN' command in Cloudy, with the multiple power law continuum characterized by a 'blue bump' temperature of $10^6$ K and a spectral energy index of $\alpha_{UV} = -0.5$ and $\alpha_x = -1.35$ in UV and X-rays, respectively, as in \citet{risaliti00}. The spectral index $\alpha_{ox}$ of the optical to X-ray SED is free to vary among the following values: $-2$, $-1.7$, $-1.4$, $-1.2$, spanning the range of observational findings, and consistent with typical values adopted in the literature \citep[e.g.,][]{groves04}. 
We also test different and more recent ionizing shapes. In particular, we consider the OXAF models\footnote{The OXAF models are available on the GitHub page: \url{https://github.com/ADThomas-astro/oxaf}.}, which are physically based AGN continuum emission models introduced by \citet{thomas16} and further described in \citet{thomas18}. While we refer to those papers for an exhaustive discussion, we briefly present their main properties. These models are specifically designed for photoionization modeling. Indeed, they are simplified enough to easily reproduce the diversity of observed AGN spectral shapes with only three main parameters: the energy at the peak of the accretion disk emission (E$_\text{peak}$), the power-law index of the non-thermal emission ($\Gamma$), and the fraction of the total flux coming from the non-thermal component (p$_\text{NT}$). 
The first is a crucial parameter, as it incorporates a dependence on the black hole mass, AGN luminosity, and 'coronal' radius, and it is sensitive to contamination from shocks or HII region emission (the higher their contribution to the AGN, the lower the value of E$_\text{peak}$). Moreover, it is not affected by dust screening effects. In our simulations, we vary E$_\text{peak}$ from $20$ to $100$ eV, following Fig. 5 of \citet{thomas16}. The parameter p$_\text{NT}$ affects the fraction of photons in the X-ray regime (see Fig. \ref{ionizing_spectra}), and we consider three possible values: $0.1$, $0.25$, and $0.4$, covering the entire physical range typically found in AGNs \citep{thomas16}.
Finally, we set $\Gamma$ to a median value of $+2.0$, as it was shown that this parameter does not significantly affect the line ratios (also confirmed in the near-IR regime).

The brightness of the radiation field is set through the ionization parameter $\log(U)$ varying from $-4$ to $-1$, as for the star-forming case.
Regarding the physical properties and chemical content of the NLR, we vary the gas density from $10^2$ to $10^4$ $cm^{-3}$, and the gas-phase metallicity from $0.3$ to $3$ times solar. The element abundances relative to hydrogen are set as for the star-forming galaxies. The helium abundance is set slightly higher (by $0.1$-$0.2$ dex) than for star-forming galaxies, following the recent work by \citet{dors22}. % Dors et al. (2022). 
As in \citet{feltre16}, we adopt an 'open geometry' model because it is more appropriate for the low covering fraction of the narrow-line region. 

Finally, in both the star-forming and AGN case, we consider dust in the calculation and assume for simplicity the standard grain properties implemented in Cloudy \citep{mathis77}, with grains to hydrogen mass ratio scaling linearly with metallicity. 
We stop the Cloudy calculation when reaching a temperature in the gas of $500$ K. %   Lyman continuum optical depth $\log \tau=10$.
In table \ref{table_models}, we present the full list of parameters used to model the star-forming galaxies and the NLR of AGNs. We show instead in Fig. \ref{ionizing_spectra} a representative ionizing spectrum and the transmitted continuum radiation for star-forming galaxies and AGNs. We can see that the spectral shape obtained with the standard 'AGN' command in Cloudy, and adopting the parameter set from \citet{risaliti00} (black line), does not differ significantly from more recent SEDs, at least up to the soft X-ray regime ($\sim 1$ keV), and can be considered as fairly representative of an AGN with median properties (E$_\text{peak} \simeq 60$ eV, p$_{NT} =0.25$, and $\Gamma=2.0$ in the OXAF models). In the following of the paper, we consider this as our default AGN model, but we discuss potential changes of the line ratios as due to different SEDs. For convenience, we also release the emission line predictions for the entire grid of photoionization models (both star-forming and AGN) analyzed in this paper \footnote{The star-forming models can be retrieved on GitHub: \url{https://github.com/Anthony96/star-forming_models.git}. \\ The AGN models are available at the following page: \url{https://github.com/Anthony96/AGN_models.git}}.

\subsection{Predictions for optical lines: the BPT diagram}\label{optical_lines} 

\begin{figure}[ht!]
    \centering
    \includegraphics[angle=0,width=1\linewidth,trim={0.1cm 0cm 0.1cm 0.1cm},clip]{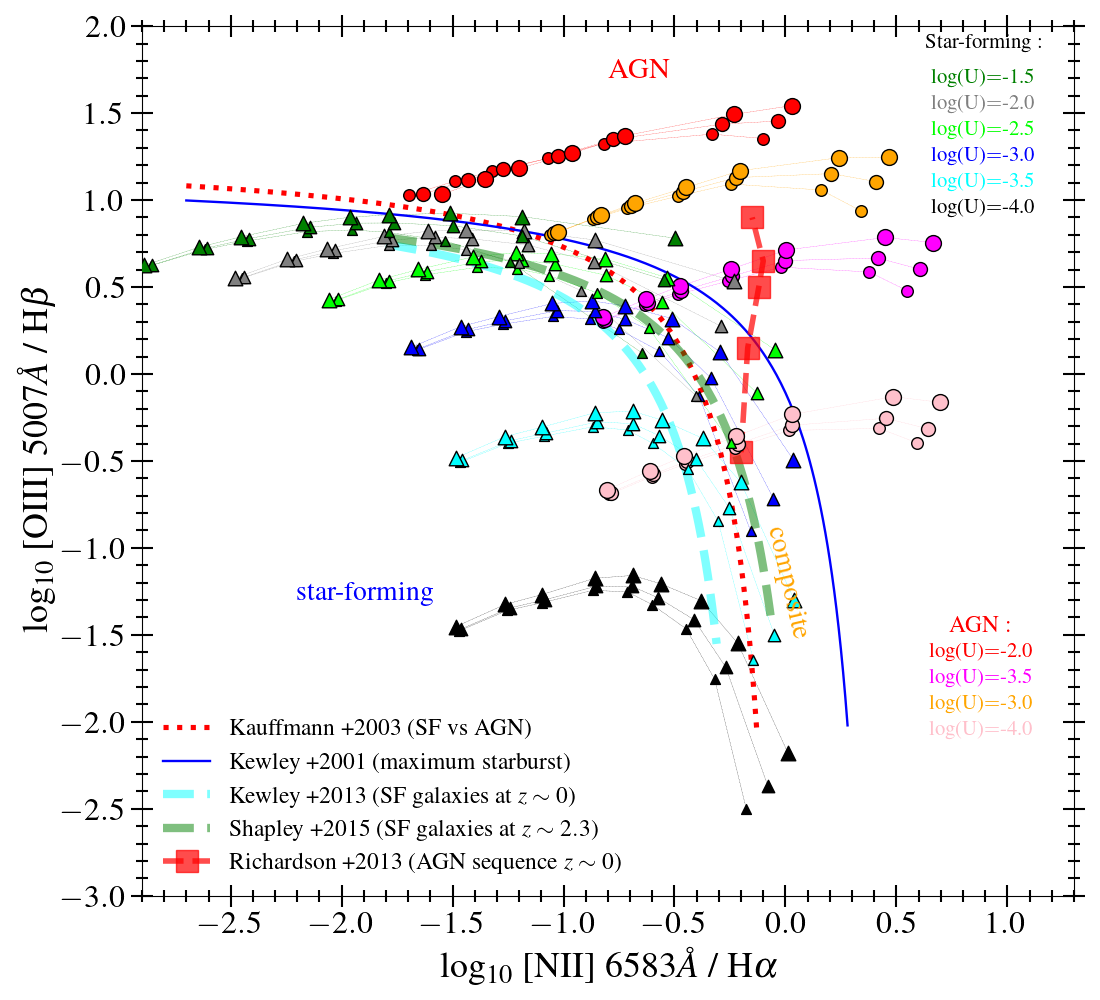}
    \caption{Prediction from our Cloudy models of the emission line ratios \OIII/H$\beta$ vs \NII/H$\alpha$ (BPT diagram). Triangles and circles represent the star-forming and AGN models, respectively. The gas-phase metallicity increases from left to right. The blue continuous line represents the separation line by \citet{kauffmann03}, while the red dotted line represents the maximum starburst model by \citet{kewley01}. The cyan and green dashed lines show the average location of star-forming galaxies at $z=0$ and $z=2.3$, respectively, from \citet{kewley13} and \citet{shapley15}. The big red squares and dashed line highlight the AGN sequence by \citet{richardson14}, with increasing ionization from bottom to top. 
    }\label{BPT_CLOUDY}
\end{figure}

We initially test the ability of our models to reproduce the classical BPT diagram, introduced originally by \citet{baldwin81}, and widely used to separate AGN and star-formation driven ionization from the local Universe to higher redshift. This diagram compares rest-frame optical emission line ratios, namely \OIII/H$\beta$ and \NII/H$\alpha$. 
The predictions of our Cloudy models are shown in Fig. \ref{BPT_CLOUDY}, where the circles are the AGN models for different values of $\log U$, metallicity, and gas density. % $\alpha$ index,
Higher gas densities are represented with bigger markers, while the marker colors are indicative of different $\log (U)$, as shown in the legend. Triangles represent the star-forming models with varying parameters of the gas shell. 

\begin{figure*}[t!]
    \centering
    \includegraphics[angle=0,width=0.98\linewidth,trim={0.1cm 5cm 5.5cm 0.1cm},clip]{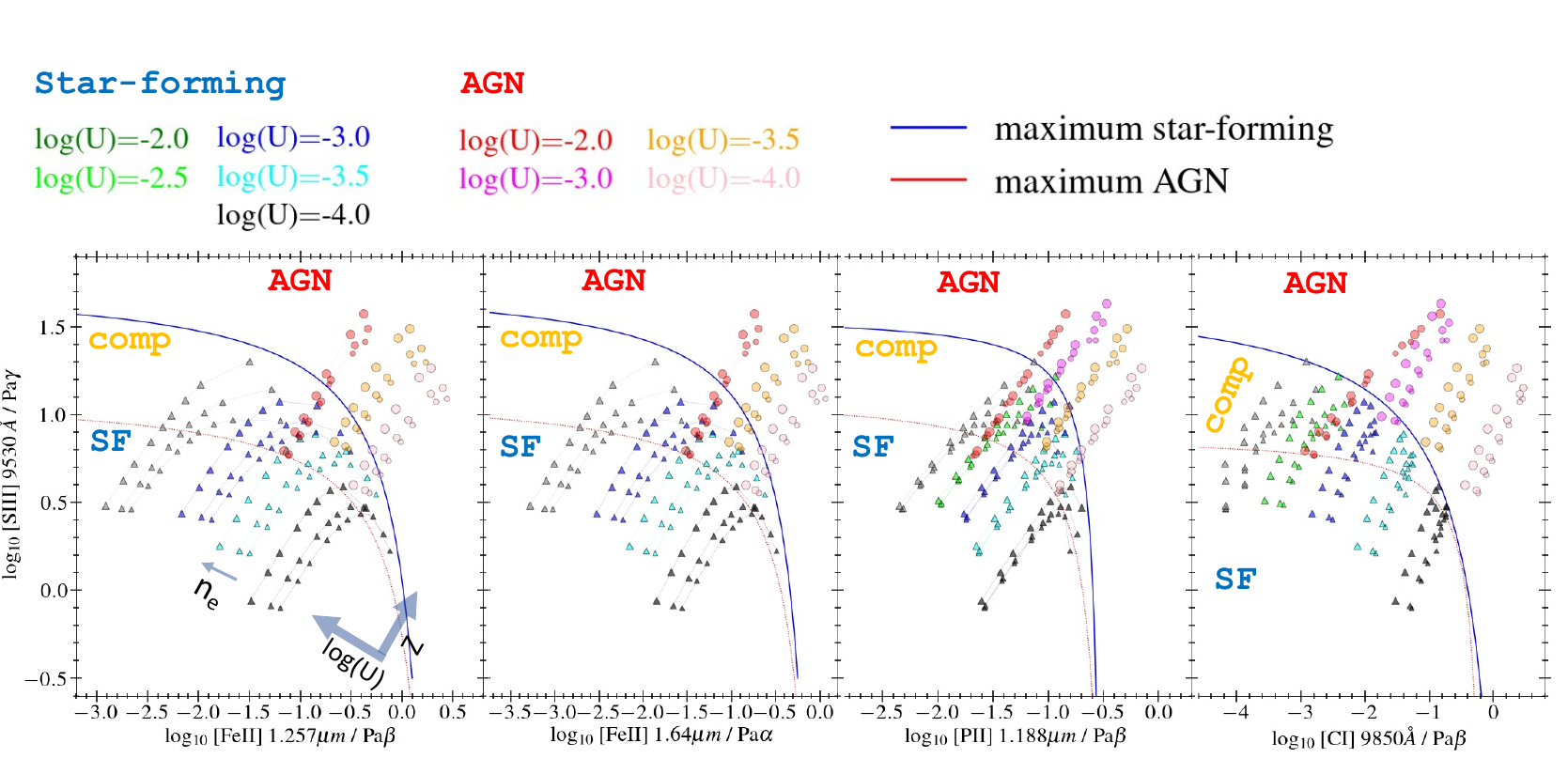}
    \includegraphics[angle=0,width=0.99\linewidth,trim={0.1cm 5cm 4.5cm 4.3cm},clip]{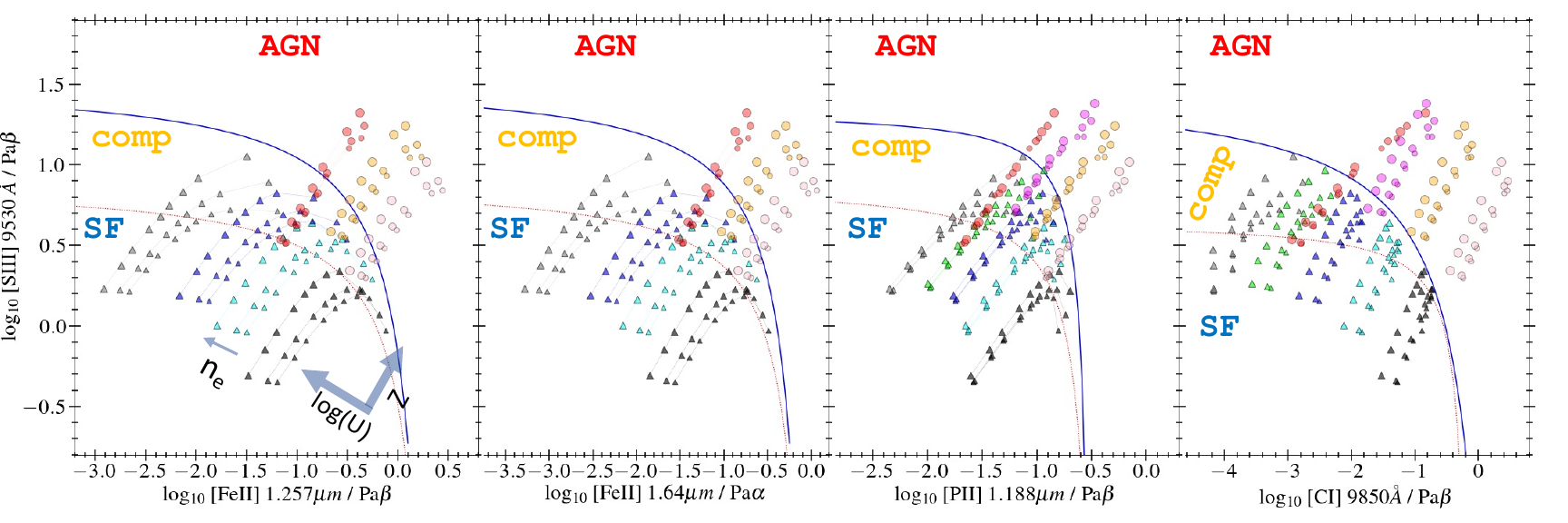}
    \caption{Figure showing the near-IR diagnostic diagrams analyzed in this paper: {Top row:} \SIII/\PaG\ as a function of \FeII/\PaB\ (Fe2S3-$\beta$), \FeII/\PaA\ (Fe2S3-$\alpha$), \PII/\PaB\ (P2S3), and \CI/\PaB\ (C1S3). AGN models are represented as circles, colored as a function of $\log(U)$, and with sizes increasing with density. Star-forming models are the triangle symbols. The maximum AGN line and the maximum starburst separation limit are defined by a dotted red line and a continuous blue line, respectively. {Bottom row:} Same as above but as a function of the \SIII/\PaB\ line ratio on the y-axis.
    }\label{nearIR_models}
\end{figure*}

Models with increasing $\log U$ tend to occupy a region toward the upper part of the diagram with a higher \OIII/\Hb\ ratio. At fixed conditions, models with higher metallicity tend instead to move toward the right part of the diagram with higher \NII/\Ha, as the nitrogen abundance also increases with $Z$. The composite region in Fig. \ref{BPT_CLOUDY} is where the AGN and star-forming models overlap, and where both the NLR and HII (star-forming) regions might contribute to the global emission.
While a more detailed analysis of the role of each parameter in the scatter of the BPT diagram has been discussed in many other works (e.g., \Citealt{brinchmann08}, \Citealt{masters16}, \Citealt{faisst18}, \Citealt{curti22}, and references therein), we want to remark here that our AGN and star-forming models are globally consistent with observations from $z=0$ to $z=3$ \citep{kewley13,shapley15,richardson14}, and with the typical separation lines adopted in the literature to constrain the AGN location \citep{kauffmann03} and the maximum starburst condition \citep{kewley01}.  %, or where we cannot be conclusive on the nature of the source. 
As expected given the similarity of ionizing shapes, the line predictions obtained with the OXAF models occupy a similar parameter space to those derived with our default AGN model. The same trends discussed above as a function of $\log U$, $Z$, and gas density, also hold for the models of \citet{thomas16}. For completeness, we show the location of these additional models in the BPT diagram in the Appendix \ref{appendix1}.
We finally note that our line predictions for AGNs in the BPT are similar to those presented by \citet{ji20}, which are also derived with Cloudy. For the star-forming models, we also checked the predictions assuming a simple black-body with temperature T $=50000$ K as ionizing spectrum, yielding similar results.

The BPT diagram is also routinely used to study the ionizing properties of galaxies well beyond the local Universe. %at redshifts significantly higher than $0$. 
As shown by \citet{topping20}, star-forming galaxies at $1<z<3$ have harder ionizing spectra at higher $z$, lower metallicity, and increased alpha element abundances (i.e., increased O/Fe abundance ratio) compared to local samples. However, the combined effect of this evolution is rather limited in the BPT diagram, with only a slight offset observed for the star-forming population toward the upper right part of the diagram compared to the local BPT (see also Fig. 1 of \Citealt{bian18}). 
%However, the effects on the emission line ratios in the BPT diagrams are rather limited. 
Most importantly, they show that normal star-forming galaxies in the above redshift range do not cross the maximum starburst condition of \citet{kewley01}. Similarly, the separation line introduced by \citet{kewley13} to classify AGN and SF sources at $z\simeq 1.5$ falls entirely on the left of our maximum starburst condition, which thus provides a more stringent condition for AGNs.

\citet{hirschmann17} also study the evolution of the BPT diagram toward higher redshifts, claiming that the ionization parameter, which is regulated by the star-formation history, is the main responsible for the enhancement of the \OIII/\Hb\ ratio compared to local samples, but star-forming galaxies at $z<3$ do not typically overcome the maximum starburst separation lines set in previous works in all the optical diagnostics (see Fig. 3 in that paper). 
Therefore, $z \sim 3$ represents the redshift limit up to which the local AGN conditions in the BPT and similar optical diagrams have been tested. Similarly, we consider that our models can also be safely used if we extend our analysis up to at least $z \sim 3$. Nevertheless, we will discuss in the following sections the possible effects of the $\alpha$-enhancement and variations of the depletion strengths when considering the emission of iron-like elements. 
With these models in hand, we analyze the predictions of AGN and star-forming models for emission line ratios in the rest-frame near-IR.

% % % -------------------------------------------------------------------------------

\section{Results}\label{results}

In this section, we show the near-IR rest-frame emission line predictions of our star-forming and AGN models. Then we compare them to the $64$+$65$ observed sources in the redshift range from $z=0$ to $z=3$.  

\subsection{Infrared rest-frame predictions}\label{nearIR_diagnostics_models}

We utilize the models described in Section \ref{CLOUDY} to analyze the emission line ratios in the rest-frame near-IR, from $\sim0.8$ to $\sim2\ \mu m$. 
Further extending the analysis to longer wavelengths dramatically reduces the number of objects targeted by CEERS. 
For example, longer wavelength transitions of the ionized hydrogen, like the Brackett emission line series, are more difficult to detect as those lines are significantly fainter than Pa$\alpha$, owing to theoretical Br$\gamma$ and Br$\delta$ to \PaA\ ratios of $0.06$ and $0.09$, respectively. 
We focus on near-IR lines in the above spectral range that are intrinsically brighter, as discussed in Section \ref{sample_selection} and \ref{mpfit}, and hence easier to detect for a larger number of galaxies. In particular, we have explored all combinations of close, bright near-IR emission line ratios and how they vary between purely star-forming systems and AGNs. 

\subsubsection{Iron-based diagnostics for AGNs}\label{iron_diagnostics}

% NOW PUT TABLE AFTER THE FIRST FIGURE : 
\begin{table*}[t!]
\centering{\textcolor{blue}{Maximum starburst lines}}
\renewcommand{\arraystretch}{1.5} 
\vspace{-0.2cm}
\small
\begin{center} { %\small
\begin{tabular}{ | m{2.5cm} | m{3.5cm} | m{3.5cm} | m{2cm} | m{2cm} | m{2cm} |} 
  \hline
  \textbf{diagram} & \textbf{Y} & \textbf{X} & \textbf{A} & \boldsymbol{$\log(X_0)$} & \textbf{B} \\ % \boldsymbol{$\lambda_{rest} [\mu m]$} \\ 
  \hline
  \textbf{Fe2S3-$\beta$} & \xSIII/\PaG & \xFeIIa/\PaB & $0.87$ & $0.47$  & $1.80$ \\
  \textbf{Fe2S3-$\alpha$} & \xSIII/\PaG & \xFeIIc/\PaA & $0.87$ & $0.12$  & $1.80$ \\
  \textbf{P2S3}    & \xSIII/\PaG & \xPII/\PaB   & $0.17$ & $-0.48$ & $1.57$ \\
  \textbf{C1S3}    & \xSIII/\PaG & \xCI/\PaB    & $2.23$ & $1.095$ & $1.95$ \\
  \hline
\end{tabular} }
\end{center}

\centering{\textcolor{red}{Maximum AGN lines}}
\renewcommand{\arraystretch}{1.5} 
\vspace{-0.2cm}
\small
\begin{center} { %\small
\begin{tabular}{ | m{2.5cm} | m{3.5cm} | m{3.5cm} | m{2cm} | m{2cm} | m{2cm} |} 
  \hline
  \textbf{diagram} & \textbf{Y} & \textbf{X} & \textbf{A} & \boldsymbol{$\log(X_0)$} & \textbf{B} \\ % \boldsymbol{$\lambda_{rest} [\mu m]$} \\ 
  \hline
  \textbf{Fe2S3-$\beta$} & \xSIII/\PaG & \xFeIIa/\PaB & $0.56$  & $0.41$  & $1.13$   \\
  \textbf{Fe2S3-$\alpha$} & \xSIII/\PaG & \xFeIIc/\PaA & $0.56$  & $0.06$  & $1.13$   \\
  \textbf{P2S3}    & \xSIII/\PaG & \xPII/\PaB   & $0.285$ & $-0.43$ & $1.115$  \\
  \textbf{C1S3}    & \xSIII/\PaG & \xCI/\PaB    & $0.405$ & $0.36$  & $0.93$   \\
  \hline
\end{tabular} }
\end{center}

\vspace{-0.1cm}
\caption{\small Table of coefficients for the maximum star-forming and AGN lines, as defined in Equation \ref{Eq:nearIR_all}. If \SIII/\PaB, \SIII/Pa$\delta$, or \SIII/Pa$\epsilon$ are used on the x-axis instead of \SIII/\PaG, the separation lines obtained with the coefficients listed in this table should be translated along the y-axis by $-0.225$, $+0.195$, and $+0.365$, respectively.
}
\vspace{-0.7cm}
\label{table_coefficients}
\end{table*}

One of the most promising near-IR diagnostic diagrams that we find is the one comparing the \xSIII/\PaG\ to the \xFeIIa/\PaB\ line ratios, which we identify as \textbf{Fe2S3-$\beta$}, and is shown in Fig. \ref{nearIR_models}-\textit{top-left}. The lines involved in each ratio are sufficiently close in wavelength that the dust corrections can be neglected in all conditions. 
In this parameter space, the AGN and star-forming models, represented respectively as circles and triangles, are clearly separated, with a narrow overlapping region. 
The star-forming galaxies occupy the bottom left region with $\log_{10}$ \xFeIIa/\PaB\ ratios spanning almost $3$ orders of magnitude from $\sim -3.0$ to $0$, and \xSIII/\PaG\ showing a similarly large variation between $-0.3$ and $1.5$. The AGN models occupy instead the upper-right part of the diagram with higher \xFeIIa/\PaB\ and \xSIII/\PaG\ line ratios, up to $\log_{10}$ \FeII/\PaB\ $\simeq 0.7$ and $\log_{10}$ \SIII/\PaG\ $\simeq 1.7$, respectively. 
There is also an intermediate (also dubbed 'composite') region where the two models overlap. The objects lying in this region, while being consistent with pure stellar or pure AGN photoionization, might represent sources with intermediate properties, where we might have the contribution from both photoionization mechanisms. 

The theoretical modeling with Cloudy also has the advantage that we can investigate how the different physical properties of the emitting source and surrounding gas cloud influence the observed line ratios. As shown in Fig. \ref{nearIR_models}-\textit{top}, both the \xSIII/\PaG\ and the \xFeIIa/\PaB\ ratios increase with metallicity, resulting from a higher abundance of coolants. However, while the former line ratio has a turning point at around Z$_\odot$ (depending on $\log (U)$), which mimics the behavior of the R2 ($= \log$ \xOII/\Hb) and S2 ($= \log$ \xSII/\Hb) optical indices \citep{curti20} with Z$_\text{gas}$, the latter ratio increases always monothonically, suggesting that it is more sensitive and a better tracer of metallicity in the high Z regime. 
The overlapping region thus includes star-forming galaxies with higher gas-phase metallicity, AGNs with more metal poor NLRs, or sources with both stellar and AGN contribution.
An increase of the ionization parameter produces instead an enhancement of \SIII/\PaG\ but a decrease of \FeII/\PaB. However, the \SIII/\PaG\ line ratio saturates at an ionization parameter $\log (U) \simeq -2$, reaching its maximum in all conditions of density and metallicity. At higher $\log (U)$, it starts to slightly decrease, as the emission from higher ionized species of sulfur is favored in these extreme conditions. For representative purposes, we do not show in the figure the predictions from models with $\log (U) = -1.5$ and $-1$, as they would overlap with the models at $\log (U) = -2$. 
Finally, a higher density has a similar effect to that seen for the ionization parameter, with \SIII/\PaG\ slightly increasing (and \FeII/\PaB\ slightly decreasing) from $n_e = 10^2$ to $10^4$ $cm^{-3}$.
These models shed light on new promising diagnostics of metallicity and ionization for dust-enshrouded systems, which will be better investigated in the future. 
On the other hand, if we can independently measure these parameters from near-IR lines, the star-forming and AGN models would be even more separated, and the classification more precise.

\begin{figure*}[ht!]
    \centering
    \includegraphics[angle=0,width=0.98\linewidth,trim={0.1cm 5.5cm 3cm 1.1cm},clip]{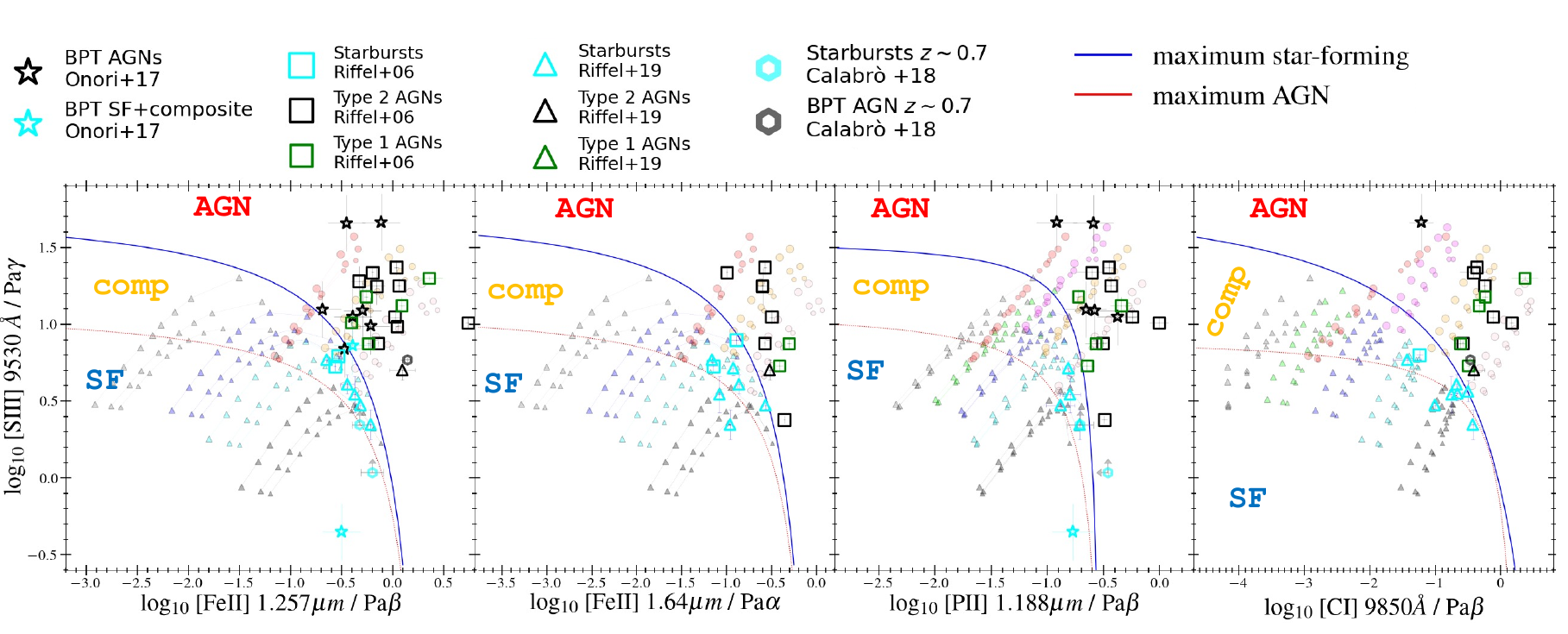}
    \includegraphics[angle=0,width=0.98\linewidth,trim={0.1cm 5.5cm 3cm 4.5cm},clip]{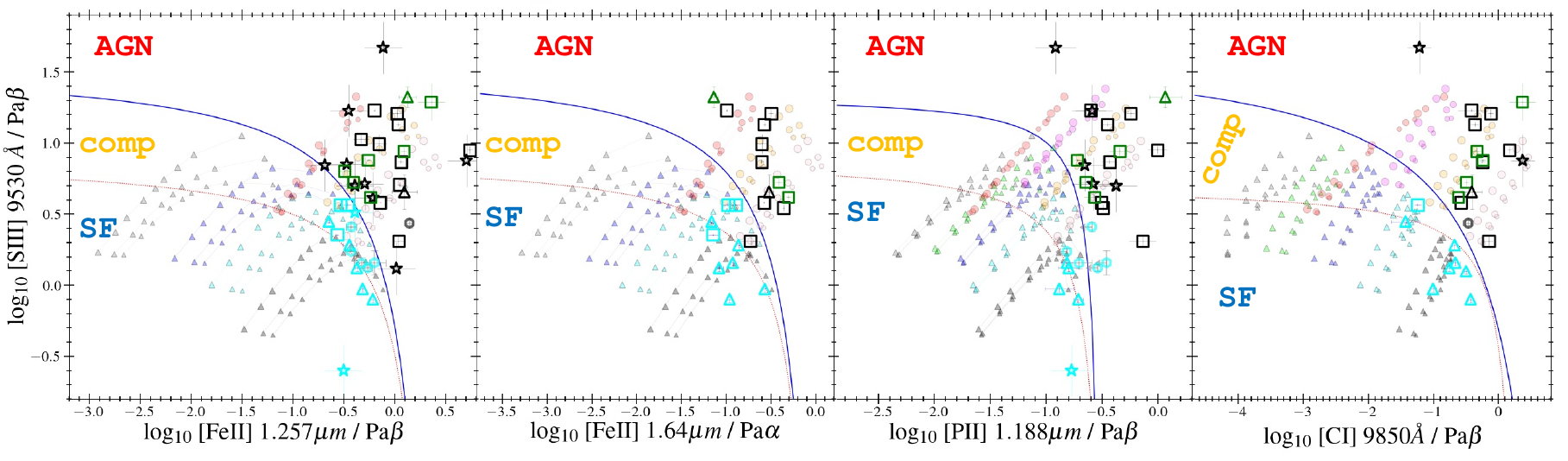}
    \caption{Figures displaying the location of local and intermediate redshift AGNs and starbursts in the near-IR diagnostics defined in the paper: the Fe2S3-$\beta$, Fe2S3-$\alpha$, P2S3, and C1S3 diagrams from left to right, as a function of \SIII/\PaG\ on the y-axis (\textit{top row}) or \SIII/\PaB\ (\textit{bottom row}).  
    Sources coming from \citet{riffel06} and \citet{riffel19} are drawn with square symbols and triangles, respectively. Those presented and measured by \citet{onori17} are identified with empty stars, while starbursts from \citet{calabro19} are shown with cyan hexagons. The underlying models are the same as described in Fig. \ref{nearIR_models}.  % overdensity flag (as defined in the text) in the left panel, 
    }\label{nearIR_observations_local}
\end{figure*}

In order to classify the galaxy spectra and provide a clear, quantitative criterion to separate sources dominated by AGN or star formation, we derive separation lines in analytical form, as already done in the optical bands in previous works. 
We first derive a line that encloses all the star-forming models that we have analyzed, by fitting a rectangular hyperbola to regions close and on the upper-right side of the highest metallicity SF models, for each different ionization parameter family, corresponding to triangles with the same colors in Fig. \ref{nearIR_models}. 
This can be dubbed as the maximum star-forming (or starburst) line, defining a clear boundary for SF models located in the bottom-left corner of the diagram.
Similarly, we also derive a boundary, dubbed the maximum AGN line, including the parameter space where the observed line ratios can be explained by an AGN. The points lying outside of the maximum starburst line in the AGN region can be identified as secure narrow-line AGNs.
The best-fit separation lines are written in general as : 
\begin{equation}\label{Eq:nearIR_all}
\log Y=\ffrac{A}{(\log X - \log X_0)}+B,
\end{equation}
where the coefficients A, B, and $\log X_0$ depend on the emission line ratios considered. A summary including all the coefficients used in Equation \ref{Eq:nearIR_all} is presented in Table \ref{table_coefficients} for all the diagnostic diagrams. 
 
We also present a diagnostic diagram (dubbed \textbf{Fe2S3-$\alpha$}) that is based on the measurement of the \xFeIIc/\PaA\ line ratio on the x-axis. Even though these two lines can be detected in a limited redshift range ($z \lesssim 1.75$) by JWST-NIRSpec (\PaA\ is detected for less than one-half of the CEERS sample), and despite the \xFeIIc\ being intrinsically $\sim15\%$ fainter than \xFeIIa, they can provide useful additional contraints on the nature of dust-enshrouded sources as they probe longer wavelengths than the previous near-IR diagnostics. Nevertheless, they could be reached at higher redshifts with MIRI observations.  

The model predictions for the Fe2S3-$\alpha$ diagnostic are shown in the second panel of Fig. \ref{nearIR_models}-\textit{top} as a function of \SIII/\PaG. Considering that the involved ionized species are the same as in the Fe2S3-$\beta$ diagnostic, the behavior of the models as a function of the various parameters is similar. Considering the intrinsic line ratios between \PaA\ and \PaB, and between \xFeIIa\ and \xFeIIc, the new separation lines can be obtained by translating the Fe2S3-$\beta$ corresponding lines along the x-axis by $-0.35$ dex, that is, by setting the $X_0$ coefficient $0.35$ dex lower (see Table \ref{table_coefficients}).

The diagnostic diagrams presented above require a measurement of \PaG. Among the recombination lines used in our diagrams, \PaG\ is usually the faintest for the majority of sources. 
For this reason, we also present an alternative set of diagrams where we use \PaB\ to compute the line ratio on the y-axis. These alternative diagrams are shown in Fig. \ref{nearIR_models}-\textit{bottom}. In contrast with the previous ones, these are slightly more dependent on dust attenuation. %, especially in case of strong obscuration.
However, we have estimated that knowing A$_V$ with an uncertainty of $\sim 0.5$ is generally enough to constrain the \SIII/\PaB\ ratio with an uncertainty of $0.05$ dex in case of strong obscuration (A$_V \simeq 5$ mag), or much lower in case of moderate and low attenuation (A$_V \leq 2$ mag). 
Therefore, this dust correction does not significantly impact the classification of the source in this diagnostic.  

As was done in the first case, for these diagrams we can also derive separation lines to distinguish between the emission from star-forming galaxies and the narrow-line region of AGNs. Given the intrinsic ratio between \PaG\ and \PaB, the new maximum starburst and maximum AGN lines are obtained by simply translating downwards the curves described in Equations \ref{Eq:nearIR_all} by $0.225$ (i.e., the intrinsic \PaG/\PaB\ ratio in logarithm) along the y-axis. In practice, we subtract this constant value from the second member of Equation \ref{Eq:nearIR_all}. The new lines are visible in Fig. \ref{nearIR_models}-\textit{bottom} with the same color scheme as before.
In principle, this approach can be applied to any other Paschen line available in the spectrum. For example, if Pa$\delta$ or Pa$\epsilon$ are detected, we can use \xSIII/Pa$\delta$ or \xSIII/Pa$\epsilon$ on the y-axis, in which case we should translate upwards the separation lines calculated for \xSIII/\PaG\ by $0.195$ and $0.365$ dex, respectively. 

Finally, if we consider the AGN models developed by \citet{thomas16}, they do not significantly alter the line ratio predictions compared to our default AGN SED in both the Fe2S3-$\beta$ and Fe2S3-$\alpha$ diagnostics, as they tend to occupy a similar parameter space to that shown for AGNs in Fig. \ref{nearIR_models} for all the physical ranges of the three model parameters E$_\text{peak}$, p$_{NT}$, and $\Gamma$ (see Table \ref{table_models}). The predicted lines show a similar dependence on metallicity, ionization parameter, and gas density to that explained above. The same conclusions also hold for the other near-IR diagnostic diagrams analyzed in this paper and presented in the following subsections. For completeness, we include in Appendix \ref{appendix1} a more detailed discussion about the effects of those parameters for all the near-infrared line ratios considered in this section.

\subsubsection{Phosphorus-based AGN diagnostics}\label{phosphorus}

In the third panels of each row in Fig. \ref{nearIR_models}, we present another useful diagnostic diagram in which we compare the \SIII/\PaG\ (or \SIII/\PaB) to the \xPII/\PaB\ line ratio, dubbed \textbf{P2S3}. As in the previous case, we can see that the separation between star-forming and AGN models is rather clear, with the former occupying the lower-left region of the diagrams with lower \SIII/\PaG\ (\SIII/\PaB) and lower \xPII/\PaB, while AGNs are located in the upper right corner as due to their higher ratios. We can also see in both panels an intermediate region in the middle where the two models overlap. 
As for the ratio between \FeII\ and Paschen lines, here the \PII/\PaB\ ratio also increases with metallicity, even though it spans a narrower range ($\sim2$ dex), between $-2.5$ and $-0.5$ (in logarithmic scale) for star-forming galaxies, and from $-1.5$ to $\sim 0$ for AGNs, suggesting that \PII\ is slightly less sensitive to metallicity than \FeII. On the other hand, sharing the same y-axis with previous diagrams, the models saturate at $\log (U) \simeq -2$.
With the same procedure adopted before, we can define both an AGN and a starburst boundary by applying Equation \ref{Eq:nearIR_all}, from which we can assess the dominant ionizing source in a galaxy. The coefficients derived for the \SIII/\PaG\ (or \SIII/\PaB) vs \PII/\PaB\ diagrams are listed in Table \ref{table_coefficients}. 

Compared to the \FeII\ based diagnostics, these diagrams require a measurement of the \xPII\ line, which is fainter than \FeII\ and thus harder to detect, especially for individual star-forming galaxies. Indeed, the \PII/\PaB\ ratios span smaller values than \FeII/\PaB. %, with a range between $-2.1$ and $-0.6$ for star-forming galaxies, and from $-1.4$ to $\sim 0$ for AGNs.  
Despite this, an upper limit on this ratio would still be very useful to assess a possible AGN contribution. % As an alternative, we can stack multiple spectra to facilitate the detection of \PII\ for samples of star-forming galaxies, as we will see later for the CEERS sample.
% Furthermore, as we will explain in the discussion section, \PII\ is fundamental to verify a possible major contribution from shocks to the observed spectrum of a galaxy.

\subsubsection{Carbon-based AGN diagnostics}\label{carbon}

Finally, we identify a third diagnostic diagram which is based on the measurement of \xCI. This diagram compares the \SIII/\PaG\ (or \SIII/\PaB) on the y-axis to the \xCI/\PaB\ line ratio on the x-axis, and can be identified as the \textbf{C1S3} diagram (fourth panels in each row of Fig. \ref{nearIR_models}). The latter quantity can span six orders of magnitudes if we include both star-forming and AGN models, more than any other ratio that we have introduced before. However, in contrast to previous diagrams, this spread is mostly driven by the ionization parameter rather than the gas-phase metallicity. In star-forming galaxies, we can find values of $\log_{10}$ \CI/\PaB\ $\simeq -4$ for models with higher ionization ($\log (U) \sim -2$), which increase up to $\simeq - 0.5$ for the lowest ionization models considered ($\log(U)=-4$). In AGNs, we can have \CI\ from $\sim 0.1\%$ of \PaB\ at $\log(U) \sim -2$ up to $\sim 2.5$ times brighter than \PaB\ at $\log(U)=-4$. Similarly to the previous diagrams, the line ratio in both axis saturates at $\log(U) \sim -2$. To avoid overcrowding, we do not represent the extreme ionization models, which overlap with those at $\log(U) = -2$.
Overall, while still showing a small secondary dependence on metallicity (i.e., higher Z$_\text{gas}$ slightly increase \CI/\PaB), \CI\ is a poor tracer of the metal content and a better tracer of $\log (U)$ than \FeII\ and \PII, at least in the range of $\log (U)$ from $-4$ to $-2$. Similarly to the previous elements, also \CI/\PaB\ decreases at higher gas densities.

Even considering all possible combinations of chemical composition, ionization parameter, and gas density, the star-forming and the AGN models are in general well separated at $\log (U) \leq -3$, more than in any other near-IR diagnostic diagram, while overlap between SF and AGN is present at higher ionization ($\log (U) \sim -2$). However, as we will see later, AGNs from low to intermediate redshift do not typically populate this parameter space, preferring lower ionization parameters and higher \CI/\PaB\ ratios, hence this is not an important limitation for this work. 
As for \PII, \CI\ is mostly detected for individual sources if they are AGNs, while in purely star-forming galaxies it can be too faint with \CI/\PaB\ $<< 0.1$. 
Even though a small correction should be preferably applied to the \CI/\PaB\ (e.g., using the A$_V$ estimated from available Balmer and Paschen lines), the separation between the AGN and star-forming models allows us to assess with enough confidence the ionizing type of the source at $\log (U) \leq 3$. 

\begin{figure*}[t!]
    \centering
    \includegraphics[angle=0,width=0.95\linewidth,trim={0.1cm 9cm 9.3cm 0.1cm},clip]{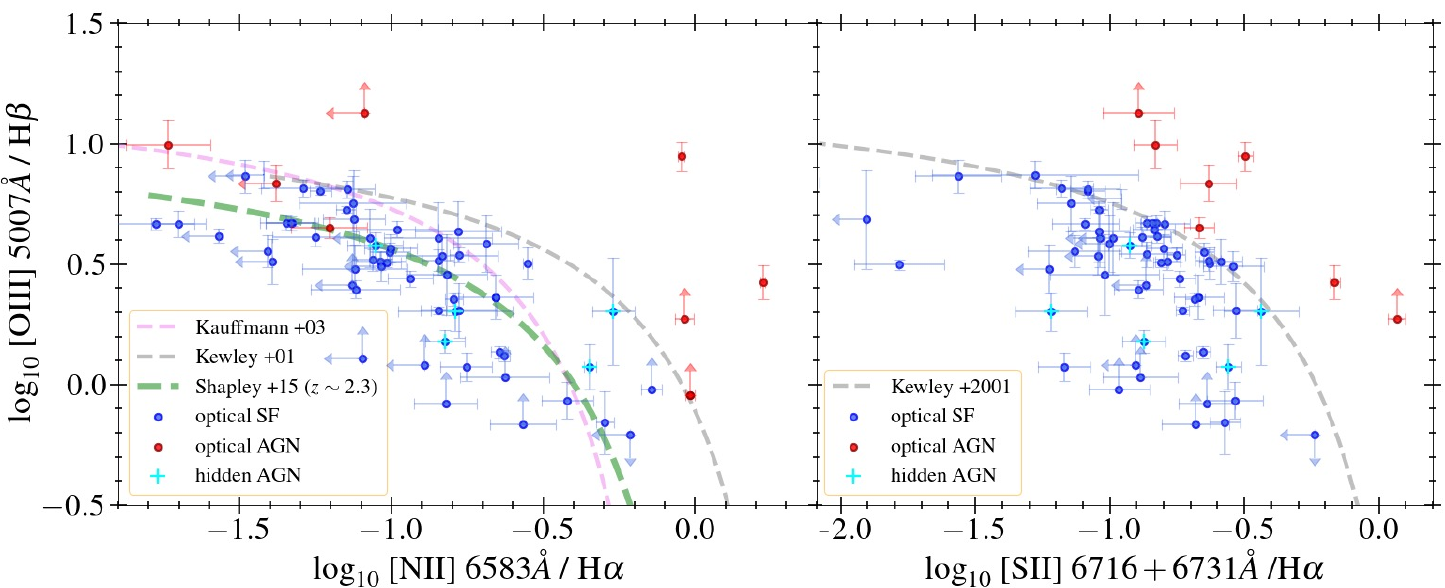}
    \caption{Figure showing the position of CEERS galaxies selected at $1<z<3$ in the BPT diagram (\textit{left}) and in the \SII-based optical diagnostic (\textit{right}). Optical star-forming galaxies and AGNs are represented with blue and red circles, respectively. To be idenfitied as an AGN, a source must lie in the AGN region (within its $1\sigma$ uncertainty) in at least one of the two optical diagnostics. Cyan crosses are included to highlight hidden AGNs, that is, optical star-forming but near-IR AGNs, as explained in the text. The green dashed line is a relation representative of star-forming galaxies at $z=2.3$, from \citet{shapley15}. The violet and gray dashed lines are the classification criterion and the maximum starburst condition from \citet{kauffmann03} and \citet{kewley01}, respectively. 
    }\label{BPT_highz}
    \vspace{-0.3cm}
\end{figure*}

\subsection{Near-infrared AGN diagnostics at low and intermediate redshifts}\label{nearIR_diagnostics_local}

We first check how the new near-IR diagnostic diagrams perform at $z \leq 1$ and their ability to distinguish between local AGNs and sources powered by star formation. To this aim, we use the observations and the line measurements described in Sections \ref{Magellan_starbursts} and \ref{local_galaxies_AGNs}. 
For line ratios including the \PaB\ line, we correct the fluxes for dust attenuation, using the value of A$_V$ estimated from a combination of Paschen and Balmer lines (i.e., from the \PaA/\PaB\ and \PaB/\Ha\ line ratios), or the \xFeIIa\ and \xFeIIc\ lines, as in \citet{riffel06}. We adopt the median of the three estimates if all those lines are available. 

In Fig. \ref{nearIR_observations_local}, we present the comparison between the observed \xFeIIa/\PaB, \xFeIIc/\PaB, \xPII/\PaB, and \xCI/\PaB\ line ratios as a function of \xSIII/\PaG\ (top row) and \xSIII/\PaB\ (bottom row), on top of our Cloudy model predictions. 
The observed galaxies are divided into three classes of objects, depending on their classification in the optical range: type 1 AGNs, type 2 AGNs, and starburst (i.e., star-forming, infrared bright) galaxies, which are drawn, respectively with green, black, and cyan colors. 

We can see that, globally, the separation lines defined in Section \ref{nearIR_diagnostics_models}  are able to separate the two classes of sources. Indeed, all the AGNs, either type 1 or type 2, have narrow line properties consistent with AGN models, and nearly all of them lie beyond the maximum starburst limit, in a region that cannot be explained by star formation alone. These local and intermediate-$z$ AGNs span the entire parameter range allowed by the models with different Z$_\text{gas}$ and $\log (U)$, even though the lowest ionization parameters are preferred ($-4 \lesssim \log (U) \lesssim -3$) in all the four diagnostics. 
On the other hand, optically selected starbursts, both at $z\sim0$ and at $z\sim0.7$, are overall consistent with the star-forming models, even though they tend to occupy the outer envelope of the SF region toward higher gas-phase metallicities (i.e., close to solar). This is somehow expected, given that they are selected as far-infrared bright sources, therefore they trace a more star-forming, dustier, and more metal-enriched population compared to typical star-forming galaxies at the same redshift. As for the AGN subset, also the starbursts are more in agreement with low ionization models, that is, $-4 \lesssim \log (U) < -3$. 
Even though we do not have statistical samples of normal, more metal-poor star-forming galaxies at $z \leq 1$ with near-IR spectral coverage, the model predictions suggest that their line ratios would span a larger region in the star-forming part toward the lower left, corresponding to lower metallicity models. We can study the normal star-forming galaxy population at higher redshifts with JWST, as we will show in the following section.  
We finally notice that these results are not significantly affected by the exact dust attenuation correction applied. % on line ratios where \PaB\ is present. 
Indeed, they would remain essentially unchanged even if we assume A$_V=0$ for all the sources.

\subsection{Near-infrared AGN diagnostics at higher redshifts}\label{nearIR_diagnostics_highz}

\begin{figure*}[ht!]
    \centering
    \includegraphics[angle=0,width=0.88\linewidth,trim={0.1cm 2.5cm 8.7cm 0.2cm},clip]{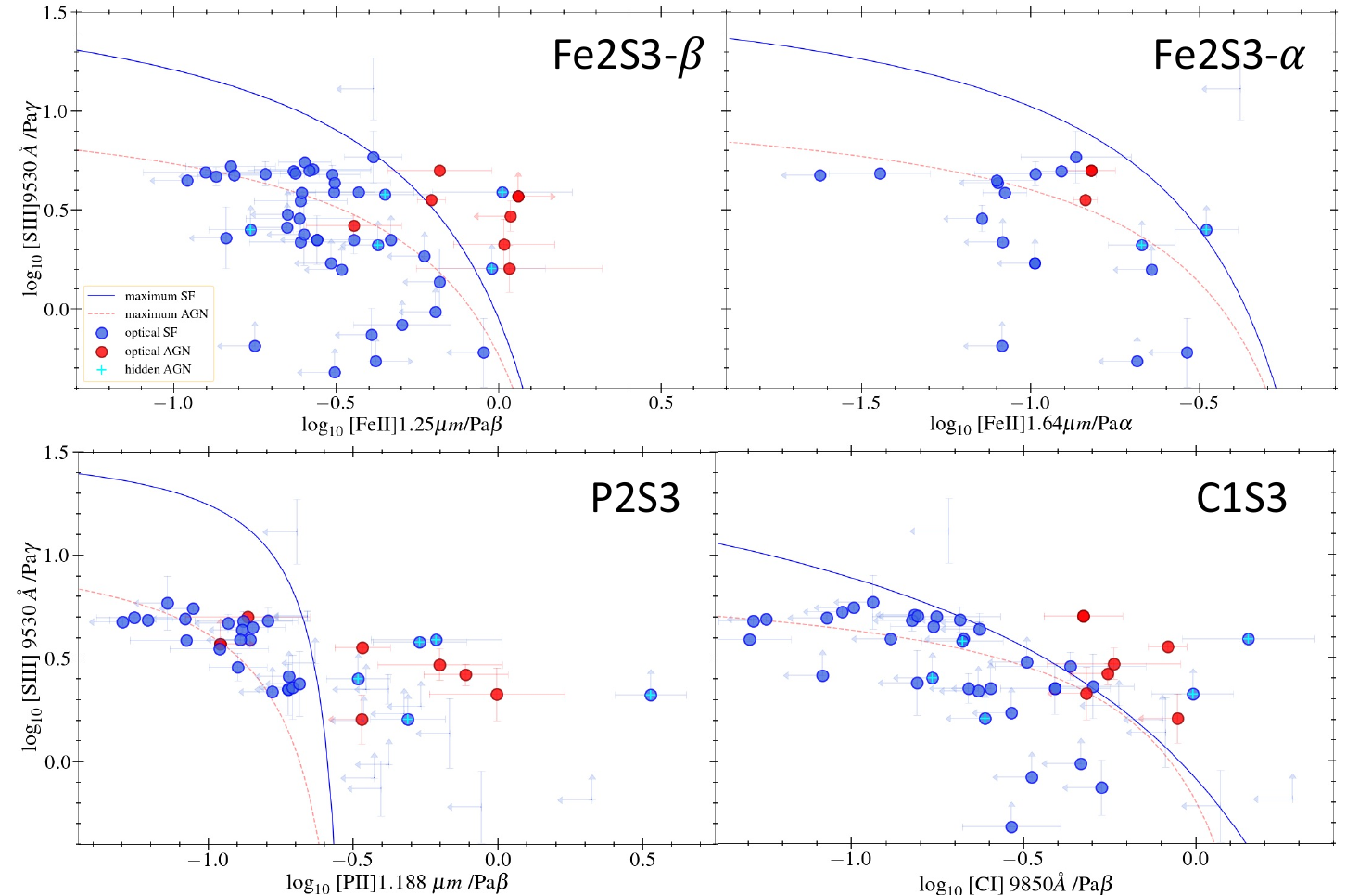}
    \includegraphics[angle=0,width=0.914\linewidth,trim={0.1cm 2.5cm 9cm 0.2cm},clip]{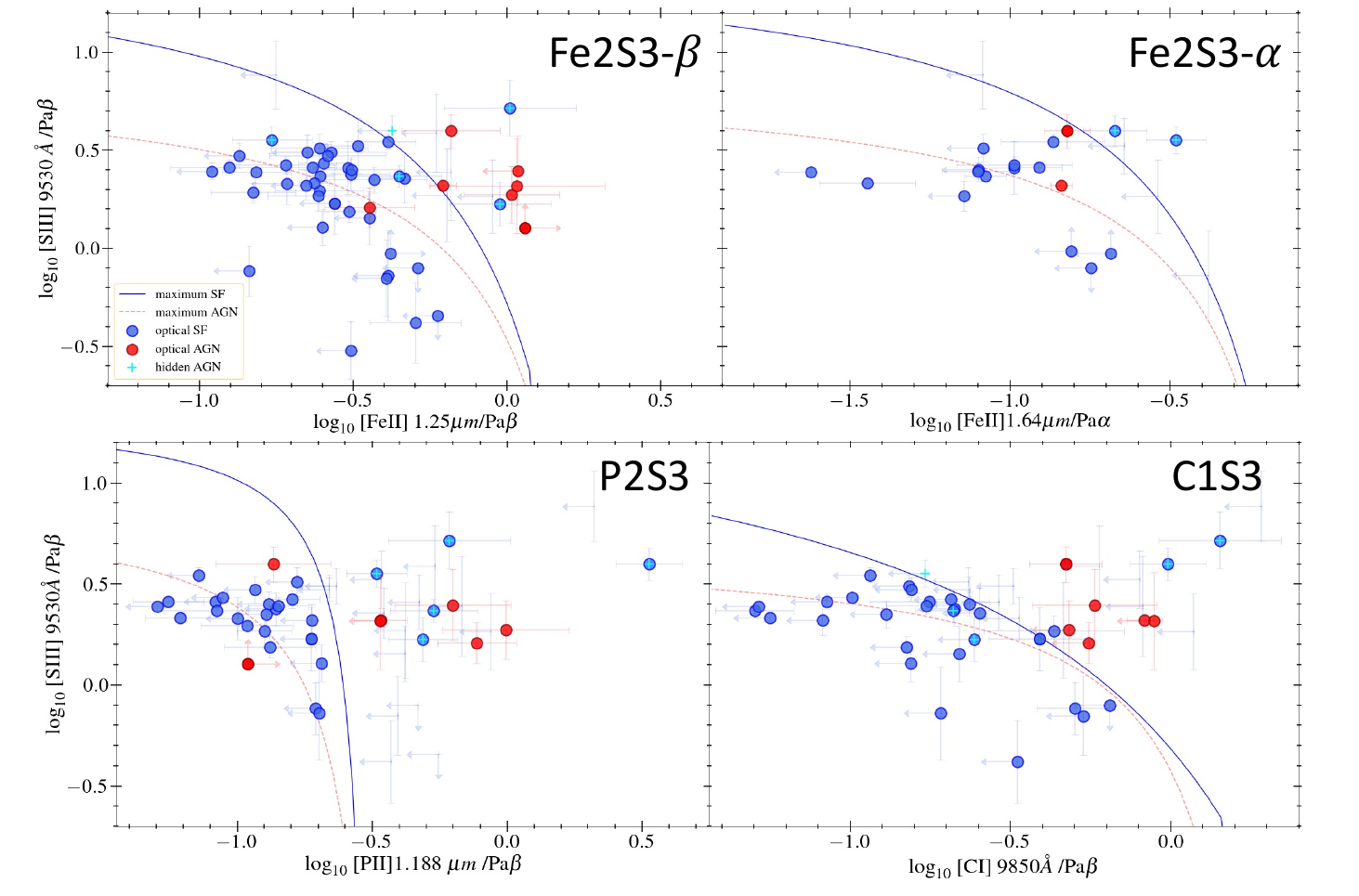}
    \vspace{-0.4cm}
    \caption{Near-IR diagnostic diagrams applied to our CEERS selected sample at $1<z<3$. Optical star-forming galaxies and AGNs are drawn with blue and red circles, respectively. Hidden AGNs (i.e., optical star-forming but near-IR AGN) are identified as cyan crosses.
    Optical star-forming galaxies with upper limits are drawn without symbols in blue if their line ratios are still consistent with being star-forming.
    The red dashed and blue continuous lines represent the maximum AGN and starburst limits (respectively) according to our models. The analytical form of the separation lines is described in Equation \ref{Eq:nearIR_all}, using the coefficients listed in Table \ref{table_coefficients}. 
    }\label{nearinfrared_highz}
\end{figure*}

\begin{figure}[ht!]
    \centering
    \includegraphics[angle=0,width=0.82\linewidth,trim={0.1cm 8cm 17cm 0.2cm},clip]{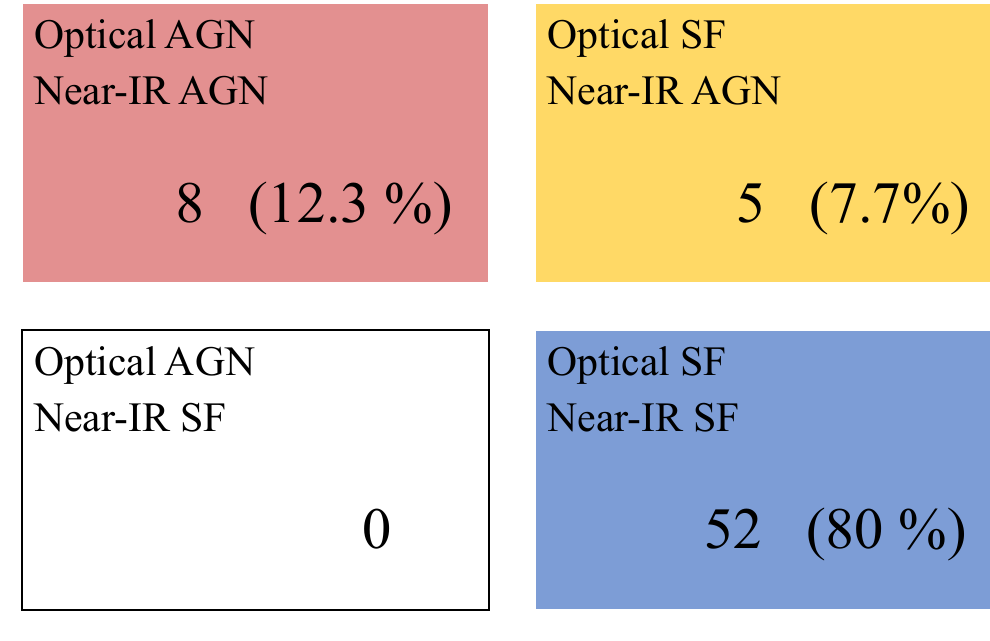}
    \caption{Summary of the AGN and star-forming galaxy classification obtained from the rest-frame optical and near-IR spectroscopic diagnostics for the sample of $65$ sources selected in CEERS. % overdensity flag (as defined in the text) in the left panel, 
    }\label{Fig:classification}
\end{figure}

We now study the properties of sources at $z>1$. This is the redshift range where JWST-NIRSpec is revolutionizing our understanding of galaxy evolution, in particular of dusty galaxies, thanks to its longer wavelength coverage compared to its predecessors. JWST surveys are observing more and more galaxies at cosmic noon with unprecedented sensitivity and spatial resolution, and CEERS provides by far the largest spectroscopic sample of galaxies at this epoch to allow the first statistical studies in the near-IR. 

Before investigating our new near-IR spectral diagnostics, we analyze the location of CEERS sources in the classical BPT and \xSII\ based diagram, to get a first understanding of their nature from the optical.
The results are shown in Fig. \ref{BPT_highz} on the left and right sides, respectively. The sources are color-coded based on their position relative to the separation lines of \citet{kewley01}: blue circles identify the sources consistent with star-forming models (within their $1 \sigma $ uncertainty), while red circles represent secure AGNs (i.e., not consistent with the star-forming region within $1 \sigma$) in at least one of the two optical diagnostics. 
Most of the CEERS galaxies are purely star-forming. However, we also identify $8$ \textit{optical AGNs} ($\sim 12 \%$ of the whole sample): all these sources lie in the AGN region of the \SII\ based diagram, except one system for which \SII\ is not available. Six of them are consistently identified as AGNs in the BPT. For the remaining two systems, one lies very close to the separation limit.  
Among the optical AGNs, we note that $3$ of them also have a broad component detected at S/N $>3$ in H$\alpha$ or H$\beta$ (CEERS ID 2919, 3129, 2904), thus should be considered as Type 1 AGNs, while the remaining systems are likely Type 2 AGNs in the optical (CEERS ID 2754, 5106, 12286, 16406, and 17496). %The ID 2754 does not have \SII\ available, so it only appears in the BPT. 
Two of the broad-line AGNs (2919 and 2904) are also detected by Chandra in X-rays \citep{nandra15}. 

It is interesting to notice that CEERS sources, both star-forming galaxies and AGNs with $z$ ranging $1 < z < 3$, span almost the entire parameter range of the models. Indeed, we find sources, located in the bottom right of the SF and AGN regions, that are in agreement with low ionization models (down to $\log(U) \simeq -3.5$) and consequently have higher metallicity, while the other galaxies and AGNs have increasingly higher $\log(U)$. Broad-line AGNs show a preference for lower ionization models compared to narrow-line AGNs, which are mostly residing in the upper left part of the BPT diagram. We note that the location of our star-forming galaxies in the BPT is broadly consistent with the relation found for typical star-forming systems at $z \sim 2.3$ \citep{shapley15}. In this star-forming sample, galaxies with upper limits in the \NII/\Ha\ ratio might have lower metallicity than those with similar \OIII/\Hb\ ratios. 

Moving now to the near-IR rest-frame diagnostics, we analyze the position of our galaxies in the \textbf{Fe2S3-$\beta$}, \textbf{Fe2S3-$\alpha$}, \textbf{P2S3}, and \textbf{C1S3} diagrams presented in Section \ref{nearIR_diagnostics_models}. 
All the results are shown in Fig. \ref{nearinfrared_highz}, where we color code the sources according to their classification in the optical (red $=$ optical AGN, blue $=$ optical star-forming). 

In all the $4$ possible versions of the \FeII\ based diagnostics (first and third rows of Fig. \ref{nearinfrared_highz}), most of the optical star-forming galaxies are consistent with star-forming models in these diagrams, while also the optical AGNs are globally residing in a parameter space covered by the AGN models, that is, above the maximum AGN line in red. Therefore, these new diagnostics provide in general a consistent picture to that seen at shorter wavelengths. %A smaller number of optical AGNs lie below the maximum starburst line, but almost all of them are still consistent with AGN models, residing in the intermediate region. 
We also notice that, for optical AGNs, while the majority lies in the pure AGN region in the Fe2S3-$\beta$ diagram, they tend to move toward the left to the intermediate region (below the maximum starburst line) in the Fe2S3-$\alpha$ diagnostic, possibly due to a larger contribution in the \PaA\ line from obscured star-formation. However, we have fewer statistics in this case, as we are limited in redshift.

In the \PII\ and \CI\ based diagnostics (second and fourth rows of Fig. \ref{nearinfrared_highz}), those lines are undetected for the majority of the optical star-forming galaxies, which is due to their intrinsic faintness compared to the \FeII\ emission lines. On the other hand, we detect them for most of the optical AGNs. Also in these diagrams, the near-IR classification is mostly consistent with the optical one. Even more, we can say that for those sources in which we detect \PII\ or \CI, there is a more clear separation between optical star-forming galaxies and optical AGNs, preferentially falling in the pure star-formation or AGN region, respectively, with less overlap compared to the \FeII\ based diagrams.

Similarly to the optical case, the CEERS sample spans a wide parameter space in the near-IR diagrams, with star-forming galaxies showing a large variety of \SIII/\PaG\ or \SIII/\PaB\ line ratios over almost $2$ orders of magnitude, from $-0.5$ to $1.5$ in logarithmic scale. This result further corroborates the presence of sources with different ionization parameters, from $\log (U)$ $= -4$ to higher values. %, even though the above line ratios start to saturate at $\log (U) \gtrsim -2$, as seen in Section \ref{nearIR_diagnostics_models}. 
In principle, the \CI\ to Paschen line ratios can better distinguish among the highest ionization parameters, but the upper limits that we have are not sufficiently low to probe different values of $\log (U)$ in that regime. 
Also, the variety of \FeII\ and \PII\ to Paschen lines flux ratios of star-forming galaxies might reflect the different metallicity properties of the sample, even though the non-detection of \PII\ for a relatively large subset hampers a more precise assessment of their properties.

For the subset of optical AGNs, even though they prefer models with low ionization parameters ($\leq 3$), some of them only have upper limits on \PaG, \PaB, or \CI, suggesting that they have higher $\log (U)$. The narrower range and the higher values of their \PII/\PaB\ ratios (compared to star-forming galaxies) indicate instead that they have more uniform metallicity properties, consistent with Z$_\text{gas}$ at least $\geq 0.7$ times solar on average.

\subsubsection{Hidden AGNs displaying from optical to near-infrared}\label{changing_look}

In analogy with the criteria adopted in the optical, we classify as near-IR AGNs those systems that entirely reside in the pure AGN region in at least one of the $8$ near-IR diagnostics, while the remaining systems, which are always consistent with star-forming models within $1 \sigma$, are identified as near-IR star-forming galaxies. 
We find that all optical AGNs are also classified as near-IR AGNs. 
However, we also find that $5$ optical star-forming galaxies become AGNs in the near-IR diagnostics (CEERS ID 2900, 8515, 8588, 8710, 9413). We identify them as \textit{hidden} AGNs, and they are represented as cyan crosses in all the panels in Fig. \ref{nearinfrared_highz}. 

All of these hidden AGNs are classified as pure AGNs in the P2S3 diagram. Three of them also lie in the pure AGN region in the C1S3 diagnostic, while four of them have an enhanced \FeII/\PaB\ line ratio (either the \FeII\ at $1.257$ or $1.64$ $\mu m$) and thus are identified as pure AGNs in at least one among the Fe2S3-$\beta$ and Fe2S3-$\alpha$ diagrams. Even the sources that do not satisfy the pure AGN condition in all four diagrams, lie close to the maximum starburst line in the overlapping region between SF and AGN. 

In each of the near-IR diagnostics, the optical SF galaxies whose upper limits do not allow us to put constraints on their near-IR nature (i.e., they could be consistent with either SF or AGN models), we have performed emission line measurements on their spectral stacks, and we have found that their \FeII, \CI, \PII, and \SIII\ line properties are similar to the other near-IR star-forming galaxies, and that they are placed in the pure star-forming region in all the new diagnostics. This suggests that we are likely not missing a significant population of hidden AGNs below our line detection limit.

In Fig. \ref{Fig:classification}, we show a summary of how our CEERS sources are classified both in the rest-frame optical and the near-IR. Overall, the near-IR classification is in agreement with the optical one, but with $5$ optical star-forming systems that show properties similar to AGNs in the near-IR diagnostics. 
These hidden AGNs represent a fraction of $8.8 \%$ of the population of optical star-forming galaxies, with a $1 \sigma$ confidence interval of $5.6 \%$-$11.8 \%$, estimated following \citet{gehrels86}. 
Even though they represent a small fraction of the entire population, they nearly double the total AGN sample going from $8$ to $13$ objects. 
We also note that the optical, narrow-line AGN with CEERS ID 5106 is likely a hidden broad-line AGN, showing a broad \PaB\ component detected at S/N $>2$.

%%%%%%%%%%%%%%%%%%%%%%%%%%%%%%%%%%%%%%%%%%%%%%%%%%%%
%%%%%%%%%%%%%%%%%%%%%%%%%%%%%%%%%%%%%%%%%%%%%%%%%%%%
%%%%%%%%%%%%%%%%%%%%%%%%%%%%%%%%%%%%%%%%%%%%%%%%%%%%

\section{Discussion}\label{discussion}

The results presented in Section \ref{results} show that we can distinguish between AGN and star-forming powered sources by looking at rest-frame near-IR emission lines. In this section, we discuss the physical reason why the new diagnostics are working well for this classification task. We also discuss alternative mechanisms that might be responsible for the emission lines analyzed in this paper, and how the line ratios would evolve at high redshift. We then compare the near-IR diagnostics to the optical ones and provide possible explanations for understanding the class of hidden AGNs. Finally, we discuss further improvements that can be made with forthcoming spectroscopic facilities.

\subsection{Origin of the line ratio enhancement in AGNs}\label{discussion1}  

The comparison between optical and near-IR predictions of Cloudy models shows us that we can consistently identify AGNs and star-forming powered sources across one order of magnitude in wavelength, using several diagnostics that involve different chemical elements, from the local Universe up to redshift $\sim 3$. This also indicates that the observed lines, both in SF galaxies and in AGNs, are all consistent with being produced by photoionization. 

All the near-IR diagrams analyzed here are based on the \xSIII\ to Paschen line ratios. The \xSIII\ line is one of the brightest available at around $\lambda_{rest} \simeq 0.95 \mu m$. It comes from the doubly ionized sulfur (S$^{2+}$), which is created by photons with energies between $23.3$ and $34.8$ eV, thus it is produced more efficiently by the harder ionizing radiation of an AGN (see Figure \ref{ionizing_spectra}), reaching higher \SIII/\PaG\ ratios compared to purely star-forming galaxies.  

AGNs also have enhanced \FeII, \PII, and \CI\ emission compared to the typical star-forming population. The origin of this enhancement is much discussed in the literature. It is due in part to the higher metallicity of the narrow line regions, both at low and high redshift \citep{kewley02,groves04}, as little evolution is observed up to $z \sim 3$ \citep{nagao06,matsuoka09}. As claimed by \citet{shields10}, the strength of the optical \FeII\ line increases almost linearly with the gas-phase iron abundance, which is observed in our models also for the \PII\ line. 

However, as the AGN and SF predictions differ even at fixed Z$_\text{gas}$, the ionization mechanism also plays an important role. 
The Fe$^+$, P$^+$, and C$^0$ require low ionization or excitation energies, thus usually trace transition regions of partially ionized hydrogen, coexisting with H$^0$, O$^0$, S$^+$, and free electrons, where electron collisions are an important excitation mechanism. As shown in several works \citep{mouri90,oliva01}, photoionization by non-thermal continuum radiation, especially by soft X-ray photons, can create such extended regions where the \FeII\ emission and that of other low ionization species can proliferate. The radiation field of an AGN can thus provide such favorable conditions (see also \Citealt{ferland09}, \Citealt{shields10}, \Citealt{gaskell22}).
We have verified that by switching off the soft X-ray emission in the incident radiation of the AGN template, also the emission from \FeII, \PII, and \CI\ is significantly reduced to the level of star-forming sources.
Overall, this suggests that photoionization models can reproduce the observed emission line ratios.

\subsection{Shock-models predictions}\label{shocks}

\begin{figure}[ht!]
    \centering % # 96 for the two column version  , 0.86 for the referee version 
    \includegraphics[angle=0,width=0.96\linewidth,trim={0.1cm 0.1cm 23.5cm 0.1cm},clip]{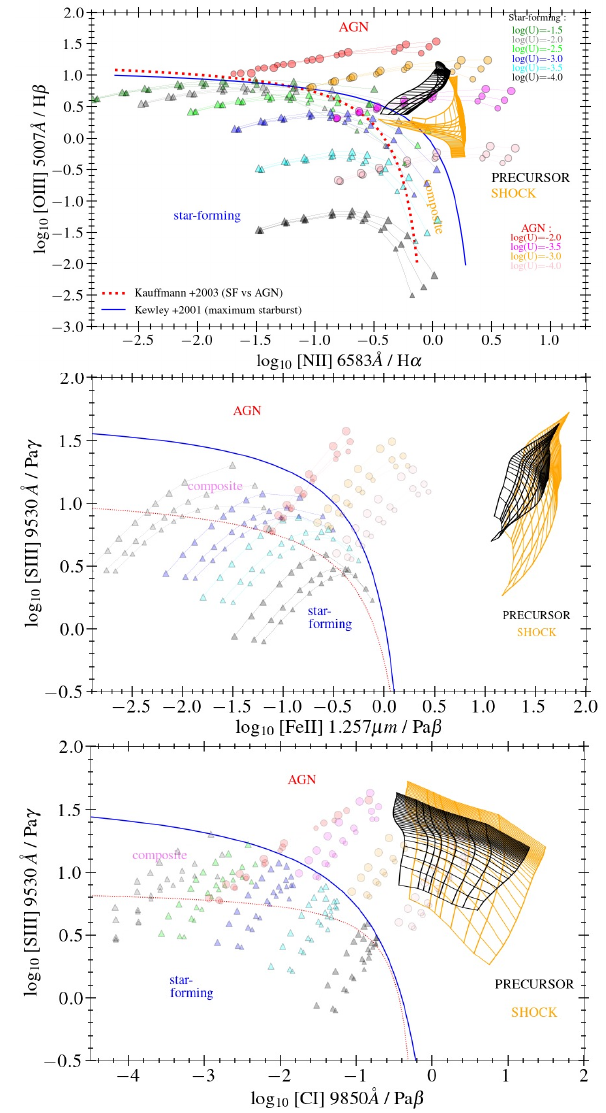} % for referee version
    \caption{Predictions of the shock models described in the text for the BPT diagram, the Fe2S3-$\beta$, and the C1S3 diagrams, from top to bottom. The AGN and SF models from Cloudy and the separation lines are the same as in Fig. \ref{nearIR_models}.
    }\label{Fig:shocks}
    \vspace{-0.3cm}
\end{figure}

It is interesting to understand whether shocks can equally reproduce the observed line ratios in our new near-IR diagnostics. To this aim, we have explored the predictions of shock models as provided by the Mexican Million Models database (3MdB, \Citealt{morisset15}). These are produced with the Mappings V code, version 5.1.13 \citep{sutherland18}, using the shock parameters of \citet{allen08}, and assuming a solar abundance for the gaseous phase, a gas density of $1 cm^{-3}$, magnetic parameter $B/n^{1/2}$ of $10^{-4}$ to $10$  $\mu G cm^{3/2}$, and shock velocity between $200$ and $1000$ km/s. These are usually identified as fast shocks, in which the ionized front (photoionized by the cooling of hot gas) expands more rapidly than the shock itself, forming a detached HII-like region called precursor. 
We have analyzed the emission line predictions of both shocks and shock+precursor models in the BPT, in the Fe2S3-$\beta$, and in the C1S3 diagrams. However, we note that our star-forming galaxies are sufficiently rich in gas that we likely observe both the shock and its precursor. Seyfert galaxies are also found to be more consistent with shock+precursor models, as opposed to LINERS \citep{allen08}. 

As we can see in Fig. \ref{Fig:shocks}, the shock models produce emission line ratios almost completely overlapping with the AGN models in the BPT and the C1S3 diagnostics. In contrast, they significantly enhance the \FeII/\PaB\ line ratio by almost $1$ or $2$ orders of magnitude compared to AGNs and SF galaxies, respectively. This is because shocks efficiently destroy dust grains by sputtering. Since iron is one of the elements most depleted on dust in the ISM (despite being one of the most abundant in total), this process releases a large amount of iron in the gas phase, which is available to cool. 

The phosphorus and \PII\ line predictions are not included in the original shock models of \citet{allen08}. However, since it is not heavily depleted, we expect a similar behavior to \NII\ and \CI. 
Indeed, \citet{storchi-bergmann09} and \citet{oliva01} have proposed the \FeII/\PII\ as a main tracer of shocks. This ratio can rise to $\sim 20$ in shock-dominated regions, significantly higher than the value of $2$ observed in photoionized regions. Our CEERS observations and the sample of AGNs and SF galaxies at lower redshifts thus suggest that shocks can be excluded as the main contributors of the global emission. % for the vast majority of them. % sourcesobjects where we only have lower limits on \FeII/\PaB. 

\subsection{Redshift evolution of near-infrared AGN diagnostics}\label{redshift_evolution}

Our analysis spans a large redshift range, corresponding to almost $12$ Billion years of cosmic time. It is therefore worth considering possible evolutionary effects on our near-IR emission line properties.
The first effect to take into account is the enhancement from low-$z$ to high-$z$ of the fraction of $\alpha$ elements (which include, among all, oxygen, carbon, and sulfur) compared to iron. This is due to galaxies at $z>1$ being younger on average than those at $z \sim 0$, so that type 1a supernovae, the main producers of iron, did not have time enough to enrich the surrounding ISM.

% From the oxygen abundance derived from the optical spectrum and based on oxygen lines, 
Previous studies have found abundance ratios $\left[\frac{O}{Fe}\right]$ a factor of $2$ (or more) higher than the solar value \citep{steidel16,topping20,cullen21}. Assuming a conservative $\alpha$-enhancement of $2\times$ [$\alpha$/Fe]$_\odot$, we find that it would shift the maximum starburst line in the Fe2S3-$\beta$ and Fe2S3-$\alpha$ diagrams by $-0.3$ dex along the x-axis. However, at $z>1$, the gas-phase metallicity of typical star-forming galaxies with stellar mass ranging $9 < \log(M_\ast/M_\odot) < 11$ is lower than at $z=0$ by an amount of $\sim 0.3$ dex \citep{wuyts16,sanders20}, while for AGNs it is not completely assessed. 
This means that the depletion factor of iron (the most depleted element) should be also lower, as the dust content is directly proportional to metallicity. 
Assuming a half solar gas-phase metallicity, one can estimate that the depletion factor is lower by approximately the same order of magnitude \citep{vladilo11}, which counterbalances the previous effect, enhancing the \FeII/\PaB\ and \FeII/\PaA\ ratios by $\sim 0.3$ dex. 
For this reason, as a first approximation, and considering the lack of tighter constraints at high-$z$, we keep at $z>1$ the same local relation for separating star-forming galaxies from AGNs. 

For the other near-IR diagrams, since phosphorus and carbon are almost non-refractive elements, the P2S3 and the C1S3 diagnostics are not significantly altered by this effect. Moreover, phosphorus behaves similarly to an $\alpha$ element like sulfur \citep{caffau11}. As a result, we adopt at $z>1$ the same separation criteria between SF galaxies and AGNs as in the local Universe. 

Overall, we remark that, despite the \FeII\ lines being usually brighter than \PII\ and \CI, the Fe2S3-$\beta$ and Fe2S3-$\alpha$ diagnostics have also the highest uncertainties among all near-IR diagrams. 
A higher $\alpha$ enhancement factor than our conservative value, as suggested by some previous works (e.g., \Citealt{steidel16} report a ratio of $4$-$5$ $\times$ [O/Fe]$_\odot$), would explain the fraction of optical AGNs at $z>1$ that in the Fe2S3-$\beta$ and Fe2S3-$\alpha$ diagrams fall in the intermediate region.
On the other hand, the ISM depletion of iron can vary by $1$ order of magnitude or more \citep{shields10}, sufficient to explain the wide range of \FeII\ strengths in all types of sources compared to the other diagnostics based on \PII\ or \CI.
It is important to remind that each diagnostic has its pros and cons, thus we recommend combining all the available near-IR diagnostic diagrams to obtain a more accurate and complete understanding of the nature of sources at $0<z<3$. The presence of coronal lines and broad components in permitted lines can also help in identifying AGNs. % properly taking into account broad Paschen lines (if detected), 

Finally, we also note that near-IR star-forming galaxies in CEERS have typical \FeII/\PaB\ ratios that are lower (with \FeII\ undetected for a large fraction of them) compared to local starbursts. 
This is likely driven by a combination of the above-mentioned $\alpha$-enhancement and selection effects: while star-forming galaxies targeted in CEERS are representative of normal star-forming galaxies at mostly M$_\ast < 10^{10}$ M$_\odot$, local samples are selected as massive (M$_\ast > 10^{10}$ M$_\odot$) infrared bright sources with higher SFRs and higher specific SFRs, for which we might expect an enhancement of the iron abundance as due to supernovae remnants \citep{lester90}.

\subsection{Optical versus near-infrared diagnostics and hidden AGNs}\label{discussion2}

In the near-IR versions of the AGN diagnostics, \xSIII\ replaces \xOIII\ on the y-axis, while \xFeIIa, \xFeIIc, \xPII, and \xCI\ are used on the x-axis instead of \xNII\ and \xSII.
Comparing the \OIII\ to the \SIII\ emission, the production of O$^{2+}$ requires photons with energies ranging 35.1-54.9 eV, significantly higher than those required to create S$^{2+}$. At those high energies indeed, a large fraction of sulfur in the gas phase would be triply ionized (S$^{3+}$), lowering the abundance of S$^{2+}$. 
This implies that \SIII/\PaB\ is not a good tracer of the ionization parameter as the \OIII/\Hb\ line ratio, as it does not have a monotonic trend over the full range of possible $\log (U)$ from $-4$ up to $= -1$. The models are thus degenerate, especially at high ionization.  However, an advantage is that star-forming galaxies at higher redshifts, despite having typically higher $\log (U)$, would not have also extreme \SIII/\PaB\ line ratios, hence they likely would not cross the maximum starburst lines even at $z>3$. The remaining drawback is that the AGNs can move to the intermediate part of all the diagnostics (below the maximum star-forming line) if they have metallicities lower than one-half solar. 
As far as the other lines are concerned, both \FeII\ and \PII\ (except \CI), are very sensitive to metallicity, as \NII\ and \SII\ in the optical regime. 
In addition, all the metal emission lines used in the near-IR diagnostics on the x-axis likely have a similar excitation mechanism and a common origin to the low-ionization line \xSII, as discussed in Section \ref{discussion1}, while \NII\ traces slightly higher ionization regions \citep{mouri90}. This explains why all AGNs identified with the \SII\ diagram are also AGNs in near-IR diagnostics.

Overall, using the new near-IR diagrams has several advantages compared to the optical ones. 
First, the Fe2S3-$\beta$ and Fe2S3-$\alpha$ diagrams, or the \FeII/\PII\ line ratio, can distinguish between emission lines produced by photoionization or by shocks \citep{oliva01}, which is not possible through the classical BPT diagram (see section \ref{shocks}). The \FeII, \SIII, and \PaB\ lines, being among the brightest features in the rest-frame range $0.8$-$1.3 \mu m$, open the possibility to potentially characterize large samples of sources and galactic regions as shock, AGN, or star-forming driven. 
Secondly, the wavelength separation among the near-IR lines considered in this paper is always larger than the one between \Ha\ and \xNII, or \Hb\ and \xOIII. As a consequence, while the optical lines can be resolved only with medium to high-resolution spectrographs (R$\gtrsim600$), we can measure individual near-IR lines also at lower spectral resolutions ($100 \leq R \leq 600$). 
Finally, and most importantly, \xSIII, \xCI, \xFeII, \xPII, \PaG, and \PaB\ are significantly less affected by dust attenuation than optical lines. For example, assuming a Calzetti dust law, the attenuation (in magnitudes) at $\lambda \geq 1 \mu m$ is a factor of $4$ lower than in the range $0.4 < \lambda < 0.65 \mu m$ \citep{calzetti00}. In case of optically thick dust that is well mixed with the emitting gas, the situation is even worse, with optical lines probing only $\leq10 \%$ of the systems, while Paschen lines can recover up to $30$-$40\%$ more of the total unobscured emission \citep{calabro18}. 

This last point provides a clue to the physical interpretation of the class of hidden AGNs. In Section \ref{changing_look} we have found $5$ of these systems, with $4$ being AGNs in the P2S3 diagram and at least one Fe-based diagnostic, while the remaining one is identified as such in P2S3 and C1S3. 
The host galaxies of all hidden AGNs have stellar masses greater than $10^{9.3}$ M$_\odot$. Having a median $\log$ M$_\ast/$M$_\odot = 9.8$, they are on average more massive than both the $5$ narrow-line optical AGNs (M$_{\ast,med}=10^{9.25}$ M$_\odot$) and than the parent population of star-forming galaxies selected at $1<z<3$ (M$_{\ast,med}=10^{9.26}$ M$_\odot$). 
Moreover, we find that in all these systems, the attenuation inferred from the Balmer decrement in the optical (adopting the Calzetti law for simplicity) is systematically lower than the one derived using the Paschen decrement (i.e., \PaA\ + \PaB), or the \PaG/\Ha\ and \PaB/\Ha\ ratio if the longest wavelength line is not available. The A$_V$ estimated from Paschen lines is on average $\sim0.3$ dex higher than those obtained in the optical.
This suggests that there might be optically thick dusty regions in or around the narrow-line region of the AGNs, or the NLR itself might be partially covered by the dusty torus. 
In particular, we find that the hidden AGN ID 2900 is a Compton thick AGN with a luminosity $\log (L_X/erg/s) = 44.07 \pm 0.54 $ and a hydrogen column density N$_\text{H}=10^{25}$ $cm^{-2}$ \citep{buchner15}. The high central density and obscuration might be the reason why it does not show broad lines and it is harder to identify compared to the other X-ray AGNs in our sample (ID 2919 and 2904), which have broad lines and a lower N$_\text{H} < 10^{23}$ $cm^{-2}$. 
To fully understand these trends and the nature of hidden systems, we would need larger statistics.
A higher S/N of the spectra would also be necessary to check whether there are hidden broad-line AGNs (emerging in the Paschen lines) in the hidden AGN population, as found for the ID 5106.

\subsection{Comparison with previous near-infrared diagnostics}\label{previous_diagnostic}

\begin{figure}[ht!]
    \centering
    \includegraphics[angle=0,width=0.9\linewidth,trim={0.1cm 0.8cm 1.cm 0.1cm},clip]{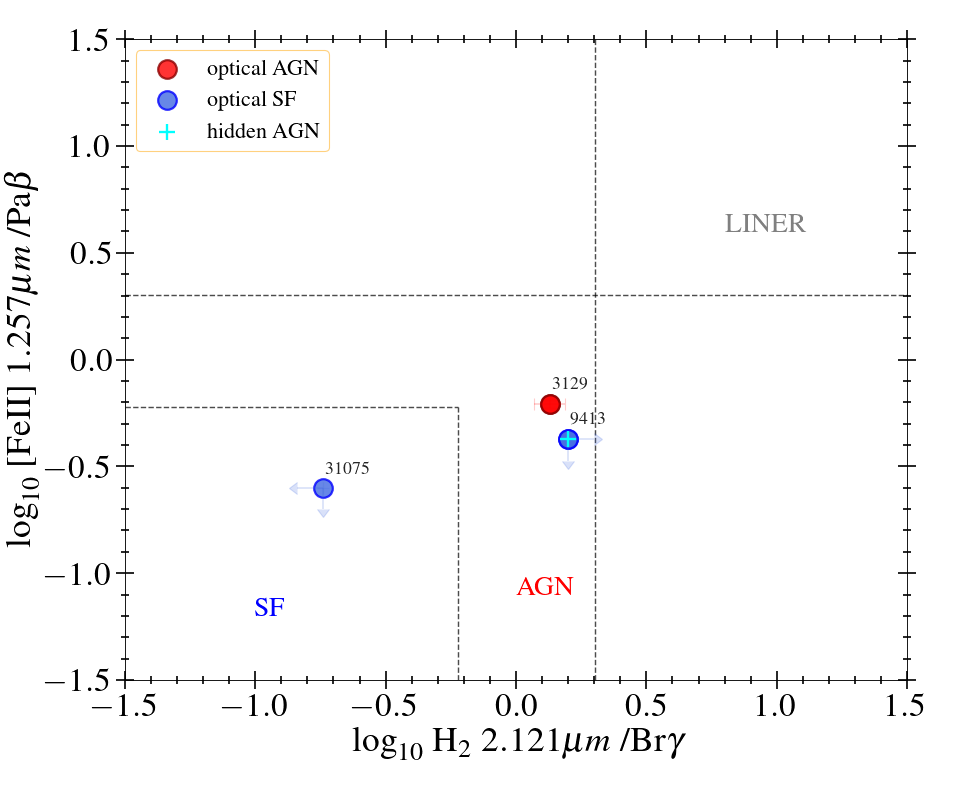} % for referee version
    \includegraphics[angle=0,width=0.9\linewidth,trim={0.1cm 0.8cm 1.cm 0.1cm},clip]{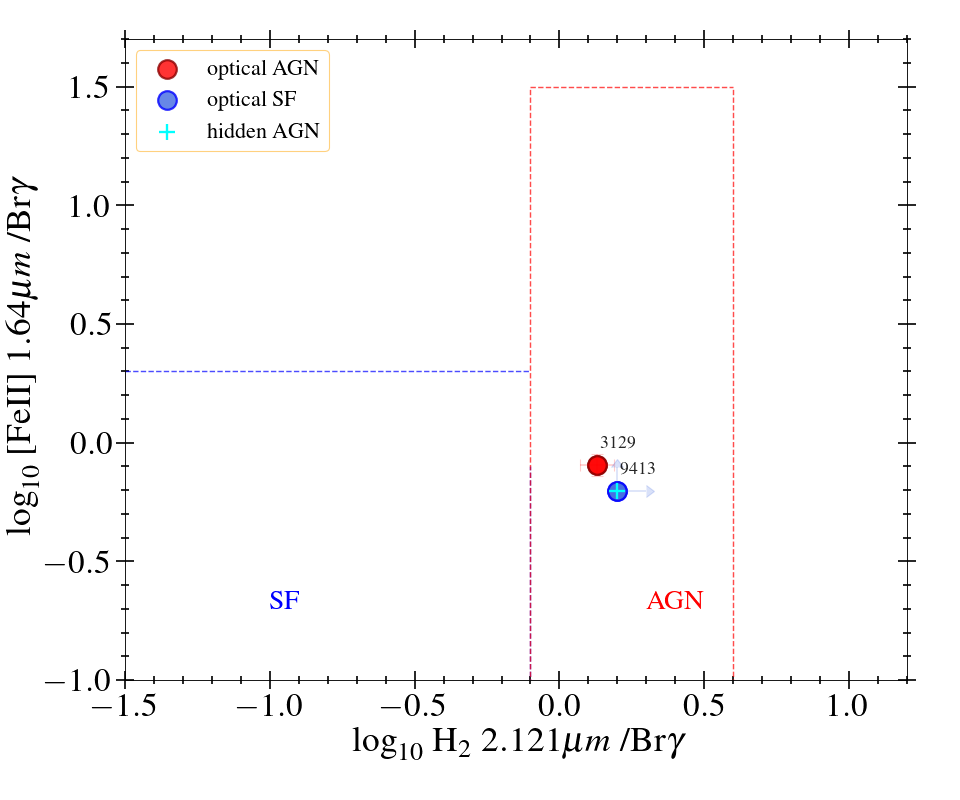}
    \caption{\textit{Top:} Diagnostic diagram comparing the \xFeIIa/\PaB\ and \xHtwo/\BrG\ line ratios for CEERS galaxies in the redshift range $1<z<1.4$ with detection of either \Htwo\ or \BrG. The black dashed lines represent the separation lines to distinguish among star-forming sources, AGNs, and LINERs, according to \citet{rodriguezardila05}. \textit{Bottom:} \xFeIIc/\BrG\ vs \xHtwo/\BrG\ diagram \citep{colina15} for the same sample.  
    }\label{Fig:Fe2H2}
    \vspace{-0.3cm}
\end{figure}

The idea of using near-infrared rest-frame diagnostics to mitigate the problem of dust attenuation in obscured sources is not new in the literature. \citet{rodriguezardila04} proposed the diagram comparing \xFeIIa/\PaB\ and \xHtwo/\BrG\ line ratios as a useful diagnostic to disentangle AGNs from star-forming galaxies and LINERs. Since then, this diagram (that we indicate as \textbf{Fe2H2}) has been applied to analyze the ionizing sources in statistical samples of galaxies in the local Universe \citep[e.g.][]{rodriguezardila04,riffel13,lamperti17}. A different version of this diagnostic compares the \xFeIIc/\BrG\ and \xHtwo/\BrG\ line ratios \citep{colina15}.  % \Citealt{morisset15} 

The main difference of these diagnostics compared to those previously analyzed in this paper is that they are based on emission lines (\Htwo\ and \BrG) that are in general fainter than \PaA. For example, assuming Case B recombination \citep[e.g.][]{osterbrock89}, the intrinsic \BrG\ luminosity is $8 \%$ that of \PaA\ and $17 \%$ that of \PaB. Furthermore, being at longer wavelengths, we can detect them in our spectra up to redshift $\sim 1.4$, required to include \BrG\ in the reddest NIRSpec channel. 
In our sample, we have the coverage of \BrG\ in $6$ galaxies at $1 < z < 1.4$, of which $1$ is an optical AGN (ID 3129), $1$ is a hidden AGN (ID 9413), and the remaining $4$ are normal star-forming galaxies from all our previous classifications. 

Despite the small sample size, we have checked their position in the two Fe2H2 diagrams, which are shown in Fig. \ref{Fig:Fe2H2}. 
We can see that the optical AGN and the hidden AGN fall in the AGN parameter space identified by \citet{rodriguezardila04,rodriguezardila05} in the first diagram (top panel), and by \citet{colina15} in the bottom panel, with the \Htwo/\BrG\ ratios significantly higher than $1$. Regarding the star-forming galaxies, one has lower \Htwo/\BrG\ placing it in the star-forming region (\xFeIIc/\BrG\ is not available), while for the remaining systems \Htwo\ and \BrG\ are both undetected. 
To conclude, the Fe2H2 diagrams give results that are consistent to those obtained with the four main near-infrared diagrams introduced in this paper, confirming our previous classification. This also suggests that the \Htwo\ and \BrG-based diagnostics can be reliably used also at $z>1$, even though a larger statistics and deeper observations are needed to confirm their effective strengths.

%Mi raccomando, ultima cosa da fare aggiungi referenze BASS anche in Introduction.

\subsection{Future observations and facilities}\label{discussion3}

At low redshift, we have been able to characterize large, representative, and almost complete samples of AGNs with various degrees of activity, and sizable subsets of near-IR bright starburst galaxies. 
In the high redshift Universe instead, the number of spectroscopically confirmed AGNs with near-IR observations is still small and does not cover all possible conditions of spectral type, black hole mass and obscuration, and level of activity, being limited to a few objects targeted with JWST/NIRSpec from ERS and GO public programs.

Surveys such as FRESCO \citep{oesch23} will soon observe near-IR rest-frame lines in galaxies and AGNs at redshifts $<3$ with NIRSpec in slitless mode. At higher redshifts, up to the reionization epoch, the brightest near-IR lines can be observed instead with JWST-MIRI, which represents a challenging frontier. 
Euclid will also perform slitless spectroscopy in a wavelength range between $1.25$ and $1.85$ $\mu m$ \citep{costille16}, allowing to map the near-IR lines from \SIII\ to \PaB\ at $z<0.45$ with much larger statistics than JWST. Thanks to the resolving power of $\sim 450$ in the red grism \citep{scaramella22}, star-forming galaxies and AGNs can be identified from near-IR diagnostics and broad line components in the brightest cases.
% With R=450 you have FWHM resolution of 666 km/s, which is enough to identify broad line AGNs. 

Another opportunity will be offered by the MOONS near-IR ($0.6$-$1.8 \mu m$) spectrograph at the VLT \citep{cirasuolo20,maiolino20}, which is scheduled to start operations in $2024$. This spectrograph will take spectra over a sky area of $\sim0.15$ $\deg^2$ with $\sim1000$ fibers simultaneously, with a multiplexity performance significantly larger than previous near-IR instruments. %an area $60$ times bigger than NIRSpec and an improved multiplexity. 
This will allow us to detect the near-IR spectral range of thousands of galaxies up to $z \sim 0.4$ (for \PaB\ coverage), identify all types of AGNs, and map their 3D distribution, shedding light on their environment. 
Finally, the Prime Focus Spectrograph (PFS) at the Subaru telescope will carry out multi-fiber spectroscopy in the range $0.38$-$1.3 \mu m$ of $2400$ targets simultaneously, within an area of $1.3$ square degree \citep{takada14}. Despite the shorter wavelength coverage, it has a higher resolution ($R\simeq3000$), which can be used to study the kinematic properties of local, near-IR selected AGNs. 
These spectroscopic campaigns will allow us to study the properties of both broad line and narrow line AGNs even for more obscured systems that are difficult to identify at shorter wavelengths. This will finally provide a more complete census and physical characterization of AGNs beyond the local Universe, with the same statistics reached by the SDSS at $z < 0.1$. 

%In this work, we have explored the integrated galaxy properties. 
Thanks to the IFU spectroscopic mode of NIRSpec and MIRI we will be able in the future to derive spatially resolved maps of near-IR lines. 
At the VLT, we should also consider ERIS \citep{kravchenko22}, which is a near-IR imager and spectrograph, capable of medium-resolution integral field spectroscopy in J, H, and K bands, and long-slit spectroscopy in band L. It can operate in synergy with a modern adaptive optics system, necessary to spatially resolve the different galaxy components. 
This will enable a better understanding of the nuclear regions and determine the physical extension where the AGN significantly contributes to the observed line ratios. The resolved \FeII/\PII\ ratio will be used instead to trace shock-dominated regions and outflows in the outskirts, testing the stellar and AGN feedback on galactic scales. 

\section{Summary}\label{summary}

In this paper, we have investigated new diagnostic diagrams for selecting AGNs in the rest-frame near-infrared. 
We have combined model predictions with observations of star-forming galaxies and AGNs at $z<3$ coming from CEERS and previous spectroscopic surveys. 
We summarize the main findings as follows:
\begin{enumerate} 
\item using Cloudy photoionization models, we present four new diagnostic diagrams, named \textbf{Fe2S3-$\beta$}, \textbf{Fe2S3-$\alpha$}, \textbf{P2S3}, and \textbf{C1S3}, to distinguish between stellar and AGN driven photoionization in galaxies, based only on rest-frame near-IR emission lines. These diagnostics share the same \xSIII/\PaG\ (or \xSIII/\PaB) line ratio on the y-axis, while on the x-axis they have the \xFeIIa/\PaB, \xFeIIc/\PaB, \xPII/\PaB, or \xCI/\PaB\ line ratios, respectively. 
We derive in each case analytic expressions for the maximum star-forming lines and maximum AGN lines, which can be used to assess the dominant ionizing mechanism contributing to the global emission. 
\item these diagnostics are successfully applied from redshift $\sim0$ to $z\sim3$. The majority of the high redshift sample ($z>1$) is made of star-forming galaxies, which are coherently identified as such both in the rest-frame optical and near-IR. We also find $8$ optical AGNs, which are coherently classified as AGNs in the near-IR. Previously adopted diagrams including the \xHtwo\ and \BrG\ emission lines at longer wavelengths yield results consistent with our new near-IR diagnostics.
\item we find that $5$ sources, classified as optical star-forming galaxies at $z>1$, are identified as AGNs from near-IR diagnostics. All these sources reside in the pure AGN ionization region in the P2S3 diagram, while $2$ and $4$ are also identified as AGNs, respectively from the C1S3 diagram and at least one iron-based diagnostic (Fe2S3-$\beta$ or Fe2S3-$\alpha$). Within the uncertainties, all of them are consistent with AGN models in the entire set of near-IR diagnostics.
The hidden AGNs represent a consistent fraction of the AGN population, increasing their total number by $\sim 60 \%$.  
They might be systems in which the emission from the narrow-line AGN region is embedded in optically thick dust and thus can be better identified at longer wavelengths. We also find a hidden broad-line AGN candidate, classified as a narrow-line AGN in the optical but for which we tentatively detect a broad \PaB\ component with S/N $>2$. 
\end{enumerate}

The near-IR diagnostics presented in this paper are preferable to standard optical diagnostics for several reasons. They are less affected by dust attenuation, require a lower spectral resolution compared to the BPT diagram and the \SII/\Ha\ based diagram, and finally, they can distinguish shocks from photoionization. Therefore, they represent promising tools for identifying AGNs in future spectroscopic surveys. 

\noindent

\begin{acknowledgements}
We thank the anonymous referee for the thoughtful and constructive comments that improved the quality of the manuscript. AC acknowledges support from the INAF Large Grant for Extragalactic Surveys with JWST. A. F. acknowledges the support from project "VLT-MOONS" CRAM 1.05.03.07, INAF Large Grant 2022 "The metal circle: a new sharp view of the baryon cycle up to Cosmic Dawn with the latest generation IFU facilities" and INAF Large Grant 2022 "Dual and binary SMBH in the multi-messenger era”.
\end{acknowledgements}

%\clearpage
\appendix
%\renewcommand\thefigure{\thesection.\arabic{figure}}   

%\counterwithin{figure}{section}

\section{Flux calibration of NIRSpec spectra} %\label{appendix}

\subsection{Comparison of Balmer and Paschen line ratios}\label{appendix0}

\begin{figure}[h!]
    \centering
    \includegraphics[angle=0,width=1\linewidth,trim={0.3cm 0.5cm 19.5cm 0.1cm},clip]{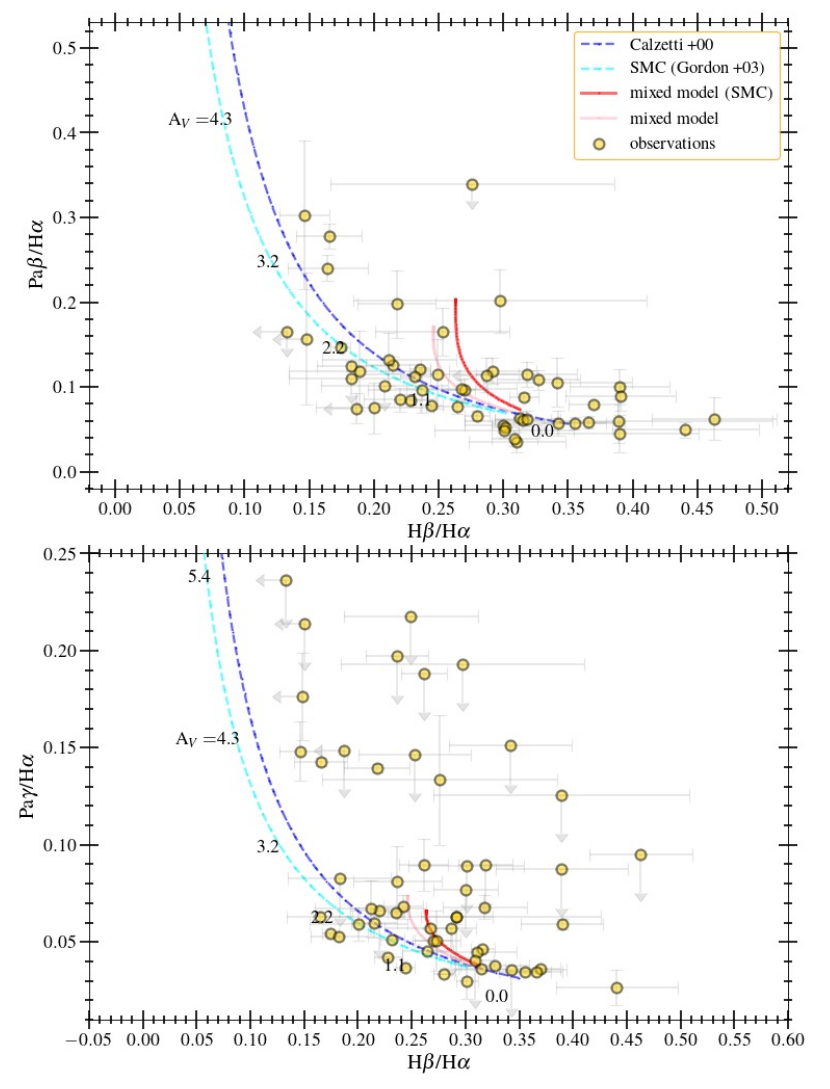}
    \caption{Diagram comparing the \Hb/\Ha\ emission line ratio to the \PaB/\Ha\ ratio (\textit{top panel}) and the \PaG/\Ha\ ratio (\textit{bottom panel}), for CEERS galaxies in the redshift range $1<z<3$. The predictions of different dust attenuation models are shown with colored lines. The attenuation A$_V$ along the line is specified for the Calzetti law.  
    }\label{balmer_paschen_comparison_1}
    \vspace{-0.3cm}
\end{figure}

In this appendix, we show a representative test performed to assess the quality of the relative flux calibration of our fully reduced spectra across the entire wavelength range from $1 \mu m$ to $5 \mu m$. In this test, we take the recombination lines in our NIRSpec data, and that are available for the highest number of sources, namely \Ha, \Hb, \PaB, and \PaG.
In Fig. \ref{balmer_paschen_comparison_1}, we compare the optical line ratio \Hb/\Ha\ and the Balmer-Paschen ratio \PaB/\Ha\ (or \PaG/\Ha) to the predictions of different dust attenuation laws, including the \citet{calzetti00} law, an SMC law from \citet{bouchet85}, and two mixed model predictions from \citet{calabro18}. 

Overall, we can see that the observed points occupy a region of the diagrams that is physically allowed by one or more of the above attenuation models, with most galaxies showing relatively low dust attenuations and located in the bottom right part. Other galaxies have slightly different properties and higher attenuations A$_V$, being consistent with either the Calzetti law, an SMC law, or the mixed model. Nearly all outliers in Fig. \ref{balmer_paschen_comparison_1} have at least one line that is undetected (usually \Hb\ or \PaG), and thus their line ratios are upper limits and cannot be used for constraining the dust attenuation. 
As a result, the absence of a significant number of unphysical line ratios, or systematic deviations from model predictions, indicates that the relative flux calibration of the spectra across $2 \mu m$ in wavelength can be trusted on average. However, we still suggest that each object should be checked individually before computing ratios of distant lines or using their line luminosities for scientific purposes.

\section{Comparison with other AGN model predictions}\label{appendix1}

\begin{figure}[ht!]
    \centering
    \includegraphics[angle=0,width=1\linewidth,trim={0.1cm 0.1cm 0.1cm 0.1cm},clip]{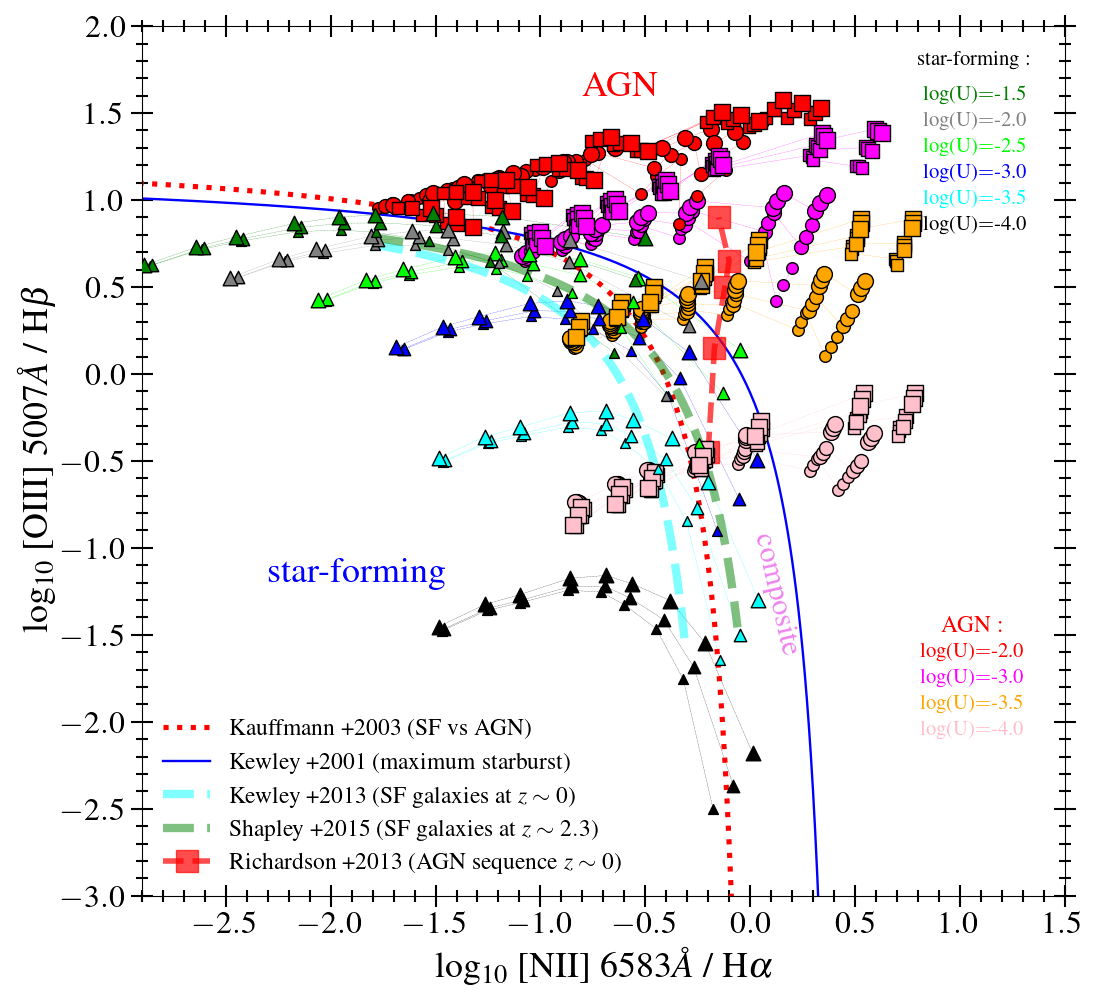}
    \caption{BPT diagram, analog of Fig. \ref{BPT_CLOUDY}, where the line ratio predictions for AGNs are obtained using the recent models of \citet{thomas16,thomas18}. For AGN models with the same ionization parameter (colored as indicated in the legend), the circles are the predictions obtained with E$_\text{peak}=20$ eV, while the square symbols are derived assuming E$_\text{peak}=100$ eV. The marker size varies as a function of gas density (from $10^2$ to $10^4$ cm$^{-3}$ from the smaller to the larger). The three points with the same marker, size, and color, are the predictions of three different values of p$_{NT}$: $0.1$, $0.25$, and $0.4$. The lines, markers, and colors for the star-forming models and observations are the same as in Fig. \ref{BPT_CLOUDY}.  
    }\label{BPT_diagram_Thomas}
    \vspace{-0.3cm}
\end{figure}

\begin{figure*}[t!]
    \centering
    \includegraphics[angle=0,width=0.98\linewidth,trim={0.cm 5cm 5.5cm 0.1cm},clip]{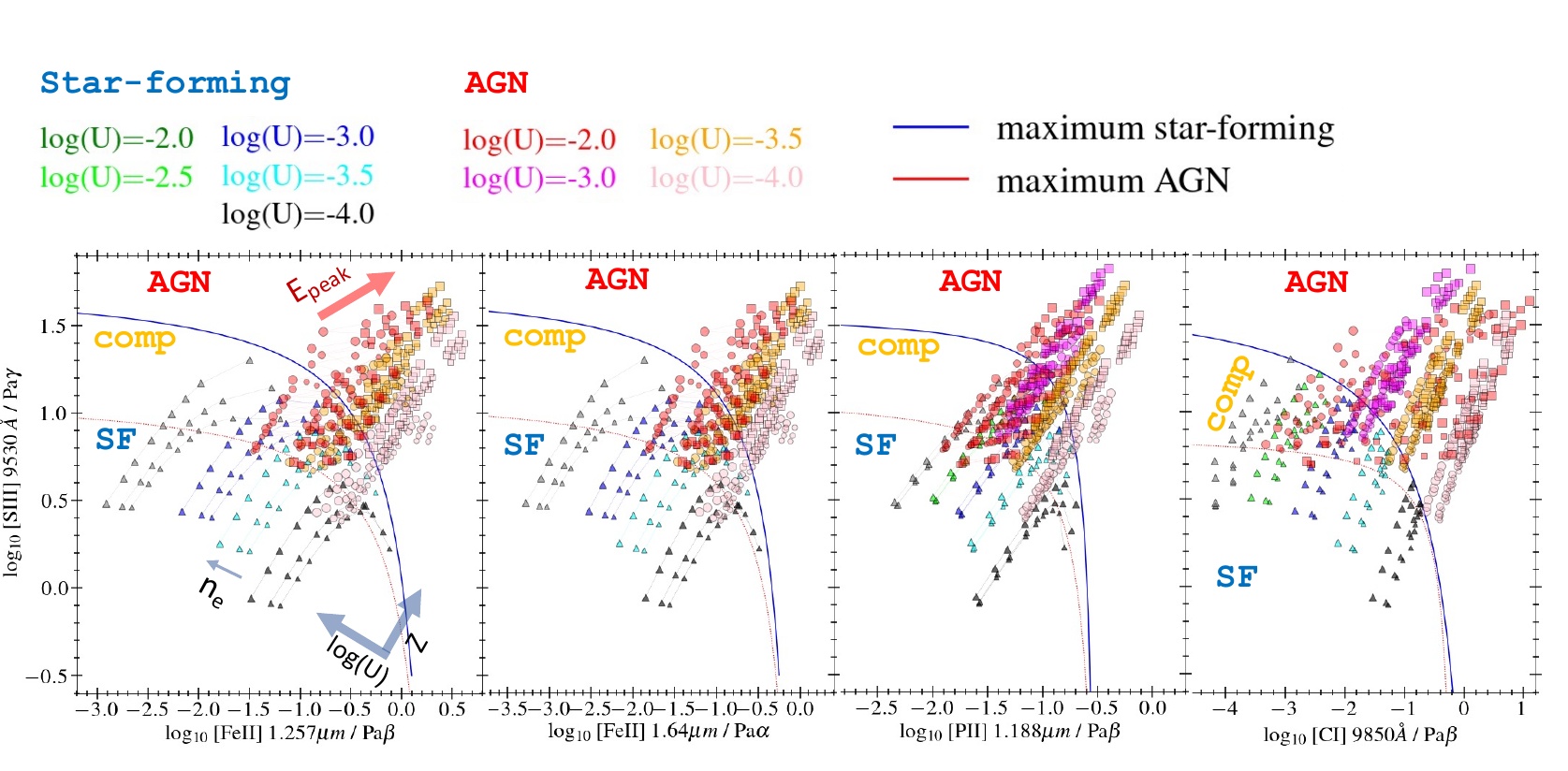}
    \includegraphics[angle=0,width=0.99\linewidth,trim={0.cm 5cm 4.7cm 4.3cm},clip]{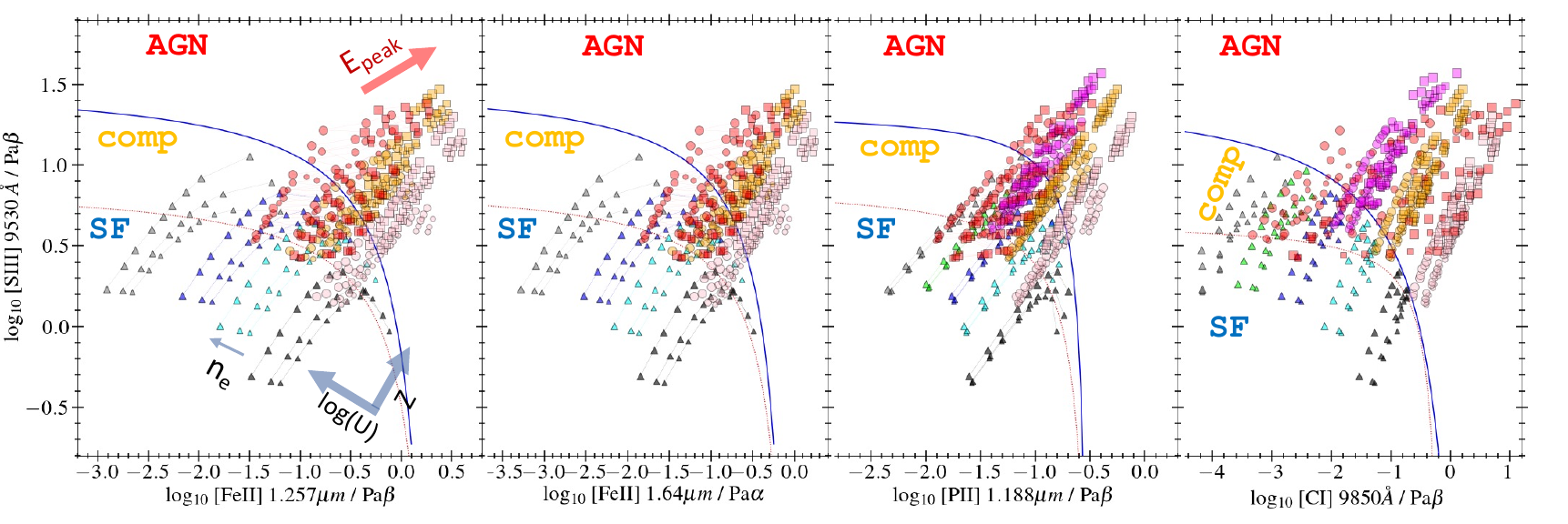}
    \caption{Near-infrared diagnostic diagrams including Fe2S3-$\beta$, Fe2S3-$\alpha$, P2S3, and C1S3. This figure is the analog of Fig. \ref{nearIR_models}, but the emission line predictions for the NLR of AGNs are obtained using the AGN models described in \citet{thomas16,thomas18}. The markers, sizes, and colors of AGN models are the same as described in Fig. \ref{BPT_diagram_Thomas}. The star-forming models are the same as in Fig. \ref{nearIR_models}. The maximum starburst and AGN lines are defined in Section \ref{nearIR_diagnostics_models} and in Table \ref{table_coefficients}.
    }\label{nearIR_models_Thomas}
    \vspace{-0.3cm}
\end{figure*}

We analyze in this appendix all the diagnostic diagrams presented in this paper, using emission line predictions obtained with alternative AGN SEDs compared to our default spectral shape \citep{risaliti00}. In particular, we analyze the more recent AGN spectral shapes introduced by \citet{thomas16}, which depend on three parameters: E$_\text{peak}$, p$_{NT}$, and $\Gamma$, as discussed in Section \ref{AGN_models}. 
We first analyze the predictions for the \OIII/H$\beta$ and \NII/H$\alpha$ line ratios (BPT diagram), which are shown in Fig. \ref{BPT_diagram_Thomas}. We can see that the model predictions, at fixed ionization parameter, gas density, and metallicity, occupy a similar region compared to those obtained with our default AGN SED (Fig. \ref{BPT_CLOUDY}). Within the same family where those three parameters ($\log U$, $Z$, and $n_e$) are fixed, we can notice smaller variations as due to the specific properties of the ionizing shape. 
The stronger variation is produced by E$_\text{peak}$, for which we show for clarity only the predictions obtained with the extreme values considered in this work (E$_\text{peak}=$ $20$ eV and $100$ eV). In detail, a higher E$_\text{peak}$ yields on average a higher \OIII/H$\beta$ and \NII/H$\alpha$ ratios, with the increase rate depending on $Z$ and $\log U$. The enhancement is almost negligible at small (i.e., subsolar) metallicities and small ionization parameters ($\log U \leq 3.5$), but it can be up to $+0.2$ dex at supersolar metallicity and higher $\log U$. This effect was already discussed in \citet{thomas16}, and is due to more energetic and ionizing photons in SEDs with higher E$_\text{peak}$, which mimics the enhancement of the ionization parameter and gas temperature. % (see also Fig. \ref{ionizing_spectra}). 

On the other hand, p$_{NT}$ produces much smaller variations in the predicted line ratios. A higher p$_{NT}$ increases the fraction of energetic photons in the X-ray part of the spectrum (Fig. \ref{ionizing_spectra}), including the soft X-rays that are more easily absorbed by the nebula. This causes a more extended partially ionized zone and more emission coming from this region, especially from low-ionization species. However, p$_{NT}$ does not significantly alter the photoionization model predictions in the BPT diagram, with variations of the line ratios typically smaller than $0.1$ dex, even smaller than the effects due to varying gas density in the nebula. In Fig. \ref{BPT_diagram_Thomas} and all subsequent figures, model predictions with different E$_\text{peak}$ are represented with different symbols (circles and squares), while predictions corresponding to different values of p$_{NT}$ are drawn with the same symbol and are almost overlapping. 

We also analyze the predictions of the OXAF models for all the near-infrared diagrams introduced in Section \ref{nearIR_diagnostics_models}. The results are presented in Fig. \ref{nearIR_models_Thomas}, which is the analog of Fig. \ref{nearIR_models}. Similarly to the BPT diagram, also in the near-IR diagnostics the models of \citet{thomas16} produce line ratios that are comparable to those obtained with our default AGN SED, occupying the same parameter space with the same limits on \SIII/\PaG, \SIII/\PaB, \FeII/\PaB, \FeII/\PaA, \PII/\PaB, and \CI/\PaB. If we exclude the influence of metallicity and ionization parameter, E$_\text{peak}$ has the third major effect on the emission line intensities. In particular, choosing models with higher E$_\text{peak}$ increases all the line ratios listed above, moving all the points toward the upper right part of the plot, by factors that depend on the metallicity, ionization parameter, and on the specific diagram. 
In detail, for models with $\log U > -2$, the line ratios on both axis increase by $0.2$-$0.5$ dex in the range $20 < E_\text{peak}/eV < 100$, because of more energetic ionizing photons available at higher E$_\text{peak}$. In the last bin of $\log U$ shown in Fig. \ref{nearIR_models_Thomas} ($\simeq -2$), the \SIII/\PaB\ and \SIII/\PaG\ ratios saturate with E$_\text{peak}$, as higher ionization states of sulfur are populated. On the other hand, the ratios on the x-axis still increase, as those low-ionization species are very sensitive to the higher contribution of soft X-ray photons from models with enhanced E$_\text{peak}$ (see Fig. \ref{ionizing_spectra}), as also discussed in Section \ref{discussion1}. As shown by \citet{thomas16}, a similar mechanism is responsible for the higher \OI\ and \SII\ emission in the optical. 

The influence of p$_{NT}$ on the same near-IR line ratios is in general smaller than that produced by E$_\text{peak}$, with variations $< 0.1$ dex for models with low and intermediate ionization ($\log U < -2$) in all the diagrams in Fig. \ref{nearIR_models_Thomas}, except \CI/\PaB, which can vary by up to $\sim 0.3$ dex from p$_{NT}=0.1$ to $0.4$. 
Also for this parameter, at the highest $\log U$ considered here, it mostly affects the line ratios on the x-axis by the same mechanism explained above for E$_\text{peak}$, with increments of up to $\sim 1$ dex for \CI/\PaB\ and $\sim0.5$ dex for the remaining ratios. We note that \CI/\PaB\ is more sensitive to variations of E$_\text{peak}$ and p$_{NT}$ because it is entirely produced in partially ionized regions where electron collisions stimulated by non-thermal continuum radiation are the main excitation mechanism. 

Finally, as shown by \citet{thomas16}, the third parameter $\Gamma$ has a similar behavior to p$_{NT}$, but produces even smaller variations of the line ratios in all the diagrams analyzed above (both in the optical and near-IR), so we have fixed it to $+2.0$ in the simulations. We note that a more exhaustive discussion of the effects of each parameter can be found in the introductory paper of the OXAF models.  
We also remark that the exact line ratio predictions of AGN models do not affect the maximum starburst separation line (derived from star-forming models), hence they do not alter the results of this paper.


\begin{thebibliography}{}
%\label{bibliography}
%\bibliographystyle{aa}

\bibitem[Acquaviva et al.(2012)]{acquaviva12} Acquaviva, V., Gawiser, E., \& Guaita, L.\ 2012, The Spectral Energy Distribution of Galaxies - SED 2011, 284, 42. doi:10.1017/S1743921312008691
\bibitem[Allen et al.(2008)]{allen08} Allen, M.~G., Groves, B.~A., Dopita, M.~A., et al.\ 2008, \apjs, 178, 20. doi:10.1086/589652
\bibitem[Antonucci(1993)]{antonucci93} Antonucci, R.\ 1993, \araa, 31, 473. doi:10.1146/annurev.aa.31.090193.002353
\bibitem[Arrabal Haro et al.(2023)]{arrabalharo23a} Arrabal Haro, P., Dickinson, M., Kartaltepe, J., et al.\ 2023, \aas
\bibitem[Arrabal Haro et al.(2023)]{arrabalharo23b} Arrabal Haro, P., Dickinson, M., Finkelstein, S.~L., et al.\ 2023, arXiv:2304.05378. doi:10.48550/arXiv.2304.05378
\bibitem[Backhaus et al.(2022)]{backhaus22} Backhaus, B.~E., Trump, J.~R., Cleri, N.~J., et al.\ 2022, \apj, 926, 161. doi:10.3847/1538-4357/ac3919
\bibitem[Baldwin et al.(1981)]{baldwin81} Baldwin, J.~A., Phillips, M.~M., \& Terlevich, R.\ 1981, \pasp, 93, 5. doi:10.1086/130766
\bibitem[Bian et al.(2018)]{bian18} Bian, F., Kewley, L.~J., \& Dopita, M.~A.\ 2018, \apj, 859, 175. doi:10.3847/1538-4357/aabd74
\bibitem[Bouchet et al.(1985)]{bouchet85} Bouchet, P., Lequeux, J., Maurice, E., et al.\ 1985, \aap, 149, 330
\bibitem[Brammer et al.(2012)]{brammer12} Brammer, G.~B., van Dokkum, P.~G., Franx, M., et al.\ 2012, \apjs, 200, 13. doi:10.1088/0067-0049/200/2/13
%\bibitem[Bressan et al.(2012)]{bressan12} Bressan, A., Marigo, P., Girardi, L., et al.\ 2012, \mnras, 427, 127. doi:10.1111/j.1365-2966.2012.21948.x
\bibitem[Brinchmann et al.(2008)]{brinchmann08} Brinchmann, J., Pettini, M., \& Charlot, S.\ 2008, \mnras, 385, 769. doi:10.1111/j.1365-2966.2008.12914.x
\bibitem[Buchner et al.(2015)]{buchner15} Buchner, J., Georgakakis, A., Nandra, K., et al.\ 2015, \apj, 802, 89. doi:10.1088/0004-637X/802/2/89
\bibitem[Bushouse et al.(2022)]{bushouse22} Bushouse, H., Eisenhamer, J., Dencheva, N., et al.\ 2022, Zenodo
\bibitem[Caffau et al.(2011)]{caffau11} Caffau, E., Bonifacio, P., Faraggiana, R., et al.\ 2011, \aap, 532, A98. doi:10.1051/0004-6361/201117313
\bibitem[Calabr{\`o} et al.(2018)]{calabro18} Calabr{\`o}, A., Daddi, E., Cassata, P., et al.\ 2018, \apjl, 862, L22. doi:10.3847/2041-8213/aad33e
\bibitem[Calabr{\`o} et al.(2019)]{calabro19} Calabr{\`o}, A., Daddi, E., Puglisi, A., et al.\ 2019, \aap, 623, A64. doi:10.1051/0004-6361/201834522
\bibitem[Calzetti et al.(1996)]{calzetti96} Calzetti, D., Kinney, A.~L., \& Storchi-Bergmann, T.\ 1996, \apj, 458, 132. doi:10.1086/176797
\bibitem[Calzetti et al.(2000)]{calzetti00} Calzetti, D., Armus, L., Bohlin, R.~C., et al.\ 2000, \apj, 533, 682. doi:10.1086/308692
\bibitem[Chabrier(2003)]{chabrier03} Chabrier, G.\ 2003, \pasp, 115, 763. doi:10.1086/376392
\bibitem[Cirasuolo et al.(2020)]{cirasuolo20} Cirasuolo, M., Fairley, A., Rees, P., et al.\ 2020, The Messenger, 180, 10. doi:10.18727/0722-6691/5195
\bibitem[Coil et al.(2015)]{coil15} Coil, A.~L., Aird, J., Reddy, N., et al.\ 2015, \apj, 801, 35. doi:10.1088/0004-637X/801/1/35
\bibitem[Colina et al.(2015)]{colina15} Colina, L., Piqueras L{\'o}pez, J., Arribas, S., et al.\ 2015, \aap, 578, A48. doi:10.1051/0004-6361/201425567
\bibitem[Costille et al.(2016)]{costille16} Costille, A., Caillat, A., Rossin, C., et al.\ 2016, \procspie, 9912, 99122C. doi:10.1117/12.2231420
\bibitem[Cullen et al.(2021)]{cullen21} Cullen, F., Shapley, A.~E., McLure, R.~J., et al.\ 2021, \mnras, 505, 903. doi:10.1093/mnras/stab1340
\bibitem[Curti et al.(2022)]{curti22} Curti, M., Hayden-Pawson, C., Maiolino, R., et al.\ 2022, \mnras, 512, 4136. doi:10.1093/mnras/stac544
\bibitem[Curti et al.(2020)]{curti20} Curti, M., Mannucci, F., Cresci, G., et al.\ 2020, \mnras, 491, 944. doi:10.1093/mnras/stz2910
\bibitem[den Brok et al.(2022)]{denbrok22} den Brok, J.~S., Koss, M.~J., Trakhtenbrot, B., et al.\ 2022, \apjs, 261, 7. doi:10.3847/1538-4365/ac5b66
\bibitem[Dom{\'\i}nguez et al.(2013)]{dominguez13} Dom{\'\i}nguez, A., Siana, B., Henry, A.~L., et al.\ 2013, \apj, 763, 145. doi:10.1088/0004-637X/763/2/145
\bibitem[Dopita et al.(2006)]{dopita06} Dopita, M.~A., Fischera, J., Sutherland, R.~S., et al.\ 2006, \apjs, 167, 177. doi:10.1086/508261
\bibitem[Dors et al.(2022)]{dors22} Dors, O.~L., Valerdi, M., Freitas-Lemes, P., et al.\ 2022, \mnras, 514, 5506. doi:10.1093/mnras/stac1722
\bibitem[Eldridge et al.(2017)]{eldridge17} Eldridge, J.~J., Stanway, E.~R., Xiao, L., et al.\ 2017, \pasa, 34, e058. doi:10.1017/pasa.2017.51
\bibitem[Elitzur et al.(2014)]{elitzur14} Elitzur, M., Ho, L.~C., \& Trump, J.~R.\ 2014, \mnras, 438, 3340. doi:10.1093/mnras/stt2445
\bibitem[Euclid Collaboration et al.(2022)]{scaramella22} Euclid Collaboration, Scaramella, R., Amiaux, J., et al.\ 2022, \aap, 662, A112. doi:10.1051/0004-6361/202141938
\bibitem[Faisst et al.(2018)]{faisst18} Faisst, A.~L., Masters, D., Wang, Y., et al.\ 2018, \apj, 855, 132. doi:10.3847/1538-4357/aab1fc
\bibitem[Feltre et al.(2016)]{feltre16} Feltre, A., Charlot, S., \& Gutkin, J.\ 2016, \mnras, 456, 3354. doi:10.1093/mnras/stv2794
\bibitem[Ferland et al.(2017)]{ferland17} Ferland, G.~J., Chatzikos, M., Guzm{\'a}n, F., et al.\ 2017, \rmxaa, 53, 385. doi:10.48550/arXiv.1705.10877
\bibitem[Ferland et al.(2009)]{ferland09} Ferland, G.~J., Hu, C., Wang, J.-M., et al.\ 2009, \apjl, 707, L82. doi:10.1088/0004-637X/707/1/L82
\bibitem[Fern{\'a}ndez et al.(2023)]{fernandez23} Fern{\'a}ndez, V., Amor{\'\i}n, R., Sanchez-Janssen, R., et al.\ 2023, \mnras, 520, 3576. doi:10.1093/mnras/stad198
\bibitem[Ferruit et al.(2022)]{ferruit22} Ferruit, P., Jakobsen, P., Giardino, G., et al.\ 2022, \aap, 661, A81. doi:10.1051/0004-6361/202142673
\bibitem[Finkelstein et al.(2023)]{finkelstein23} Finkelstein, S.~L., Bagley, M.~B., Ferguson, H.~C., et al.\ 2023, \apjl, 946, L13. doi:10.3847/2041-8213/acade4
\bibitem[Gardner et al.(2006)]{gardner06} Gardner, J.~P., Mather, J.~C., Clampin, M., et al.\ 2006, \ssr, 123, 485. doi:10.1007/s11214-006-8315-7
\bibitem[Gardner et al.(2023)]{gardner23} Gardner, J.~P., Mather, J.~C., Abbott, R., et al.\ 2023, arXiv:2304.04869. doi:10.48550/arXiv.2304.04869
\bibitem[Gaskell et al.(2022)]{gaskell22} Gaskell, M., Thakur, N., Tian, B., et al.\ 2022, Astronomische Nachrichten, 343, e210112. doi:10.1002/asna.20210112
\bibitem[Gehrels(1986)]{gehrels86} Gehrels, N.\ 1986, \apj, 303, 336. doi:10.1086/164079
\bibitem[Goddard et al.(2017)]{goddard17} Goddard, D., Thomas, D., Maraston, C., et al.\ 2017, \mnras, 466, 4731. doi:10.1093/mnras/stw3371
\bibitem[Greener et al.(2020)]{greener20} Greener, M.~J., Arag{\'o}n-Salamanca, A., Merrifield, M.~R., et al.\ 2020, \mnras, 495, 2305. doi:10.1093/mnras/staa1300
\bibitem[Grogin et al.(2011)]{grogin11} Grogin, N.~A., Kocevski, D.~D., Faber, S.~M., et al.\ 2011, \apjs, 197, 35. doi:10.1088/0067-0049/197/2/35
\bibitem[Groves et al.(2004)]{groves04} Groves, B.~A., Dopita, M.~A., \& Sutherland, R.~S.\ 2004, \apjs, 153, 75. doi:10.1086/421114
\bibitem[Hirschmann et al.(2017)]{hirschmann17} Hirschmann, M., Charlot, S., Feltre, A., et al.\ 2017, \mnras, 472, 2468. doi:10.1093/mnras/stx2180
\bibitem[Jakobsen et al.(2022)]{jakobsen22} Jakobsen, P., Ferruit, P., Alves de Oliveira, C., et al.\ 2022, \aap, 661, A80. doi:10.1051/0004-6361/202142663
\bibitem[Jenkins(2009)]{jenkins09} Jenkins, E.~B.\ 2009, \apj, 700, 1299. doi:10.1088/0004-637X/700/2/1299
\bibitem[Ji et al.(2020)]{ji20} Ji, X., Yan, R., Riffel, R., et al.\ 2020, \mnras, 496, 1262. doi:10.1093/mnras/staa1521
\bibitem[Jin et al.(2018)]{jin18} Jin, S., Daddi, E., Liu, D., et al.\ 2018, \apj, 864, 56. doi:10.3847/1538-4357/aad4af
\bibitem[Kartaltepe et al.(2015)]{kartaltepe15} Kartaltepe, J.~S., Sanders, D.~B., Silverman, J.~D., et al.\ 2015, \apjl, 806, L35. doi:10.1088/2041-8205/806/2/L35
\bibitem[Kashino et al.(2019)]{kashino19} Kashino, D., Silverman, J.~D., Sanders, D., et al.\ 2019, \apjs, 241, 10. doi:10.3847/1538-4365/ab06c4
\bibitem[Kauffmann et al.(2003)]{kauffmann03} Kauffmann, G., Heckman, T.~M., Tremonti, C., et al.\ 2003, \mnras, 346, 1055. doi:10.1111/j.1365-2966.2003.07154.x
\bibitem[Kewley \& Dopita(2002)]{kewley02} Kewley, L.~J. \& Dopita, M.~A.\ 2002, \apjs, 142, 35. doi:10.1086/341326
\bibitem[Kewley et al.(2001)]{kewley01} Kewley, L.~J., Dopita, M.~A., Sutherland, R.~S., et al.\ 2001, \apj, 556, 121. doi:10.1086/321545
\bibitem[Kewley et al.(2019)]{kewley19} Kewley, L.~J., Nicholls, D.~C., \& Sutherland, R.~S.\ 2019, \araa, 57, 511. doi:10.1146/annurev-astro-081817-051832
\bibitem[Kewley et al.(2013)]{kewley13} Kewley, L.~J., Maier, C., Yabe, K., et al.\ 2013, \apjl, 774, L10. doi:10.1088/2041-8205/774/1/L10
\bibitem[Kodra et al.(2023)]{kodra23} Kodra, D., Andrews, B.~H., Newman, J.~A., et al.\ 2023, \apj, 942, 36. doi:10.3847/1538-4357/ac9f12
\bibitem[Koekemoer et al.(2011)]{koekemoer11} Koekemoer, A.~M., Faber, S.~M., Ferguson, H.~C., et al.\ 2011, \apjs, 197, 36. doi:10.1088/0067-0049/197/2/36
\bibitem[Kravchenko et al.(2022)]{kravchenko22} Kravchenko, K., Dallilar, Y., Absil, O., et al.\ 2022, \procspie, 12184, 121845M. doi:10.1117/12.2629258
\bibitem[LaMassa et al.(2019)]{lamassa19} LaMassa, S.~M., Georgakakis, A., Vivek, M., et al.\ 2019, \apj, 876, 50. doi:10.3847/1538-4357/ab108b
\bibitem[Lamperti et al.(2017)]{lamperti17} Lamperti, I., Koss, M., Trakhtenbrot, B., et al.\ 2017, \mnras, 467, 540. doi:10.1093/mnras/stx055
\bibitem[Le F{\`e}vre et al.(2003)]{lefevre03} Le F{\`e}vre, O., Saisse, M., Mancini, D., et al.\ 2003, \procspie, 4841, 1670. doi:10.1117/12.460959
\bibitem[Lester et al.(1990)]{lester90} Lester, D.~F., Carr, J.~S., Joy, M., et al.\ 1990, \apj, 352, 544. doi:10.1086/168557
\bibitem[Lilly et al.(2007)]{lilly07} Lilly, S.~J., Le F{\`e}vre, O., Renzini, A., et al.\ 2007, \apjs, 172, 70. doi:10.1086/516589
\bibitem[Maiolino et al.(2020)]{maiolino20} Maiolino, R., Cirasuolo, M., Afonso, J., et al.\ 2020, The Messenger, 180, 24. doi:10.18727/0722-6691/5197
\bibitem[Markwardt(2009)]{markwardt09} Markwardt, C.~B.\ 2009, Astronomical Data Analysis Software and Systems XVIII, 411, 251. doi:10.48550/arXiv.0902.2850
\bibitem[Masters et al.(2016)]{masters16} Masters, D., Faisst, A., \& Capak, P.\ 2016, \apj, 828, 18. doi:10.3847/0004-637X/828/1/18
\bibitem[Mathis et al.(1977)]{mathis77} Mathis, J.~S., Rumpl, W., \& Nordsieck, K.~H.\ 1977, \apj, 217, 425. doi:10.1086/155591
\bibitem[Matsuoka et al.(2009)]{matsuoka09} Matsuoka, K., Nagao, T., Maiolino, R., et al.\ 2009, \aap, 503, 721. doi:10.1051/0004-6361/200811478
\bibitem[Mignoli et al.(2013)]{mignoli13} Mignoli, M., Vignali, C., Gilli, R., et al.\ 2013, \aap, 556, A29. doi:10.1051/0004-6361/201220846
\bibitem[Mignoli et al.(2019)]{mignoli19} Mignoli, M., Feltre, A., Bongiorno, A., et al.\ 2019, \aap, 626, A9. doi:10.1051/0004-6361/201935062
\bibitem[Morisset et al.(2015)]{morisset15} Morisset, C., Delgado-Inglada, G., \& Flores-Fajardo, N.\ 2015, \rmxaa, 51, 103. doi:10.48550/arXiv.1412.5349
\bibitem[Mouri et al.(1990)]{mouri90} Mouri, H., Nishida, M., Taniguchi, Y., et al.\ 1990, \apj, 360, 55. doi:10.1086/169095
\bibitem[Nagao et al.(2006)]{nagao06} Nagao, T., Maiolino, R., \& Marconi, A.\ 2006, \aap, 447, 863. doi:10.1051/0004-6361:20054127
\bibitem[Nandra et al.(2015)]{nandra15} Nandra, K., Laird, E.~S., Aird, J.~A., et al.\ 2015, \apjs, 220, 10. doi:10.1088/0067-0049/220/1/10
\bibitem[Noeske et al.(2007)]{noeske07} Noeske, K.~G., Weiner, B.~J., Faber, S.~M., et al.\ 2007, \apjl, 660, L43. doi:10.1086/517926
\bibitem[Oesch et al.(2023)]{oesch23} Oesch, P.~A., Brammer, G., Naidu, R.~P., et al.\ 2023, arXiv:2304.02026. doi:10.48550/arXiv.2304.02026
\bibitem[Oliva et al.(2001)]{oliva01} Oliva, E., Marconi, A., Maiolino, R., et al.\ 2001, \aap, 369, L5. doi:10.1051/0004-6361:20010214
\bibitem[Onori et al.(2017)]{onori17} Onori, F., La Franca, F., Ricci, F., et al.\ 2017, \mnras, 464, 1783. doi:10.1093/mnras/stw2368
\bibitem[Osterbrock(1989)]{osterbrock89} Osterbrock, D.~E.\ 1989, Astrophysics of Gaseous Nebulae and Active Galactic Nuclei, by Donald E. Osterbrock. Published by University Science Books, ISBN 0-935702-22-9, 408pp, 1989.
\bibitem[Pannella et al.(2015)]{pannella15} Pannella, M., Elbaz, D., Daddi, E., et al.\ 2015, \apj, 807, 141. doi:10.1088/0004-637X/807/2/141
\bibitem[Rayner et al.(2003)]{rayner03} Rayner, J.~T., Toomey, D.~W., Onaka, P.~M., et al.\ 2003, \pasp, 115, 362. doi:10.1086/367745
\bibitem[Ricci et al.(2022)]{ricci22} Ricci, C., Ananna, T.~T., Temple, M.~J., et al.\ 2022, \apj, 938, 67. doi:10.3847/1538-4357/ac8e67
\bibitem[Richardson et al.(2014)]{richardson14} Richardson, C.~T., Allen, J.~T., Baldwin, J.~A., et al.\ 2014, \mnras, 437, 2376. doi:10.1093/mnras/stt2056
\bibitem[Riffel et al.(2006)]{riffel06} Riffel, R., Rodr{\'\i}guez-Ardila, A., \& Pastoriza, M.~G.\ 2006, \aap, 457, 61. doi:10.1051/0004-6361:20065291
\bibitem[Riffel et al.(2013)]{riffel13} Riffel, R., Rodr{\'\i}guez-Ardila, A., Aleman, I., et al.\ 2013, \mnras, 430, 2002. doi:10.1093/mnras/stt026
\bibitem[Riffel et al.(2019)]{riffel19} Riffel, R., Rodr{\'\i}guez-Ardila, A., Brotherton, M.~S., et al.\ 2019, \mnras, 486, 3228. doi:10.1093/mnras/stz1077
\bibitem[Risaliti et al.(2000)]{risaliti00} Risaliti, G., Maiolino, R., \& Bassani, L.\ 2000, \aap, 356, 33. doi:10.48550/arXiv.astro-ph/0002169
\bibitem[Rodr{\'\i}guez-Ardila et al.(2004)]{rodriguezardila04} Rodr{\'\i}guez-Ardila, A., Pastoriza, M.~G., Viegas, S., et al.\ 2004, \aap, 425, 457. doi:10.1051/0004-6361:20034285
\bibitem[Rodr{\'\i}guez-Ardila et al.(2005)]{rodriguezardila05} Rodr{\'\i}guez-Ardila, A., Riffel, R., \& Pastoriza, M.~G.\ 2005, \mnras, 364, 1041. doi:10.1111/j.1365-2966.2005.09638.x
\bibitem[Sanders et al.(2020)]{sanders20} Sanders, R.~L., Shapley, A.~E., Reddy, N.~A., et al.\ 2020, \mnras, 491, 1427. doi:10.1093/mnras/stz3032
\bibitem[Sargent et al.(2014)]{sargent14} Sargent, M.~T., Daddi, E., B{\'e}thermin, M., et al.\ 2014, \apj, 793, 19. doi:10.1088/0004-637X/793/1/19
\bibitem[Savage \& Sembach(1996)]{savage96} Savage, B.~D. \& Sembach, K.~R.\ 1996, \apj, 470, 893. doi:10.1086/177919
\bibitem[Schreiber et al.(2015)]{schreiber15} Schreiber, C., Pannella, M., Elbaz, D., et al.\ 2015, \aap, 575, A74. doi:10.1051/0004-6361/201425017
\bibitem[Seifert et al.(2003)]{seifert03} Seifert, W., Appenzeller, I., Baumeister, H., et al.\ 2003, \procspie, 4841, 962. doi:10.1117/12.459494
\bibitem[Shapley et al.(2015)]{shapley15} Shapley, A.~E., Reddy, N.~A., Kriek, M., et al.\ 2015, \apj, 801, 88. doi:10.1088/0004-637X/801/2/88
\bibitem[Shields et al.(2010)]{shields10} Shields, G.~A., Ludwig, R.~R., \& Salviander, S.\ 2010, \apj, 721, 1835. doi:10.1088/0004-637X/721/2/1835
\bibitem[Simcoe et al.(2013)]{simcoe13} Simcoe, R.~A., Burgasser, A.~J., Schechter, P.~L., et al.\ 2013, \pasp, 125, 270. doi:10.1086/670241
\bibitem[Stefanon et al.(2017)]{stefanon17} Stefanon, M., Yan, H., Mobasher, B., et al.\ 2017, \apjs, 229, 32. doi:10.3847/1538-4365/aa66cb
\bibitem[Steidel et al.(2016)]{steidel16} Steidel, C.~C., Strom, A.~L., Pettini, M., et al.\ 2016, \apj, 826, 159. doi:10.3847/0004-637X/826/2/159
\bibitem[Storchi-Bergmann et al.(2009)]{storchi-bergmann09} Storchi-Bergmann, T., McGregor, P.~J., Riffel, R.~A., et al.\ 2009, \mnras, 394, 1148. doi:10.1111/j.1365-2966.2009.14388.x
\bibitem[Storey \& Hummer(1988)]{storey88} Storey, P.~J. \& Hummer, D.~G.\ 1988, \mnras, 231, 1139. doi:10.1093/mnras/231.4.1139
\bibitem[Sutherland et al.(2018)]{sutherland18} Sutherland, R., Dopita, M., Binette, L., et al.\ 2018, Astrophysics Source Code Library. ascl:1807.005
\bibitem[Takada et al.(2014)]{takada14} Takada, M., Ellis, R.~S., Chiba, M., et al.\ 2014, \pasj, 66, R1. doi:10.1093/pasj/pst019
\bibitem[Thomas et al.(2016)]{thomas16} Thomas, A.~D., Groves, B.~A., Sutherland, R.~S., et al.\ 2016, \apj, 833, 266. doi:10.3847/1538-4357/833/2/266
\bibitem[Thomas et al.(2018)]{thomas18} Thomas, A.~D., Dopita, M.~A., Kewley, L.~J., et al.\ 2018, \apj, 856, 89. doi:10.3847/1538-4357/aab3db
\bibitem[Topping et al.(2020)]{topping20} Topping, M.~W., Shapley, A.~E., Reddy, N.~A., et al.\ 2020, \mnras, 495, 4430. doi:10.1093/mnras/staa1410
\bibitem[Trump et al.(2011)]{trump11} Trump, J.~R., Weiner, B.~J., Scarlata, C., et al.\ 2011, \apj, 743, 144. doi:10.1088/0004-637X/743/2/144
\bibitem[Veilleux(2002)]{veilleux02} Veilleux, S.\ 2002, IAU Colloq. 184: AGN Surveys, 284, 111. doi:10.48550/arXiv.astro-ph/0201118
\bibitem[Veilleux \& Osterbrock(1987)]{veilleux87} Veilleux, S. \& Osterbrock, D.~E.\ 1987, \apjs, 63, 295. doi:10.1086/191166
\bibitem[Vito et al.(2018)]{vito18} Vito, F., Brandt, W.~N., Yang, G., et al.\ 2018, \mnras, 473, 2378. doi:10.1093/mnras/stx2486
\bibitem[Vladilo et al.(2011)]{vladilo11} Vladilo, G., Abate, C., Yin, J., et al.\ 2011, \aap, 530, A33. doi:10.1051/0004-6361/201016330
\bibitem[Wang et al.(2017)]{wang17} Wang, W., Faber, S.~M., Liu, F.~S., et al.\ 2017, \mnras, 469, 4063. doi:10.1093/mnras/stx1148
\bibitem[Whitaker et al.(2014)]{whitaker14} Whitaker, K.~E., Franx, M., Leja, J., et al.\ 2014, \apj, 795, 104. doi:10.1088/0004-637X/795/2/104
\bibitem[Wuyts et al.(2016)]{wuyts16} Wuyts, E., Wisnioski, E., Fossati, M., et al.\ 2016, \apj, 827, 74. doi:10.3847/0004-637X/827/1/74
\bibitem[Yung et al.(2021)]{yung21} Yung, L.~Y.~A., Somerville, R.~S., Finkelstein, S.~L., et al.\ 2021, \mnras, 508, 2706. doi:10.1093/mnras/stab2761



























%ENDREF
\end{thebibliography}
\end{document}